\documentclass[aps,preprint,nofootinbib,floatfix]{revtex4-1}
\usepackage{graphicx,color,hyperref}
\usepackage{amsmath,amssymb}
\usepackage{url}
\usepackage{epstopdf}
\newcommand{\lsim}{\mathrel{\mathop{\kern 0pt \rlap
  {\raise.2ex\hbox{$<$}}}
  \lower.9ex\hbox{\kern-.190em $\sim$}}}
\newcommand{\gsim}{\mathrel{\mathop{\kern 0pt \rlap
  {\raise.2ex\hbox{$>$}}}
  \lower.9ex\hbox{\kern-.190em $\sim$}}}

\begin{document}

\title{Global Constraints on Effective Dark Matter Interactions: \\
 Relic Density, Direct Detection, Indirect Detection, and Collider}

\author{Kingman Cheung$^{1,2}$, Po-Yan Tseng$^{2,3}$, Yue-Lin S. Tsai$^{4}$
  and Tzu-Chiang Yuan$^5$}

\affiliation{
$^1$Division of Quantum Phases \& Devices, School of Physics, 
Konkuk University, Seoul 143-701, Republic of Korea \\
$^2$Department of Physics, National Tsing Hua University, 
Hsinchu 300, Taiwan\\
$^3$Department of Physics, University of Wisconsin, Madison, WI 53706, USA\\
$^4$National Centre for Nuclear Research, Hoza 69, 00-681 Warsaw, Poland \\
$^5$Institute of Physics, Academia Sinica, Nankang, Taipei 11529, Taiwan
}

\date{\today}

\begin{abstract}
  An effective interaction approach is used to describe the
  interactions between the spin 0 or spin 1/2 dark matter particle and
  the degrees of freedom of the standard model. 
 This approach is applicable to those models in which the dark matter
particles do not experience the standard-model interactions, e.g., 
hidden-sector models.
 We explore the
  effects of these effective interaction operators on (i) dark matter
  relic density, (ii) spin-independent and spin-dependent dark
  matter-nucleon scattering cross sections, (iii) cosmic antiproton
  and gamma ray fluxes from the galactic halo due to dark matter
  annihilation, and (iv) monojet and monophoton production plus
  missing energy at the Tevatron and the Large Hadron Collider (LHC).
  We combine the experimental data of relic density from WMAP7,
  spin-independent cross section from XENON100, spin-dependent cross
  section from XENON10, ZEPLIN-III, and SIMPLE, cosmic antiproton flux
  from PAMELA, cosmic gamma-ray flux from ${\it Fermi}$-LAT, and the
  monojet and monophoton data from the Tevatron and the LHC, to put
  the most comprehensive limits on each effective operator.
\end{abstract}

\maketitle
\section{Introduction}
The presence of cold dark matter (CDM) in our Universe is now well established
by a number of observational experiments, especially the very precise 
measurement of the cosmic microwave background radiation
in the Wilkinson Microwave Anisotropy Probe (WMAP) experiment \cite{wmap}.
The measured value of the CDM relic density is
\begin{equation}
\label{wmap}
 \Omega_{\rm CDM}\, h^2 = 0.1126 \;\pm 0.0036 \;,
\end{equation}
where $h$ is the Hubble constant in units of $100$ km/Mpc/s.
Though the gravitation nature of the dark matter (DM) is 
commonly believed to be well established, 
its particle nature remains alluring except that it is
nonbaryonic and to a high extent electrically neutral. 

One of the most appealing and natural CDM particle candidates is the
{\it weakly-interacting massive particle} (WIMP).  
If the dark matter, generically denoted by $\chi$ here, is thermally 
produced in the early Universe, the required annihilation cross section 
is right at the order of weak interaction.  There may be a dynamical 
connection between the dark matter and weak-scale physics.
The relation between the fractional relic
density of $\chi$ relative to the critical density and its thermal
annihilation cross section can be given by the following simple
formula \cite{hooper}
\begin{equation}
\label{rate}
\Omega_\chi h^2 \simeq \frac{ 0.1 \;{\rm pb} }{\langle \sigma v \rangle} \;,
\end{equation}
with $\langle \sigma v \rangle$ being the annihilation cross section
of the dark matter around the time of freeze-out, at which the
annihilation rate could no longer catch up with the Hubble expansion
rate of the Universe.  Assuming the measured $\Omega_{\rm CDM} h^2$ to
be saturated by a single component WIMP, its annihilation cross
section should be about $1$ pb or $3\times 10^{-26}\;{\rm cm}^3 \,
{\rm s}^{-1}$.  This is exactly the size of the cross section that one
expects from a weak interaction process, which implies an appreciable
size of production rate of the WIMP at the Large Hadron Collider (LHC)
as well as the event rates for direct and indirect searches that reach
the sensitivities of dark matter experiments like XENON100 \cite{xenon} and
$Fermi$-LAT \cite{fermi-up,fermi-lat} respectively.

There have been many proposed candidates for the dark matter.  Without
committing to any particular DM model so as to perform a model
independent analysis, we adopt an effective interaction approach to
describe the interactions of the dark matter particle with the
standard model (SM) particles.  Recently, there have been a number of
works in this approach that deals with different observable signals in
various experiments \cite{cao,lhc-1,lhc-2,lhc-25,lhc-3,lhc-4,lhc-5}, 
\cite{pbar-1,pbar-2,pbar-3,pbar-4},
\cite{gamma-1,gamma-2,gamma-3,gamma-4,gamma-5,gamma-6,gamma-7,gamma-8},
\cite{fan,effective-1,effective-2,ding-liao}.
One simple realization of the effective interaction approach is that
the dark matter particle exists in a hidden sector, which communicates
to the SM sector via a heavy degree of freedom in the connector
sector. At energy scale well below this heavy mediator the
interactions can be conveniently described by a set of effective
interactions.  The strength of each interaction depends on the nature
of the dark matter particle and the mediator.  
A few models that can give rise to some of the operators in this
analysis are described in the appendix.  However, note that some
popular dark matter models, such as supersymmetry, cannot be correctly
described by this effective-interaction appraoch, because the dark
matter particles themselves also experience the SM interactions.
In this work, we will consider various spin nature of the dark matter
particle including Dirac and Majorana for fermionic dark matter,
as well as real and complex scalar.
The most important set of interactions among the fermionic dark matter 
$\chi$ and the light fermions $f$ are described by the effective 
operators $(\bar \chi \Gamma \chi)(\bar f \Gamma^\prime f)$, where $\Gamma$ and
$\Gamma^\prime$ are general Dirac matrices contracted with appropriate
Lorentz indices. 
We will discuss these and other operators in more
details in the next section. 

One of the most anticipated signals of dark matter at hadronic
colliders is a large missing energy in association with jets, photons,
or leptons, such as monojet and monophoton plus large missing-energy
signatures.  For example, if we take one of the operators, $(\bar \chi
\chi)(\bar q q)$, and attach a gluon or a photon to a quark leg, it
will give rise to a monojet or a monophoton plus missing energy
event. The Tevatron experiments and the LHC experiments have been
actively searching for these signatures in some other context, such as
large extra dimensions \cite{exd-review}.  We will use the most
updated data on monojet and monophoton production from the LHC
\cite{atlas} and the Tevatron \cite{cdf,d01,d02} to constrain each
effective operator.  It turns out that the limits from the LHC and
Tevatron are comparable to those obtained from indirect detection data
(PAMELA \cite{pamela-e,pamela-p} and ${\it Fermi}$-LAT
\cite{fermi-up,fermi-lat}), but inferior to those obtained from direct
detection data (XENON100 \cite{xenon} and CDMS \cite{cdms}) if a
particular operator contributes to spin-independent cross sections.
Some recent works in this direction have been in
Refs. ~\cite{cao,lhc-1,lhc-2,lhc-25,lhc-3,lhc-4,lhc-5,pbar-1,pbar-2,pbar-3,pbar-4,gamma-1,gamma-2,gamma-3,gamma-4,gamma-5,gamma-6,gamma-7,gamma-8,fan,effective-1,effective-2,ding-liao}.

Dark matter annihilation in the galactic halo gives rise to a number of
observable signals, including excess in positron flux, 
antiproton flux, and gamma-ray
over the corresponding cosmic backgrounds. The most current positron flux
and antiproton flux data come from PAMELA \cite{pamela-e,pamela-p}.  
The positron-fraction spectrum showed an uprising trend up to about 100 GeV
\cite{pamela-e}.
\footnote
{A very recent result from ${\it Fermi}$-LAT \cite{fermi-up} showed that the 
uprising trend continues to about 150 GeV.}
Nevertheless, the antiproton flux is consistent with the expected
cosmic background \cite{pamela-p}.  The effective operators such as
$(\bar\chi \chi) ( \bar qq)$ can give rise to dark matter annihilation into
light quarks, which will eventually hadronize into antiprotons.
In Refs.~\cite{pbar-1,pbar-2,pbar-3,pbar-4} the effects 
of dark matter annihilation on positron flux and antiproton flux were
studied. It was shown that the antiproton flux data can give a better 
constraint on the effective dark matter interactions 
than the positron flux data \cite{pbar-1}.  
So in this work we focus on antiproton flux
data when we use antimatter search experiments to constrain the
effective dark matter interactions. 

Another powerful set of indirect detection data comes from the gamma ray
due to dark matter annihilation in the galactic halo.  The data from
the extragalactic sources contain large uncertainties, such that we 
concentrate on the galactic data in this work.  Currently, the best data
come from the ${\it Fermi}$-LAT experiment \cite{fermi-lat}.  
It detects gamma rays in sub-GeV 
region to hundreds of GeV from all directions 
$(0^\circ < |b| < 90^\circ, \; 0^\circ < l < 360^\circ )$, 
i.e., including Galactic Center (GC) $(0^\circ < |b| < 10^\circ)$, 
low-latitude $(10^\circ < |b| < 20^\circ)$, 
mid-latitude $(20^\circ < |b| < 60^\circ)$, and 
high latitude $(|b| > 60^\circ)$.
The data on the photon spectrum from the low-latitude 
$(10^\circ < |b| < 20^\circ, \; 0^\circ < l < 360^\circ )$ \cite{fermi-lat}  
recorded by the $Fermi$-LAT indicated a continuous spectrum and mostly 
consistent
with the known backgrounds.  We can therefore use the data to constrain
on additional sources of gamma-ray, namely, the annihilation of the
dark matter into quarks, followed by fragmentation into neutral pions,
which further decay into photons.  
Some recent works in using the ${\it Fermi}$-LAT to constrain
various models or effective dark matter interactions can be found in
Refs.~\cite{gamma-1,gamma-2,gamma-3,gamma-4,gamma-5,gamma-6,gamma-7,gamma-8}.

Another important method of detecting DM is through the direct
collision between the DM particles in the halo with the nuclei of the
detecting materials.  The DM particles then lose a fraction of the
kinetic energy to the nuclei, which can be detected by a phonon-type
or scintillation-type or ionization-type signal or some combinations 
of these types.  
Since the energy transfer is only of
order $O(10-100)$ keV and the event rate is extremely low, an almost
background-free environment is needed. The most recent result comes
from the XENON100 Collaboration \cite{xenon}, which did not see any signal
events and obtained limits on the spin-independent (SI) cross sections
versus the DM mass.  The 90\% CL upper limit on $\sigma_{\rm SI} \sim 10^{-45} \,
{\rm cm}^2$ for $m_{\chi} = 50$ GeV. We are going to use the limits
presented in Ref.~\cite{xenon} to constrain the effective operators.

In this work, we are going to constrain each operator from the combined
data sets on relic density (WMAP), direct detection (XENON, ZEPLIN and SIMPLE), 
cosmic antiproton flux (PAMELA), cosmic gamma-ray flux (${\it Fermi}$-LAT), 
and monojet and monophoton production (Tevatron and LHC).
The organization of this work is as follows. We describe the set of
effective operators for fermionic and scalar DM and describe their
nonrelativistic limits in Sec.~II. In Sec.~III, we calculate the relic density
assuming production of DM from the thermal equilibrium and by solving
the Boltzmann equation. In Sec.~IV, we calculate both the spin-independent
(SI) and spin-dependent (SD) cross sections and constrain each operator by
the XENON100 \cite{xenon}, XENON10 \cite{xenon-sd},
ZEPLIN \cite{zeplin}, and SIMPLE \cite{simple} data. 
In Sec.~V and VI, we calculate the cosmic antiproton
and gamma-ray flux, respectively, and constrain each operator from the most
current data. In Sec.~VII, we calculate 
monojet and monophoton production at the Tevatron and LHC, and use 
the most current data to constrain each operator. 
In Sec. VIII, we perform a combined analysis by adding the chi-square
of each data set.  We conclude in Sec. IX.

Before we close this introduction section we would spell out the improvements
that are achieved in this work.
\begin{enumerate}
\item
The constraint from the LHC monojet and monophoton production was only recently 
done in Refs.~\cite{lhc-4,lhc-5}.  We perform an independent analysis here.
\footnote
{
There are recent works that consider light mediators between the SM 
fermions and the dark matter \cite{ian1},  the unitarity 
bound of the operators \cite{ian2}, and light dark matter \cite{yann}.
}

\item We perform a full calculation of the relic density by solving 
the Boltzmann equation.

\item In the spin-dependent cross sections, we include data sets
from XENON10 \cite{xenon-sd}, ZEPLIN \cite{zeplin}, and 
SIMPLE \cite{simple}.

\item We combine all data sets in the combined analysis.  The
resulting limits will be the most stringent so far.
\footnote
{
A generalized analysis of WIMP in nonrelativistic limit can be found in
Ref.~\cite{marc}.
}

\end{enumerate}

Based on these improvements and by demanding an operator 
not to give too much relic density to the
Universe and satisfying the current experimental constraints from 
direct and indirect detection, and from collider data, a vast number 
of effective DM
operators are indeed ruled out. The conclusion obtained here is important for
building an effective model for the dark matter.

\section{Effective Dark Matter Interactions}

For simplicity, we will assume there is only one component of dark matter 
denoted by $\chi$ and it is a standard model singlet.  Here the $\chi$
can stand for a Dirac or Majorana fermion,  real or complex scalar, depending
on the context. Also, $f$ stands for a SM fermion, including quarks and
leptons. We will include all quarks and leptons in our analysis. 
We briefly discuss in the Appendix a few hidden-sector models that 
can give rise to some of the operators used in this work in certain limits.
For dark matter of 
spin 1 and spin 3/2, the reader may refer to the works in 
Refs. \cite{effective-2,ding-liao}.

The first set of operators that we consider
is for fermionic DM.  Its effective interactions with a pair of fermions
include vector-, axial-vector, or tensor-type exchanges, 
given by the following dimension 6 operators
\begin{eqnarray}
O_1 & = & \sum_f \frac{C_1^f}{\Lambda_1^2} \left( \bar \chi \gamma^\mu \chi \right) 
\left( \bar f \gamma_\mu f \right) \; ,\\
O_2 & = & \sum_f \frac{C_2^f}{\Lambda_2^2} \left( \bar \chi \gamma^\mu \gamma^5\chi \right) 
\left( \bar f \gamma_\mu f \right) \; ,\\
O_3 & = & \sum_f \frac{C_3^f}{\Lambda_3^2} \left( \bar \chi \gamma^\mu \chi \right) 
\left( \bar f \gamma_\mu \gamma^5 f \right) \; ,\\
O_4 & = & \sum_f \frac{C_4^f}{\Lambda_4^2} \left( \bar \chi \gamma^\mu \gamma^5 \chi \right) 
\left( \bar f \gamma_\mu \gamma^5 f \right) \; ,\\
O_5 & = & \sum_f \frac{C_5^f}{\Lambda_5^2} \left( \bar \chi \sigma^{\mu \nu} \chi \right) 
\left( \bar f \sigma_{\mu\nu} f \right) \; ,\\
O_6 & = & \sum_f \frac{C_6^f}{\Lambda_6^2} \left( \bar \chi \sigma^{\mu \nu} \gamma^5 \chi \right)  
\left( \bar f \sigma_{\mu \nu} f \right)   \; ,
\end{eqnarray}
where $\Lambda_i$ is the heavy mass scale for the connector sector 
that has been 
integrated out and $C_i$ is an effective coupling constant of order $O(1)$ 
that can be
absorbed into $\Lambda_i$.
It is understood that for Majorana fermion the vector and tensor 
structures are absent.

Next set of operators are for fermionic DM associated with (pseudo) 
scalar-type exchange
\begin{eqnarray}
O_7 & = & \sum_f \frac{C_7^f m_f}{\Lambda_7^3} \left( \bar \chi \chi  \right)
\left( \bar f f \right) \; ,\\
O_8 & = & \sum_f \frac{i C_8^f m_f}{\Lambda_8^3} \left( \bar \chi \gamma^5 \chi  \right)
\left( \bar f f  \right) \; ,\\
O_9 & = & \sum_f \frac{i C_9^f m_f}{\Lambda_9^3} \left( \bar \chi \chi  \right)
\left( \bar f \gamma^5 f  \right) \; ,\\
O_{10} & = & \sum_f \frac{C_{10}^f m_f}{\Lambda_{10}^3} \left( \bar \chi \gamma^5 \chi  \right)
\left( \bar f \gamma^5 f  \right) \;.
\end{eqnarray}
The $m_f$ dependence in the coupling strength is included for 
scalar-type interactions because this factor appears naturally from 
dark matter models
with scalar exchange diagrams.
Another light degree of freedom that couples to the fermionic dark matter 
is the gluon field 
\begin{eqnarray}
O_{11} & = &\frac{C_{11}}{\Lambda_{11}^3} 
\left( \bar \chi \chi  \right)
\left( - \frac{\alpha_s}{12 \pi} G^{\mu\nu} G_{\mu \nu} \right) \; ,\\
O_{12} & = &\frac{i C_{12}}{\Lambda_{12}^3} \left( \bar \chi \gamma^5 \chi  \right)
\left( - \frac{\alpha_s}{12 \pi} G^{\mu\nu} G_{\mu \nu} \right) \; ,\\
O_{13} & = &\frac{C_{13}}{\Lambda_{13}^3} \left( \bar \chi \chi  \right)
\left( \frac{\alpha_s}{8 \pi} G^{\mu\nu} \tilde G_{\mu \nu} \right) \; ,\\
O_{14} & = &\frac{i C_{14}}{\Lambda_{14}^3} \left( \bar \chi \gamma^5 \chi  \right)
\left( \frac{\alpha_s}{8 \pi} G^{\mu\nu} \tilde G_{\mu \nu} \right) \;.
\end{eqnarray}
For operators involving gluons, the factor of strong coupling constant 
$\alpha_s(2 m_\chi)$ 
is also included because these operators are induced at one loop level 
and evaluated at the scale $2 m_\chi$ where $m_\chi$ is the dark matter mass.

Finally, we also write down the corresponding operators for complex
scalar DM.
\begin{eqnarray}
\label{l15}
O_{15} & = & \sum_f \frac{i C_{15}^f}{\Lambda_{15}^2} \left( \chi ^\dagger \overleftrightarrow
{\partial_\mu} \chi  \right) \left( \bar f \gamma^\mu f  \right) \; ,\\
\label{l16}
O_{16} & = & \sum_f \frac{i C_{16}^f}{\Lambda_{16}^2} \left( \chi^\dagger \overleftrightarrow
{\partial_\mu} \chi  \right) \left( \bar f \gamma^\mu \gamma^5 f  \right) \; ,\\
O_{17} & = & \sum_f \frac{C_{17}^f m_f}{\Lambda_{17}^2} \left( \chi^\dagger \chi  \right)
\left( \bar f f  \right) \; ,\\
O_{18} & = & \sum_f \frac{i C_{18}^f m_f}{\Lambda_{18}^2} \left( \chi^\dagger  \chi  \right)
\left( \bar f \gamma^5 f  \right) \; ,\\
O_{19} & = &\frac{C_{19}}{\Lambda_{19}^2} \left( \chi^\dagger \chi  \right)
\left( - \frac{\alpha_s}{12 \pi} G^{\mu\nu} G_{\mu \nu} \right) \; ,\\
O_{20} & = &\frac{C_{20}}{\Lambda_{20}^2} \left( \chi^\dagger \chi  \right)
\left( \frac{\alpha_s}{8 \pi} G^{\mu\nu} \tilde G_{\mu \nu} \right) \; .
\end{eqnarray}
We note that for real scalar dark matter the vector couplings in 
Eqs.(\ref{l15}) and (\ref{l16}) are absent.
In what follows, we simply focus on the complex scalar dark matter.  
Note also that we have redefined the coefficients of some of the operators, 
which
are different from our previous works \cite{pbar-1,gamma-8}, 
such that they can conform with the normalization for the nucleon matrix 
elements
used in the literature for the direct detection experiments.

In Ref.~\cite{pbar-1}, we showed that in the calculation of the 
annihilation cross section 
for the DM relic density, 
the relative importance of each operator can be understood by considering
the nonrelativistic expansion of the operator and studying the velocity
dependence. We briefly review this matter here for convenience.
In the nonrelativistic limit, 
the spinors for the Dirac DM $\chi$ and $\bar \chi$ annihilation are 
$\psi \simeq  \left( \xi , \epsilon \xi \right)^{\rm T}$ and 
$\bar \psi \simeq  ( \epsilon \eta^\dagger , \eta^\dagger ) \gamma^0$      
where $\xi$ and $\eta$ are two-components Pauli spinors
and $\epsilon = O(v/c)$.
We can expand $\bar \psi \gamma^\mu \psi$ as
\begin{eqnarray}
  \bar \psi \gamma^0 \psi &\simeq& 2 \epsilon \eta^\dagger  \xi \, \nonumber \\
  \bar \psi \gamma^i \psi &\simeq & (1+\epsilon^2)
                         \eta^\dagger \sigma_i \xi \, \nonumber 
\end{eqnarray}
where the spatial components are not suppressed by $v/c$.  On the other 
hand, $\bar \psi \gamma^\mu \gamma^5\psi$ in the nonrelativistic limit are 
\begin{eqnarray}
  \bar \psi \gamma^0\gamma^5 \psi &\simeq& (1+\epsilon^2)
                               \eta^\dagger  \xi \, \nonumber \\
  \bar \psi \gamma^i \gamma^5 \psi &\simeq& 2 \epsilon \eta^\dagger \sigma_i \xi 
  \, \nonumber 
\end{eqnarray}
where the spatial components are now suppressed by $v/c$. 
It is clear that in the nonrelativistic limit the time and spatial
components of the vector and axial vector bilinear 
behave very differently.  We can then consider them separately 
when it is contracted with the trace of the light fermion leg. 
If we look at the trace of $(\bar f \gamma^\mu f)$ or 
$(\bar f \gamma^\mu \gamma^5 f)$ in the annihilation amplitude, the
time component part after being squared gives a quantity close to
zero, while the spatial component part gives a quantity
in the order of $m_\chi^2$.  Therefore, it is clear now that 
$\bar \psi \gamma^\mu \psi$ multiplied to $(\bar f \gamma_\mu f)$ or 
$(\bar f \gamma_\mu \gamma^5 f)$ will not be suppressed, while
$\bar \psi \gamma^\mu \gamma^5\psi$ multiplied to $(\bar f \gamma_\mu f)$ or 
$(\bar f \gamma_\mu \gamma^5 f)$ will always be suppressed. 
Therefore, the operators $O_1$ and $O_3$ 
can contribute to annihilation
much more than the operators $O_2$ and $O_4$. 
All the other operators can be understood similarly \cite{pbar-1}. From 
Ref.\cite{pbar-1}, we knew that some of the operators are doubly 
suppressed by the 
velocity of the dark matter combined with either a light fermion
 mass or strong coupling constant.
Note that some of the lower limits that we obtained before
are relative low compared to the dark matter mass.
In such cases, one may question the validity of the
effective interaction approach.  The physics behind is easy to
understand. The effects of such operators are very suppressed
because of the small velocity suppression or helicity suppression,
not because of the size of the $\Lambda$.  Therefore, the $\Lambda$
has to be small enough in order to see an effect from these operators.
We argue that the effective momentum
transfer of such velocity-suppressed operators should be 
$m_{\chi} (v/c)$. With $(v/c) \sim 10^{-3}$ for the DM velocity 
at the present epoch, as long as the ratio $ m_{\chi} (v/c) / \Lambda$ 
remains small, 
we expect the effective interaction approach can still be valid. 

The above effective operators are relativistically invariant and
therefore appropriate for the calculation in the relic density of the
dark matter and its implication at collider physics. However, for
direct detection experiments, we need to have a nonrelativistic
reduction of these operators since the local dark matter velocity in
the halo is of order $(v/c) \sim 10^{-3}$.  It is straightforward to
demonstrate in the nonrelativistic limit only eight operators are
relevant for the direct detections.  These are $O_{1}$, $O_{4}$,
$O_{5}$, $O_{7}$, $O_{11}$, $O_{15}$, $O_{17}$, and $O_{19}$.  One can
further show that only $O_1$, $O_4$ and $O_7$ are independent, since
we have the following nonrelativistic reduction
\begin{eqnarray}
O_5 & \longrightarrow &O_4 \\
O_{11} & \longrightarrow &O_7 \\
O_{15} & \longrightarrow &O_1 \\
O_{17} & \longrightarrow &O_7 \\
O_{19} & \longrightarrow &O_7 
\end{eqnarray}
In Table~\ref{prop} we summarize some of the features of the operators 
discussed in this section.
At decoupling time, $v/c \sim 0.1$ and hence non-relativistic reduction is 
no longer applicable. The velocity scaling behaviours for each operator 
shown in the last column of Table~\ref{prop} for the annihilation 
cross sections 
are just merely serving the purpose to illustrate the physics. 
For our numerical work, we use the full expressions for the annihilation 
cross sections presented at the Appendix.

In our analysis in the following sections, we will treat one operator at 
a time. 
This working assumption of treating one operator at one time may seem 
unreasonable.  
However it is a matter of choosing between controlling the number of parameters 
and the assumptions involved.  If we treat each SM favor separately, 
then one operator at
a time would mean the DM only couples to one quark (say $u$ quark) but
not to the other (say $d$ quark).  It would be very strange that the
new physics only couples to up quark but not to the others.  But if we
take more than one operators at the same time, the number of
parameters will grow out of control in such an analysis.
On the other hand, we have summed over all SM fermions for each 
operator. The quantum numbers of the new interaction for the SM
fermions could be very different from one another.  It is entirely model
dependent.  There would be too many parameters if we treat them all
different.  Even if we assumed different coefficients for each SM
generation, we would still introduce more parameters.  Here in this work,
we take the {\it democratic} choice such that the coefficient for each 
SM fermion is of the same order, and we have treated them the same. 
Therefore, we sum over all SM fermions in each operator.

We also note that the effective operators studied here in this work do 
not address 
the issue of gauge invariance. Imposing $SU(2)$ gauge invariance for 
the SM fermions would impose certain relations among operators and hence their 
coefficients. Certain operators like those with an explicit factor of SM 
fermion mass $m_f$ breaking $SU(2)$ invariance explicitly 
can be made covariant by introducing the Higgs field.
Such issues have been partially addressed in the literature, 
see for example in \cite{cao}.

\begin{table}[thb!]
\caption{\small \label{prop}
List of properties of each operator that we define in this section.
``SI'' and ``SD'' stands spin-independent and spin-dependent 
cross sections for direct detection.
}
\begin{ruledtabular}
\begin{tabular}{ccccccc}
Operator & NR Limit  & SI & SD & Dirac/Complex & Majorana/Real & 
NR Limit $\langle \sigma^{\rm anni} v \rangle$\\
 & (Direct Detection)  &  &  &  &  & (Relic Density)\\
\hline
$O_1$ & Yes & Yes & No & Yes & No &  $\frac{N_C m^2_\chi}{\pi \Lambda^4_1}$ \\
$O_2$ & No & - & - & Yes & Yes & $\frac{N_C m^2_\chi v^2}{6 \pi \Lambda^4_2}$ \\
$O_3$ & No & - & - &  Yes & No & $\frac{N_C m_\chi^2}{\pi \Lambda^4_3}$ \\
$O_4$ & Yes & No & Yes &  Yes & Yes & $\frac{N_C m^2_\chi v^2}{6 \pi \Lambda^4_4}$ \\
$O_5$ & Yes & No & Yes &  Yes & No & $\frac{2 N_C m^2_\chi}{\pi \Lambda_5^4}$ \\
$O_6$ & No & - & - &  Yes & No & $\frac{2 N_C m^2_\chi}{\pi \Lambda^4_6}$ \\
\hline
$O_7$ & Yes & Yes & No &  Yes & Yes & $\frac{N_C m_f^2 m_\chi^2 v^2}{8 \pi \Lambda^6_7}$ \\
$O_8$ & No & - & - &  Yes & No & $\frac{N_C m_f^2 m_\chi^2}{2 \pi \Lambda^6_8}$ \\
$O_9$ & No & - & - &  Yes & Yes & $\frac{N_C m_f^2 m_\chi^2 v^2}{8 \pi \Lambda^6_9}$ \\
$O_{10}$ & No & - & - &  Yes & No & $\frac{N_C m_f^2 m_\chi^2}{2 \pi \Lambda^6_{10}}$ \\
\hline
$O_{11}$ & Yes & Yes & No &  Yes & Yes & $\frac{\alpha_s^2 m_\chi^4 v^2}{288 \pi^3 \Lambda^6_{11}}$ \\
$O_{12}$ & No & - & - &  Yes & No & $\frac{\alpha_s^2 m^4_\chi}{72 \pi^3 \Lambda^6_{12}}$ \\
$O_{13}$ & No & - & - &  Yes & Yes & $\frac{\alpha_s^2 m^4_\chi v^2}{128 \pi^3 \Lambda^6_{13}}$ \\
$O_{14}$ & No & - & - &  Yes & No & $\frac{\alpha_s^2 m^4_\chi}{32 \pi^3 \Lambda^6_{14}}$ \\
\hline
\hline
$O_{15}$ & Yes & Yes & No &  Yes  & No & $\frac{N_C m^2_\chi v^2}{6 \pi \Lambda^4_{15}}$ \\
$O_{16}$ & No & - & - &  Yes & No & $\frac{N_C m^2_\chi v^2}{6 \pi \Lambda^4_{16}}$ \\
\hline
$O_{17}$ & Yes & Yes & No &  Yes & Yes & $\frac{N_C m^2_f}{4 \pi \Lambda^4_{17}}$ \\
$O_{18}$ & No & - & - &  Yes & Yes & $\frac{N_C m^2_f}{4 \pi \Lambda^4_{18}}$ \\
\hline
$O_{19}$ & Yes & Yes & No &  Yes & Yes & $\frac{\alpha_s^2 m^2_\chi}{144 \pi^3 \Lambda^4_{19}}$\\
$O_{20}$ & No & - & - &  Yes & Yes & $\frac{4 \alpha_s^2 m^2_\chi}{301 \pi^3 \Lambda^4_{20}}$
\end{tabular}
\end{ruledtabular}
\end{table}

\section{Relic Density}

In the standard cosmic picture, it is assumed that the DM particles
were in thermal equilibrium with the other SM particles via various
fundamental processes such as $\bar \chi \chi \leftrightarrow P \bar
P$ where $P$ is any SM particles.  At the high temperature Early
Universe, the DM particles were kept in thermal equilibrium as long as
the reaction rate, scaled by the temperature, was faster than the
expansion rate $H$ (the Hubble parameter) of the Universe.  The
Universe cooled down as it kept on expanding. At around the
temperature that the reaction rate fell below the expansion rate $H$,
the DM particles began to decouple from the thermal bath. The DM
particles will keep on annihilation into the SM particles until the
point that they could no longer effectively find one another. The
remaining number density of the DM particles became the relic density
that we can observe today.

\begin{figure}[t!]
\centering
\includegraphics[width=3.2in]{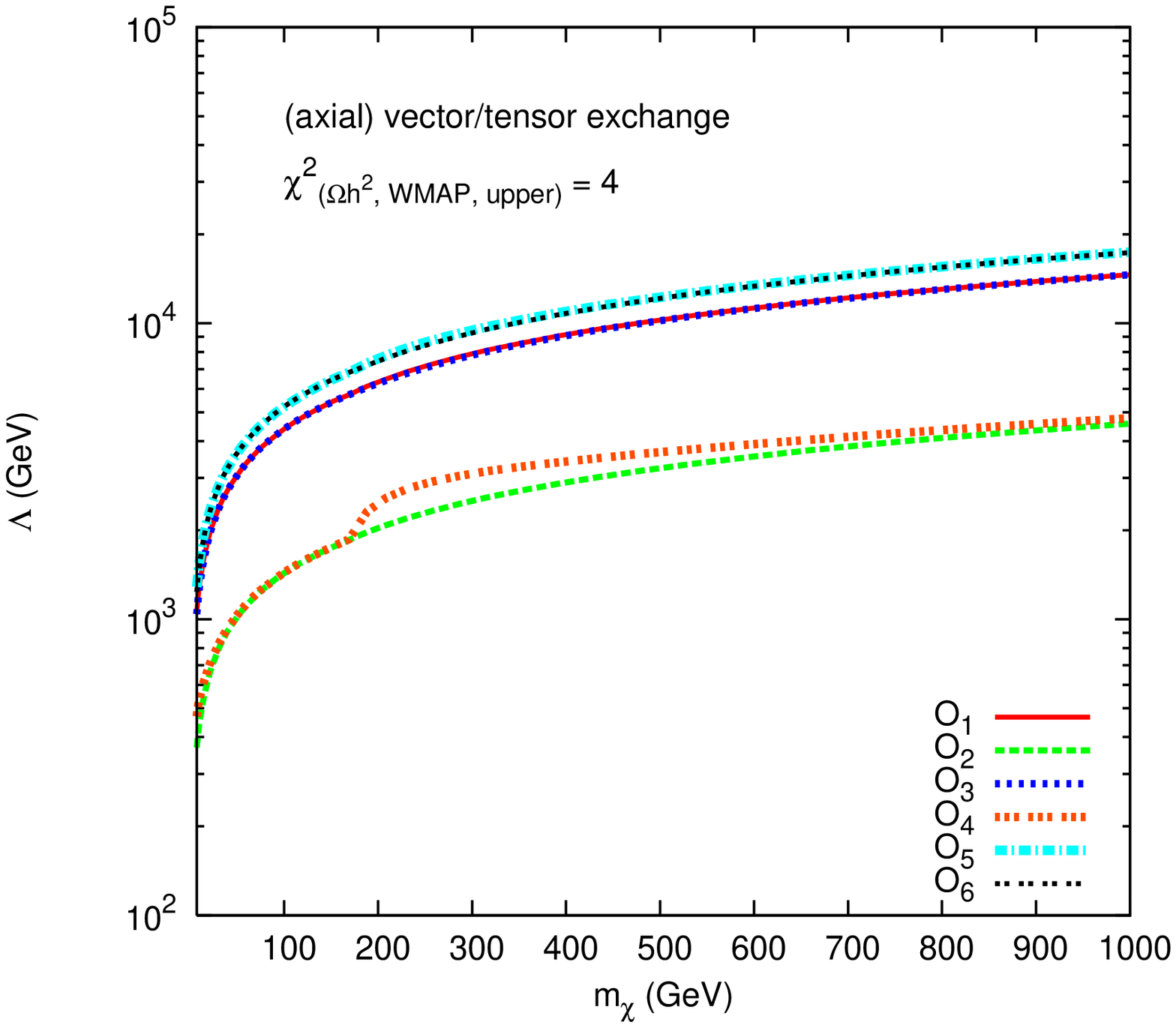}
\includegraphics[width=3.2in]{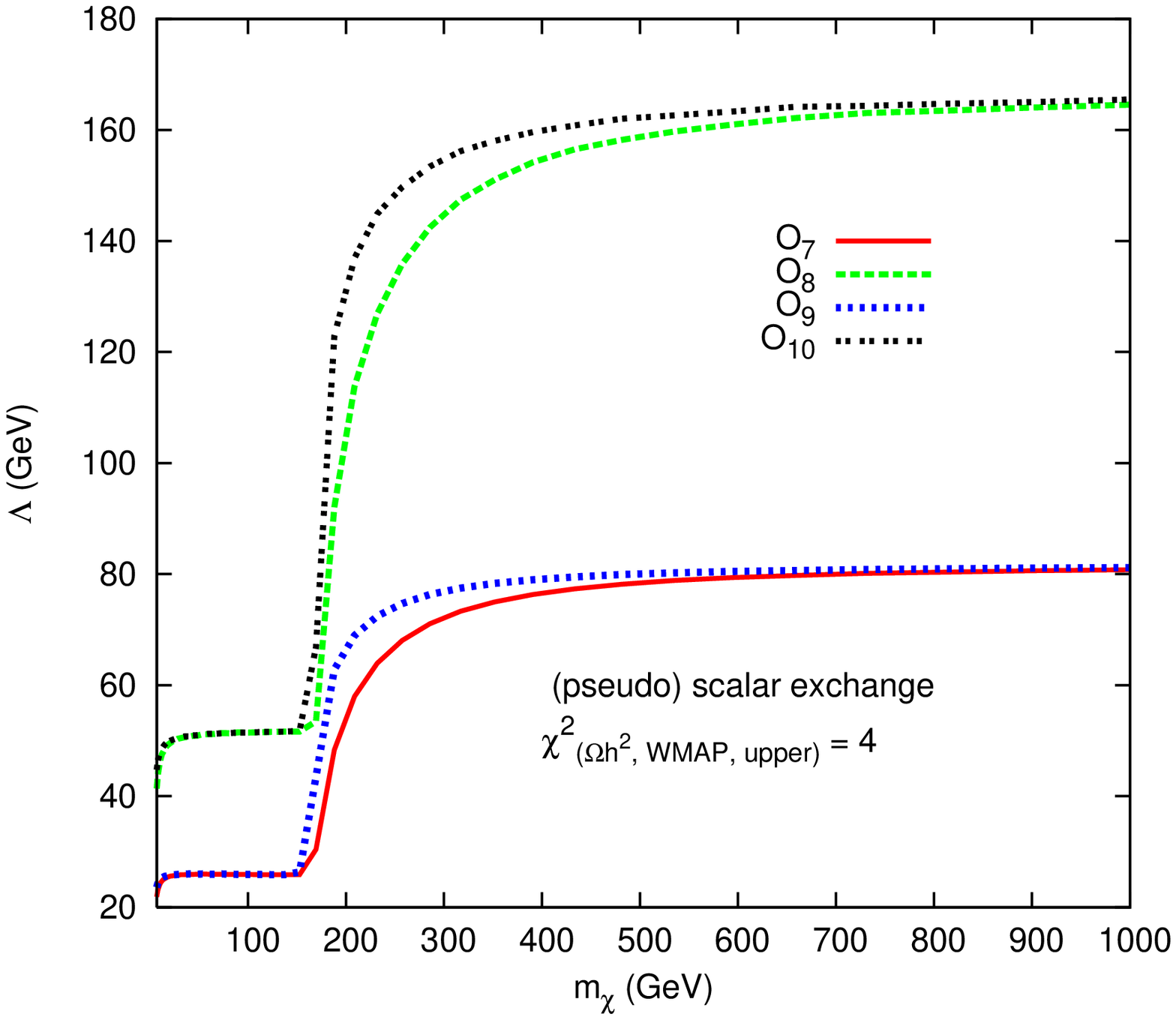}
\includegraphics[width=3.2in]{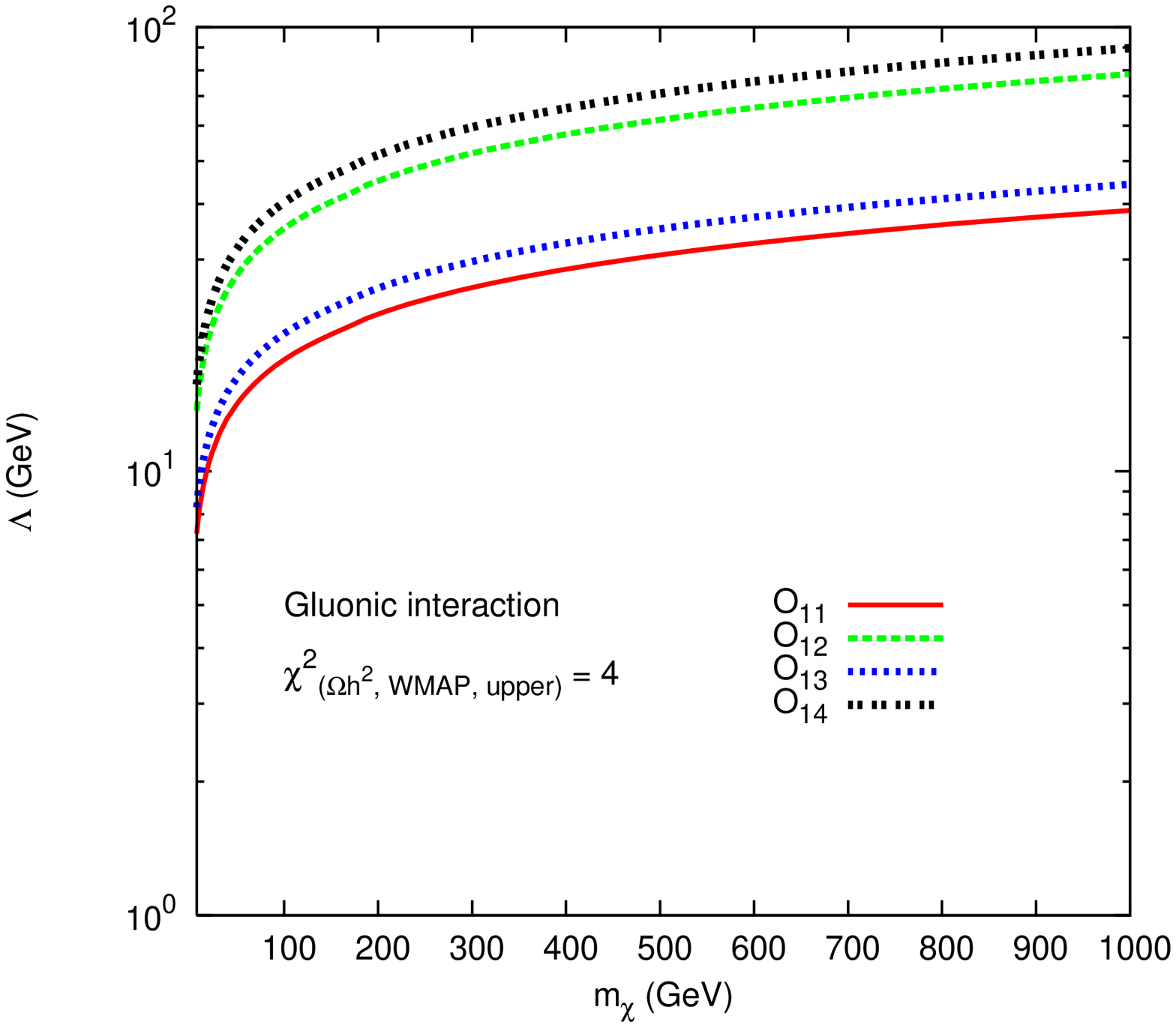}
\includegraphics[width=3.2in]{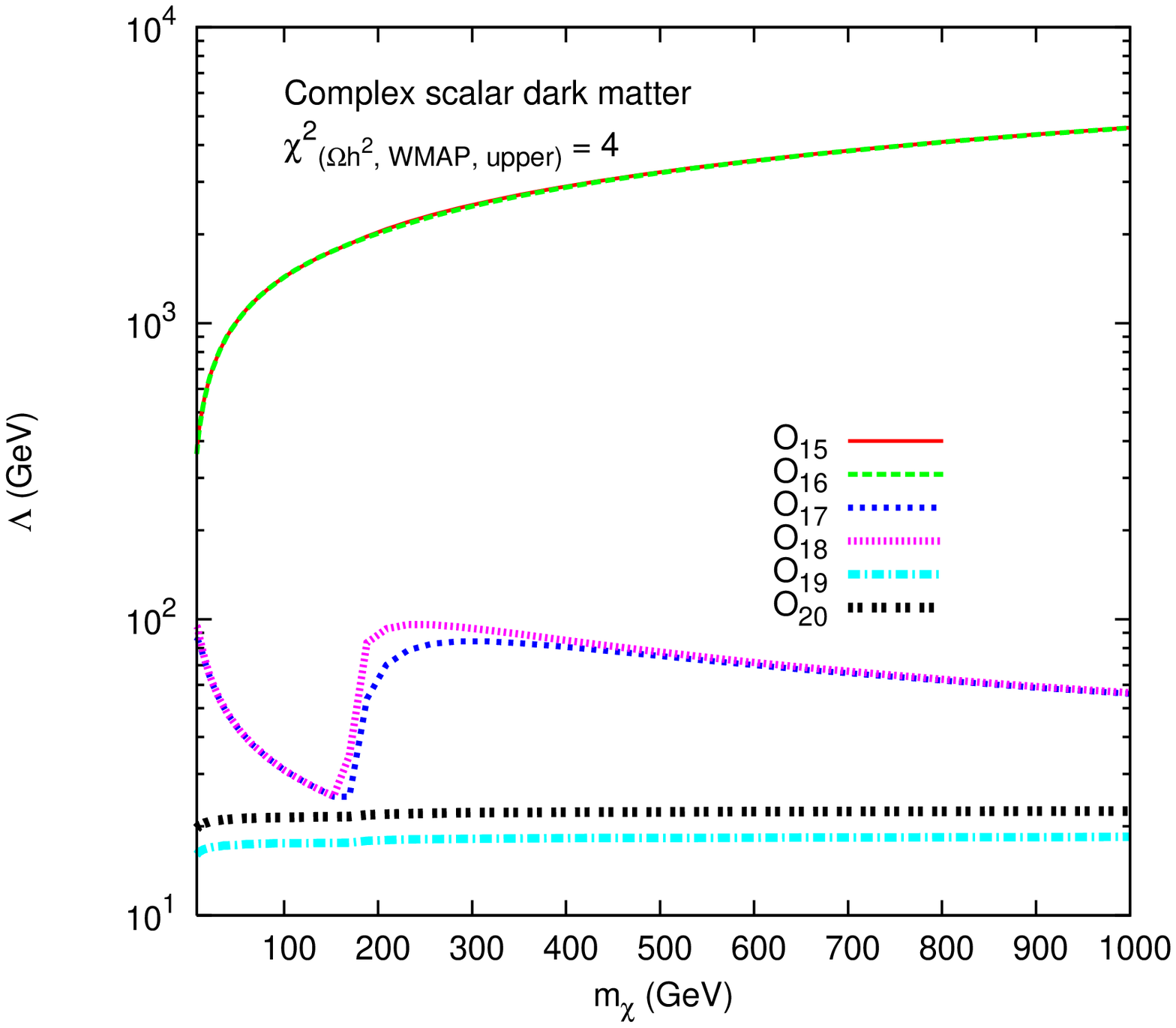}
\caption{\small \label{relic}
The upper limits on $\Lambda$ due to the WMAP7 data of Eq.~(\ref{wmap}). 
We require the resulting relic density less than the WMAP7 \cite{wmap}: central
value plus $2 \sigma$ error. 
}
\end{figure}

The input of standard computation of relic density
is $\sigma^{\rm ann}$ calculated from
each operator.  For each operator, we list the $( d \sigma^{\rm ann} /
d \cos \theta)$ in the Appendix and the nonrelativistic limit of
$\sigma^{\rm anni} v$ in the last column of Table \ref{prop}.  In our
numerical analysis we require the resulting relic density to be less
than the measured value from WMAP7 \cite{wmap} given by
Eq.~(\ref{wmap}).  When the relic density is smaller than the measured
DM density, the DM in the Universe may contain more than one
species. Since the relic density roughly scales inversely with
$\sigma^{\rm ann}$, the WMAP7 data implies an upper limit on the heavy
scale $\Lambda$ for each operator.  The results are shown in
Fig.~\ref{relic} with the requirement of the resulting density to be
less than the $2\sigma$ upper limit of the WMAP7 data.  One notices
that as the DM mass becomes larger all the curves increase gradually,
except those for $O_{17}$ and $O_{18}$ which approach a constant. This
can be understood by looking at the nonrelativistic limits of the
annihilation cross sections listed in the last column of Table
\ref{prop}.  For $O_{17}$ and $O_{18}$, they are proportional to
$1/\Lambda^4$ and independent of $m_\chi$, while for all other operators
they are proportional to either $m_\chi^2 / \Lambda^4$, 
$ m_f^2 m_\chi^2 / \Lambda^6$, or 
$m_\chi^4 / \Lambda^6$, from which we can see that
the power of $\Lambda^2$ in the
denominator is one or two higher than the power of $m_\chi^2$ in the
numerator.

\section{Direct Detection}

The solar system moves around in the Galactic halo with a
nonrelativistic velocity $v \sim 10^{-3} c$.  
When the dark matter particles move through a detector,
which is usually put under a deep mine or a mountain to reduce
backgrounds, and create collisions with the detector, 
some signals may arise in phonon-type, scintillation-type, 
ionization-type, or some combinations of them, depending on the 
detector materials.
The event rate is extremely low because of the
weak-interaction nature of the dark matter. There are controversies
among various direct detection experiments.  Both CoGeNT \cite{cogent}
and DAMA \cite{dama} observed some positive signals of dark matter detection, 
which
point to a light dark matter ($\sim 5-10$ GeV) with the $\sigma_{\rm SI} \sim
10^{-41} \;{\rm cm}^2$.  On the other hand, CDMS \cite{cdms} and the most
recent XENON100 \cite{xenon} have found nothing and disagreed with what 
were found by CoGeNT and DAMA.  In the following we will use the 
excluded regions of the XENON100 data \cite{xenon} for spin-independent cross 
sections ($\sigma_{\rm SI}$), and 
XENON10 \cite{xenon-sd}, ZEPLIN \cite{zeplin} and SIMPLE \cite{simple} data
for spin-dependent cross sections ($\sigma_{\rm SD}$) versus the 
DM mass $m_{\chi}$
in constraining the effective DM interactions.

We will be interested in the non-relativistic limit only and consider
one operator at a time. Thus possible interference effects among different
operators are ignored.

\subsection{Spin-Independent Cross Section}

Both $O_1$ and $O_7$ contribute to the
spin-independent cross section.  For a nuclei $\mathcal N$ with $Z$
protons and $(A - Z)$ neutrons, the cross section can be obtained as
\begin{equation}
\label{sixs1dirac}
\sigma^{\rm SI}_{\chi \mathcal N} (0) = \frac{\mu^2_{\chi \mathcal N}}{\pi} \vert b_{\mathcal N} \vert^2
\end{equation}
from $O_1$ for Dirac DM where 
\begin{equation}
\mu_{\chi \mathcal N} = \frac{m_\chi m_{\mathcal N}}{m_\chi + m_{\mathcal N}}
\end{equation}
is the reduced mass and
\begin{equation}
b_{\mathcal N} = Z \, b_p + (A - Z) \, b_n
\end{equation} 
with
\begin{eqnarray}
\label{sicouplings1a}
b_p & = & 2 \,\frac{C_1^u}{\Lambda_1^2} + \frac{C_1^d}{\Lambda_1^2}  \; ,\\
\label{sicouplings1b}
b_n & = & \frac{C_1^u}{\Lambda_1^2} + 2 \, \frac{C_1^d}{\Lambda_1^2} \; .
\end{eqnarray}
There is no Majorana case for  $O_1$.

For $O_7$ with Dirac DM, we have
\begin{equation}
\label{sixs7dirac}
\sigma^{\rm SI}_{\chi \mathcal N} (0) = \frac{\mu^2_{\chi \mathcal N}}{\pi} \vert f_{\mathcal N} \vert^2
\end{equation}
where 
\begin{equation}
\label{fN}
f_{\mathcal N} = Z \, f_p + (A - Z) \, f_n
\end{equation} 
with
\begin{equation}
\label{sicouplings7}
f_{p,n} = \frac{m_{p,n}}{\Lambda_7^3} 
\left\{
\sum_{q = u,d,s} C_7^q \, f^{(p,n)}_{Tq} + \frac{2}{27} f^{(p,n)}_{TG} \sum_{Q=c,b,t} C_7^Q
\right\}
\end{equation}
and
\begin{equation}
f^{(p,n)}_{TG} \equiv 1 - \sum_{q=u,d,s} f^{(p,n)}_{Tq} \; .
\end{equation}
For Majorana DM with the same effective operator, one should multiply 
the above cross section (\ref{sixs7dirac}) by a factor of 4.

For  $O_{11}$ with Dirac DM, the result is the same as $O_7$ with the
following couplings
\begin{equation}
f_{p,n} = \frac{m_{p,n}}{\Lambda_{11}^3} \frac{2}{27} f^{(p,n)}_{TG} C_{11} \; .
\end{equation}
For Majorana DM, multiply the cross section by a factor of 4.

For  $O_{15}$ with complex scalar, the result is
\begin{equation}
\sigma^{\rm SI}_{\chi \mathcal N} (0) = \frac{\mu^2_{\chi \mathcal N}}{\pi} \vert b_{\mathcal N} \vert^2
\end{equation}
which is same as $O_{1}$ with the following replacements for the couplings
in (\ref{sicouplings1a}) and (\ref{sicouplings1b}) 
\begin{eqnarray}
C^{u,d}_{1} &\longrightarrow & C^{u,d}_{15}  \; , \\
\Lambda_1 & \longrightarrow &\Lambda_{15} \; .
\end{eqnarray}

For  $O_{17}$ with complex scalar, the result is same as $\mathcal O_{7}$
\begin{equation}
\sigma^{\rm SI}_{\chi \mathcal N} (0) = \frac{\mu^2_{\chi \mathcal N}}{4 \pi} \vert f_{\mathcal N} \vert^2
\end{equation}
with $f_{\mathcal N} = Z f_p + (A-Z) f_n$  and the following replacement 
in (\ref{sicouplings7}) 
\begin{eqnarray}
C^{u,d}_{7} &\longrightarrow & C^{u,d}_{17} \; , \\
\Lambda_7 & \longrightarrow &\Lambda_{17} \; .
\end{eqnarray}

For  $O_{19}$ with complex scalar,  the result is same as $O_{7}$
\begin{equation}
\sigma^{\rm SI}_{\chi \mathcal N} (0) = \frac{\mu^2_{\chi \mathcal N}}{4 \pi} \vert f_{\mathcal N} \vert^2
\end{equation}
with $f_{\mathcal N} = Z f_p + (A-Z) f_n$ and 
\begin{equation}
f_{p,n} = \frac{m_{p,n}}{\Lambda_{19}^3} \frac{2}{27} f^{(p,n)}_{TG} C_{19}  \; .
\end{equation}
In our numerical calculations, we will use the default values for 
$f^{(p,n)}_q$ and $f^{(p,n)}_{TG}$ given in
DarkSUSY \cite{darksusy}.
\footnote
{For a recent re-evaluation of these hadronic matrix elements using
the up-to-date lattice calculation results of the strange 
quark $\sigma_s$ term and its content in the nucleon, see 
Ref.[\cite{Cheng:2012qr}].
}

\subsection{Spin-Dependent Cross Section}

For $O_4$ with Dirac DM, its contribution to the spin-dependent cross section 
can be obtained as \cite{Engel:1992bf}
\begin{equation}
\label{sdxs4dirac}
\sigma^{\rm SD}_{\chi \mathcal N} (0) = \frac{8 \mu^2_{\chi \mathcal N}}{\pi}
G^2_F {\bar\Lambda}^2 J (J+1)
\end{equation}
where $J$ is the total spin of the nuclei $\mathcal N$, $G_F$ is the 
Fermi constant and
\begin{equation}
{\bar \Lambda} = \frac{1}{J} 
\left( 
a_p \langle S_p \rangle + a_n \langle S_n \rangle
\right)
\end{equation}
with $\langle S_p \rangle$ and $\langle S_n \rangle$ 
the average of the proton and neutron spins inside the
nuclei respectively, and
\begin{equation}
\label{sdcouplings}
a_{p,n} = \sum_{q=u,d,s} \frac{1}{\sqrt 2 G_F} \frac{C_4^q}{\Lambda_4^2} \Delta q^{(p,n)}
\end{equation}
with $\Delta q^{(p,n)}$ being the fraction of the spin carried by the quark $q$ 
inside the nucleon $p$ and $n$.
The following combinations of isosinglet $a_0$ and isovector 
$a_1$ are often seen in the literature
\begin{eqnarray}
a_0  & = & a_p + a_n \; , \\
a_1 & = & a_p - a_n \; .
\end{eqnarray}
For Majorana DM with the same effective operator, one should 
multiply the cross section (\ref{sdxs4dirac}) by
a factor of 4. 

For $O_5$ with Dirac DM, its contribution to the spin-dependent cross 
section is 
the same as $O_4$ with the following replacements in (\ref{sdcouplings})
\begin{eqnarray}
C_4^q & \longrightarrow & 2 \, C_5^q \; , \\
\Lambda_4 & \longrightarrow & \Lambda_5 \; .
\end{eqnarray}
There is no Majorana case for  $O_5$.

\begin{figure}[t!]
\centering
\includegraphics[angle=270,width=3.2in]{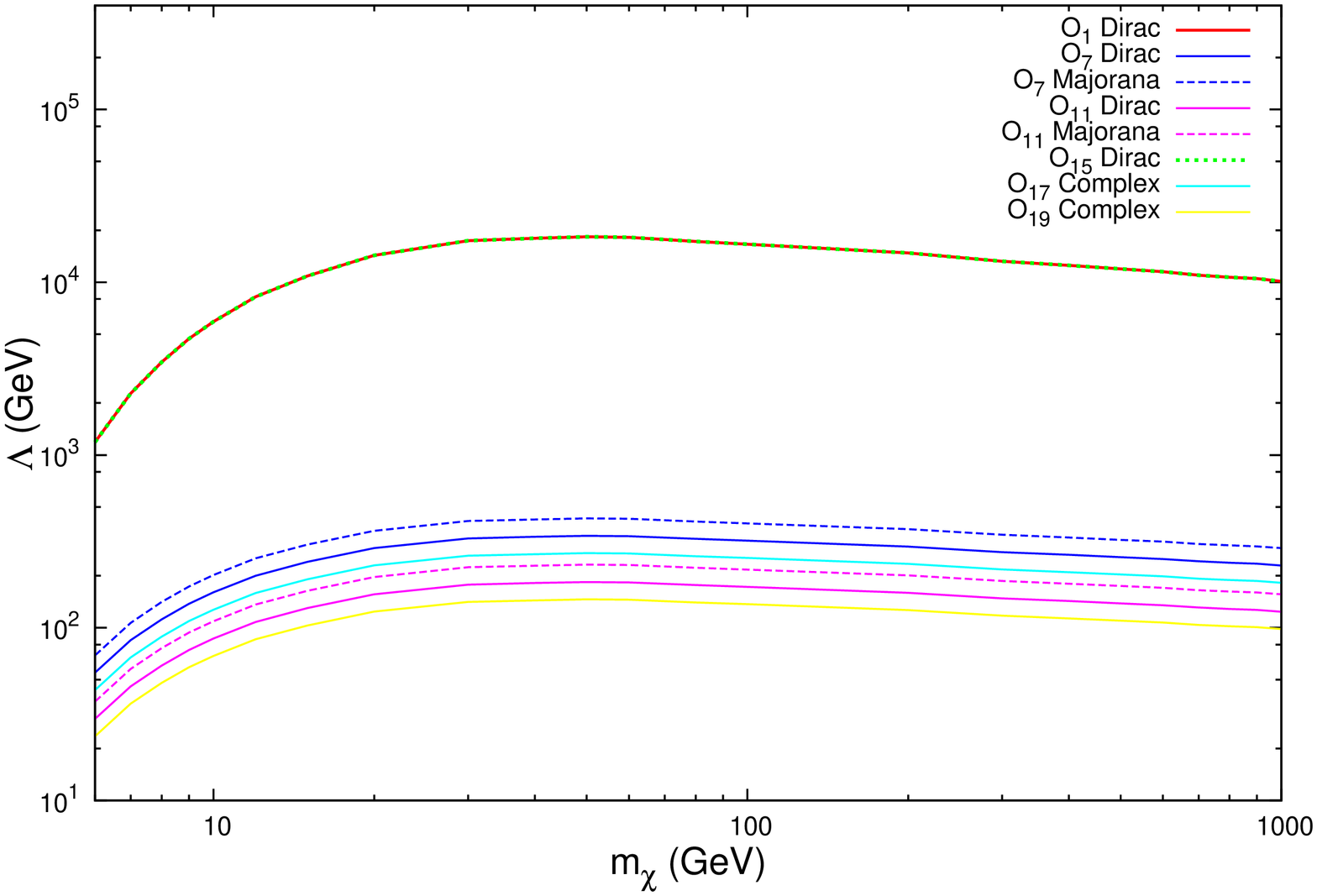}
\includegraphics[angle=270,width=3.2in]{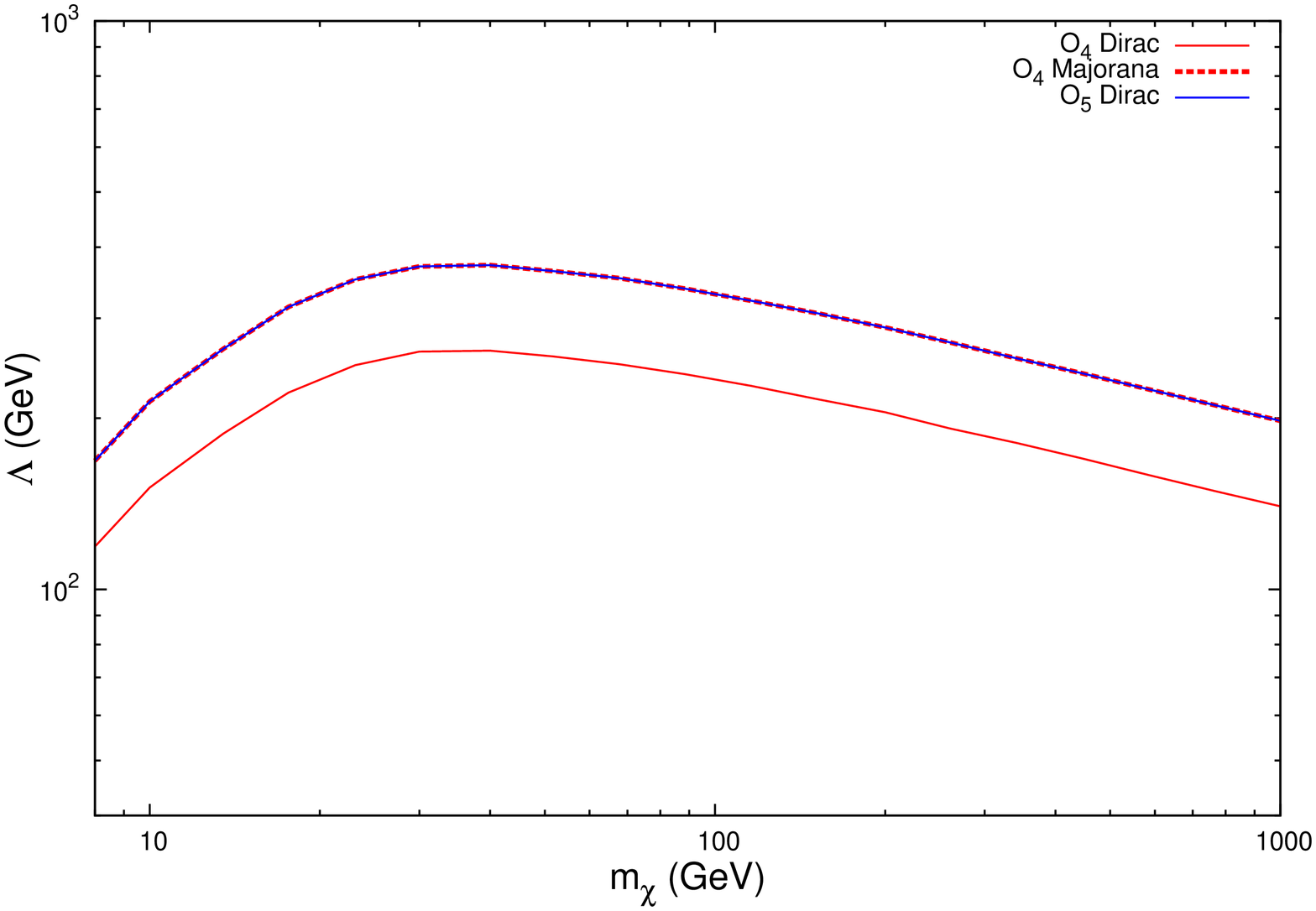}
\caption{\small \label{dd}
The lower limits on $\Lambda$ due to (a) spin-independent cross section
limits from XENON100 \cite{xenon}, and (b) spin-dependent cross section
limits from XENON10 \cite{xenon-sd}, ZEPLIN \cite{zeplin} 
and SIMPLE \cite{simple}.
}
\end{figure}

The current best limits on spin-independent WIMP-nucleon cross sections 
come from
XENON100 \cite{xenon}.  In Ref.~\cite{xenon}, the collaboration searched 
for DM candidates in their pre-defined signal region, but only 
found 3 signal events with an expected background of $1.8 \pm 0.6$. Based
on that they obtained the most stringent limits on DM spin-independent 
elastic WIMP-nucleon scattering cross sections, which already exceed
those of CDMS \cite{cdms} in almost the whole mass range of $m_\chi$.
Therefore, we only use the XENON100 data in this analysis. Since the
XENON100 result was presented by a 90\% CL upper limit curve, we take
the conservative choice that the central value of the WIMP-nucleon 
cross section for each $m_\chi$ to be zero and the $1\sigma$ error to 
be the 90\% CL curve divided by $1.645$ 
(assuming a Gaussian distribution that 90\% CL is equivalent to $1.645\sigma$.)
We obtain the $2\sigma$ limits on $\Lambda$ for each relevant operator
(note only some operators in our list can 
contribute to SI cross section, see Table \ref{prop} for a summary) 
and show the results in Fig.~\ref{dd}(a).
For SD WIMP-nucleon scattering cross sections we use the data from
XENON10~\cite{xenon-sd}, ZEPLIN \cite{zeplin}, 
and SIMPLE \cite{simple}.
We treat the SD data in the way as how we
treat the XENON100 SI data. We take the central value for the signal
cross section to be zero and the $1\sigma$ error for each $m_\chi$ is
obtained by dividing the 90\% CL curve by $1.645$.
We combine the chi-squares from all three experiments. 
The $2\sigma$  results for $\Lambda$ of each relevant operator that 
contributes to SD cross section are shown in Fig.~\ref{dd}(b). 
In both SI and SD cases, we apply our formulas for the proton and 
neutron separately.
Since the DM mass only enters in the SD and SI cross sections through the
reduced mass $\mu_{\chi \mathcal N}$, which is close to the nuclei 
mass $m_{\mathcal N}$ for large $m_\chi$, 
one expects that the limits should be weaker as the DM mass 
grows larger following merely the constraints given by the experiments.
This is evidently true for both SI and SD cases 
in Figs.~\ref{dd}(a) and \ref{dd}(b), respectively.
The most stringent experimental constraint for the SI case is 
located at $m_\chi \approx 50$ GeV, while for the 
SD case, it is about 35 GeV. These features are also 
reflected in our figures.

\section{Monojet and Monophoton production at Colliders}

In principle, dark matter particles can be directly produced in
hadronic collisions.  However, it would only give rise to something
missing in the detection.  We therefore need some additional visible
particles for trigger.  One of the cleanest signatures is monojet or
monophoton production, which has only a high $p_T$ jet or photon
balanced by a large missing transverse momentum.  Both CDF \cite{cdf}
and D\O\ \cite{d01,d02} at the Tevatron and the ATLAS \cite{atlas} at
the LHC have searched for such signals, though in other context such
as large extra dimensions.

\begin{figure}[h!]
\centering
\includegraphics[width=4in]{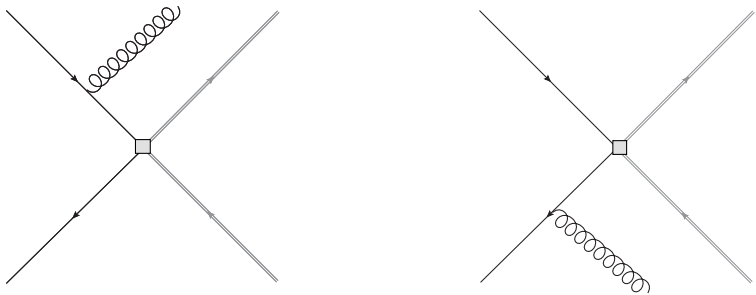}
\caption{\small \label{feyn}
One of the contributing Feynman diagrams for monojet or monophoton
production.}
\end{figure}

In our approach of effective DM interactions, we can attach either a
gluon or a photon to one of the quark legs of the relevant operators.
For example, in $O_{1-10,15-18}$ we can attach a gluon or a
photon line to the fermion line, such as the one shown in Fig.~\ref{feyn}.  For
gluonic operators we can either attach a gluon line to the gluon leg
or attach the whole 4-point diagram to a quark line such that it
becomes a $q g$-initiated process.  We then calculate the $2 \to 3$
process using FORM \cite{form}, and convolute the amplitude squared with parton
distribution functions.  The final state consists of a pair of DM
particles and a gluon or a photon. We require the jet or photon to
have a large transverse momentum according to the $p_T$ requirement of each
experiment.

\begin{figure}[t!]
\centering
\includegraphics[angle=270,width=3.2in]{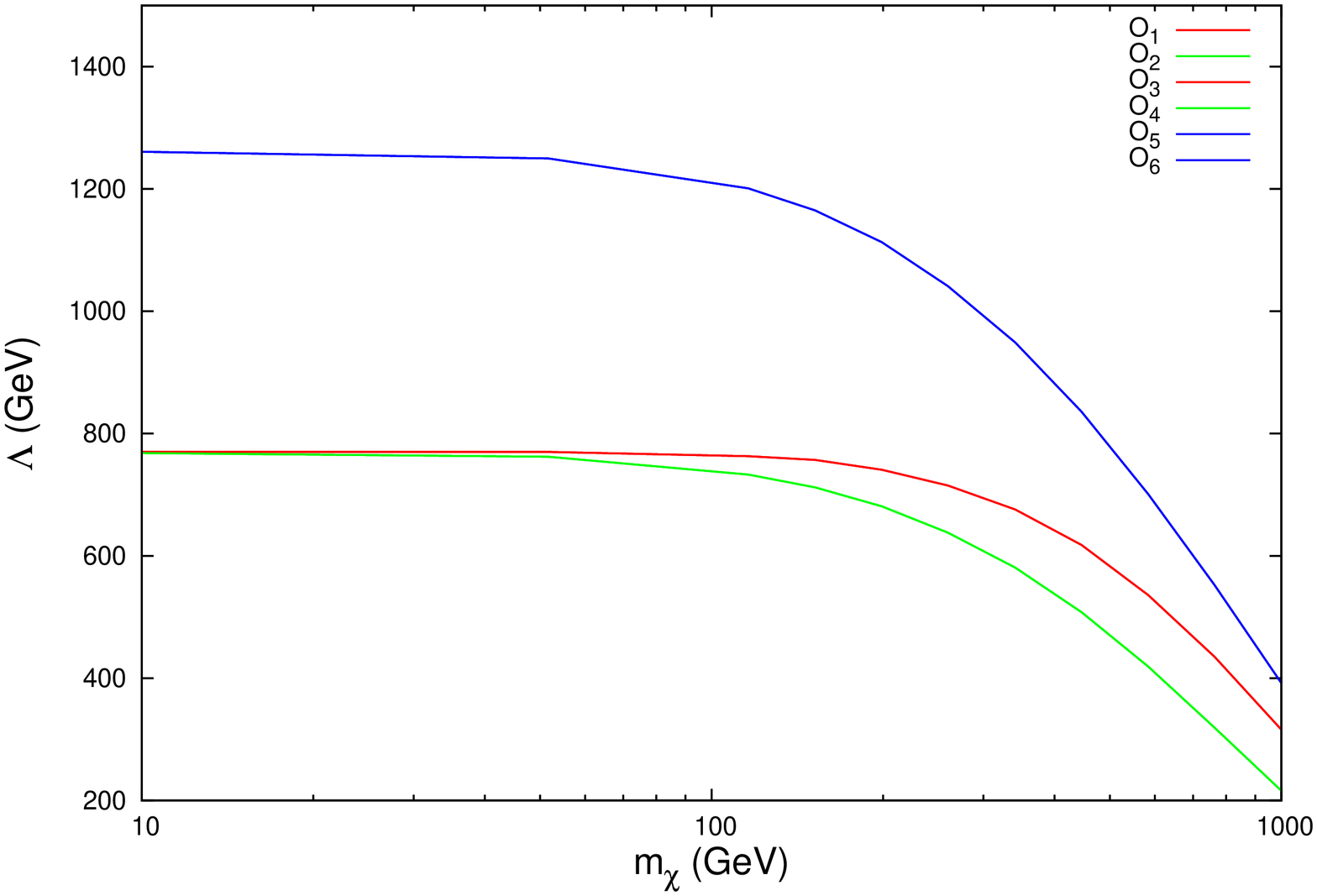}
\includegraphics[angle=270,width=3.2in]{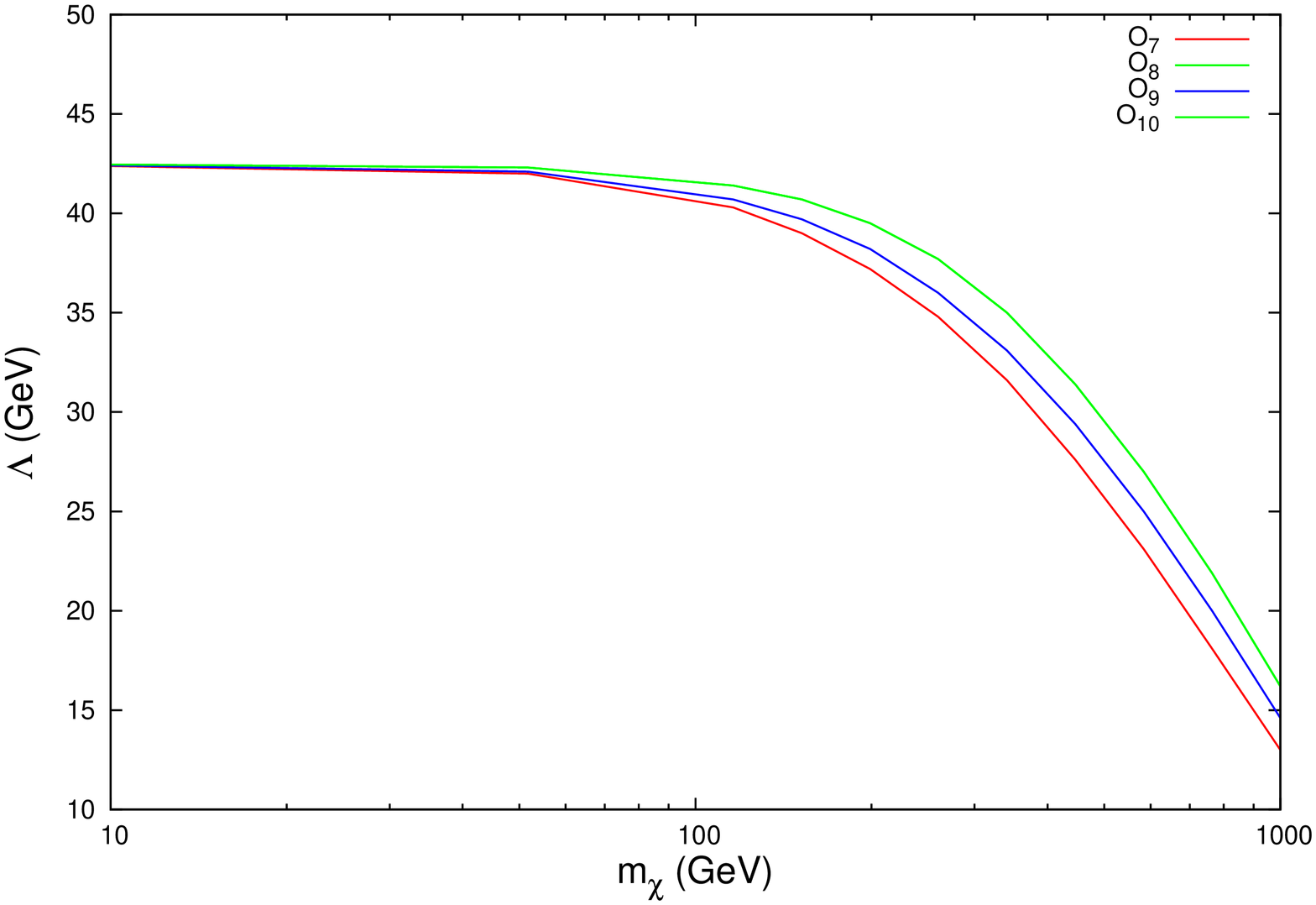}
\includegraphics[angle=270,width=3.2in]{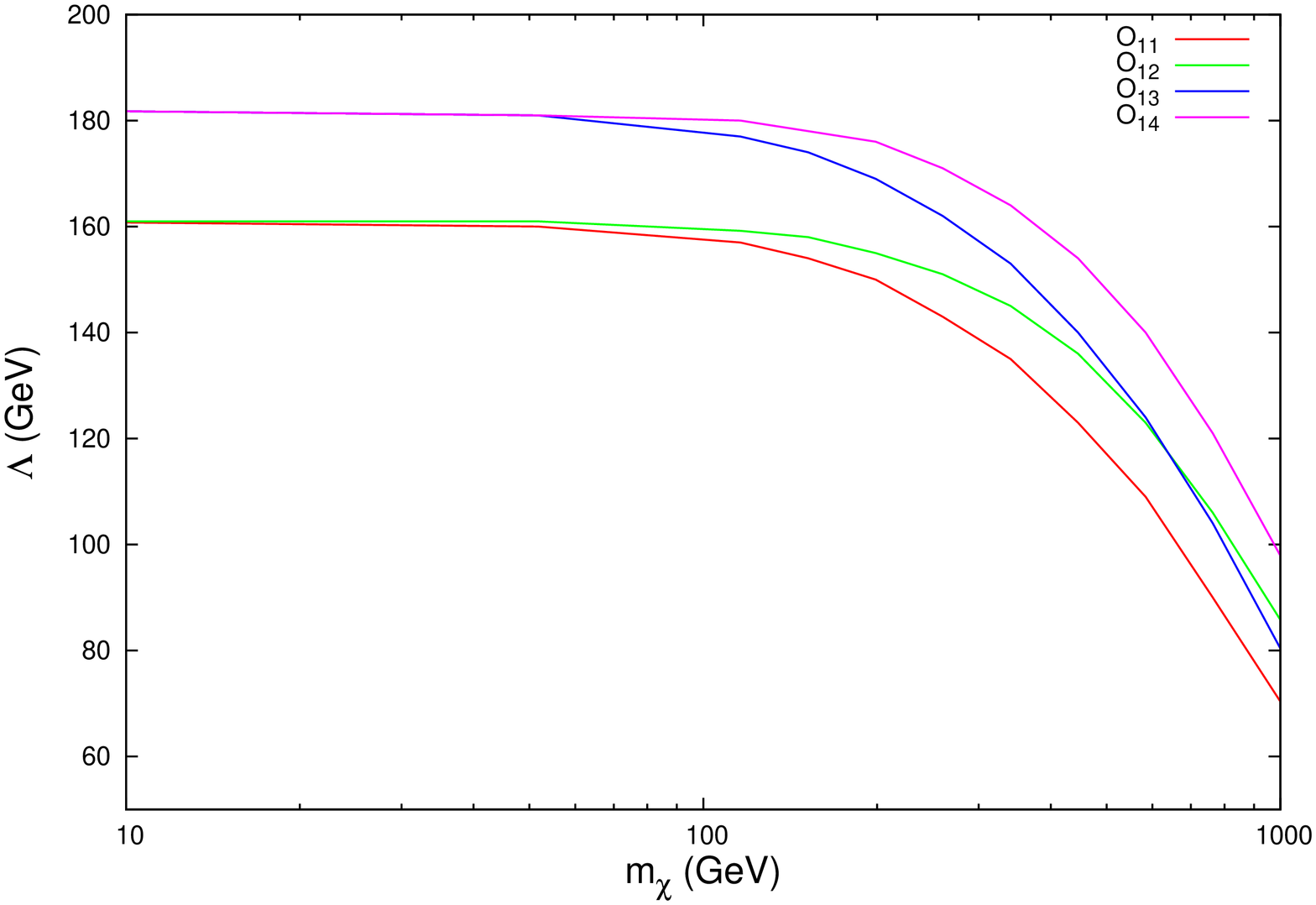}
\includegraphics[angle=270,width=3.2in]{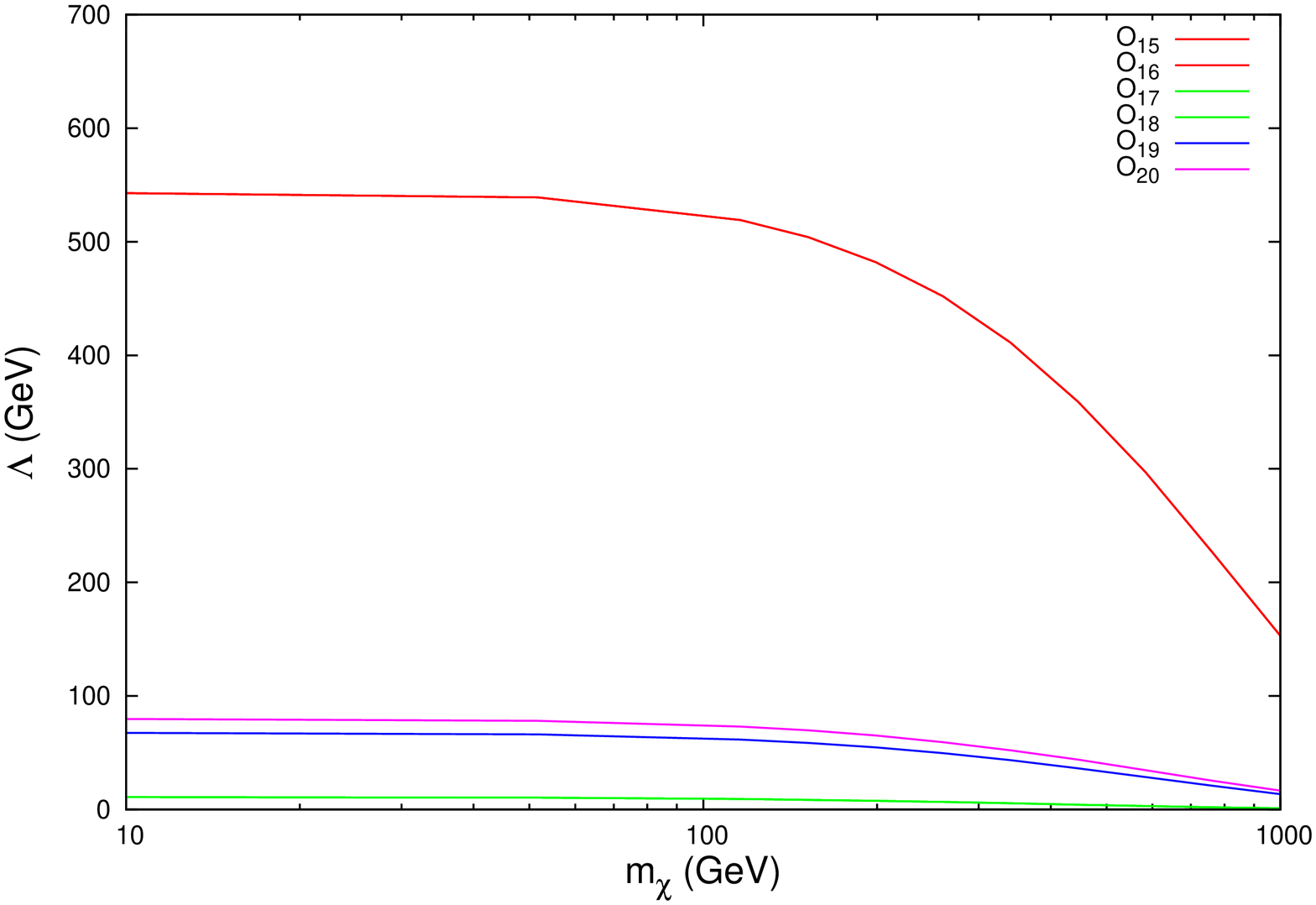}
\caption{\small \label{col}
Lower limits on $\Lambda$ due to monojet and monophoton data from
the Tevatron \cite{cdf,d01,d02} and the LHC \cite{atlas}.
}
\end{figure}

The data sets that we used in this analysis include:
(i) monojet and monophoton data from CDF \cite{cdf},
(ii) monophoton from D\O\ \cite{d01}, 
(iii) monojet from D\O\ \cite{d02}, and
(iv) monojet data from ATLAS \cite{atlas}.
Since the observed number of events are very close to the SM expectation
in each experiment, we use the number of observed events and the systematic
and statistical errors given by each experiment. For example, the observed 
number of monojet events in the very high $p_T$ selection region defined by 
ATLAS 
was 167, while the expected number from
the SM background with the errors is $193 \pm 15 \pm 20$ (see the last column 
in Table I of Ref.~\cite{atlas}.)  
Since the contribution from each operator does not interfere with the SM
background, we simply add it to the SM background, and so the chi-square is
\begin{equation}
                \chi^2 = \frac{ \left( N (\Lambda) + N_{SM} - N_{\rm obs} \right )^2 }
               {  (15^2 + 20^2 + 167 ) } \; ,
\end{equation}
where $N(\Lambda)$ is the contribution to the event number from an operator.
We show the $2\sigma$ limits for each operator in Fig.~\ref{col}.
We found that the chi-square is dominated by the monojet data of the ATLAS.
Note that since the production cross section decreases as the DM mass 
increases, one expects
the lower limit on the effective scale $\Lambda$ becomes weaker. 
This feature is clearly
reflected in Fig.~\ref{col}.

The operators involving electrons can also give rise to monophoton events 
at LEP, e.g., $(\bar \chi \gamma^\mu \chi)(\bar e \gamma_\mu e)/\Lambda^2$,
by attaching an external photon line to either the electron or positron leg.
Based on the LEP data, Refs.~\cite{yann,fox} obtained limits on the scale
$\Lambda$. For vector-type interaction they obtained $\Lambda > 470 - 400$
GeV for $m_\chi = 10 - 80$ GeV, while our limit for the same operator
is about 800 GeV (see Fig. ~\ref{col}(a)).
Since the cross section scales as $1/\Lambda^4$ for vector-type
interaction, the monophoton data from LEP would have negligible effects
to our limits.  For the scalar-type interaction the limits obtained in 
Refs.~\cite{yann,fox}, converted to our convention, are about 5 GeV, 
while our limit is about 44 GeV (see Fig.~\ref{col}(b)). Again, including the
monophoton data from LEP would have negligible effects on our results.

\section{Indirect Detection: Gamma-ray flux}

An important method to detect the dark matter is by measuring its
annihilation products in Galactic halo. Current experiments can detect
the positron, antiproton, gamma ray, and deuterium from dark matter
annihilation.  The Milky Way halo may contain clumps of dark matter,
from where the annihilation of dark matter particles may give rise to
large enough signals.  There are a number of experiments dedicating 
to measuring the gamma-ray flux from DM annihilation.

The Galactic diffuse gamma rays originate primarily from the interactions
of high energy charged particles contained in cosmic rays with the nuclei
in the interstellar medium and the associated radiation fields of the 
charged particles, via a few mechanisms briefly described below.
While most of them are well understood, the extra-galactic component
has a larger uncertainty.  We will choose a normalization such that
the total background diffuse gamma-ray flux is consistent with
the $Fermi$-LAT measurement of diffuse gamma-ray flux in the low-latitude.
This approach is the same as the $Fermi$-LAT when they estimated the
extra-galactic diffuse component \cite{fermi-lat}.

The data on the photon spectrum from the low-latitude 
$(10^\circ < |b| < 20^\circ, \; 0^\circ < l < 360^\circ )$ 
\cite{fermi-lat}  recorded by
the $Fermi$-LAT indicated a continuous spectrum and mostly consistent
with the known backgrounds.  We can therefore use the data to constrain
on additional sources of gamma-ray, namely, the annihilation of the
dark matter into quarks, followed by fragmentation into neutral 
pions, which further decay into photons.  
The production of photons via
neutral pions is the dominant mechanism for gamma-rays 
at higher energies.
The quarks can also fragment into charged pions, which in turn decay
into muons and eventually electrons.  These electrons undergo the
inverse Compton scattering  and bremsstrahlung, which give rise to photons.
The photon flux coming off inverse Compton scattering and bremsstrahlung
tends to be dominant at lower photon energies (e.g.
$\lesssim 10$ GeV for large DM mass and $\lesssim 1$ GeV for small DM mass).
On the other hand, the synchrotron radiation mostly falls outside the
photon energy range of the ${\it Fermi}$-LAT. 

The choice of the data beyond the Galactic Center is simply
because the gamma-ray in the outside region is dominated by local sources 
(within our Galactic halo)
and we have clarity in understanding the background flux and point sources
within the low-latitude. 
On the other hand, 
the Galactic Center is supposed to have a number of known and known-unknown 
point sources, including a supermassive black hole near the Center, 
and perhaps some unknown sources too.  
Given the purpose
of constraining the new DM interactions it is better to pick the data from the 
low-latitude region that we understand the background better, rather than 
from the Galactic Center region with less control background despite having a 
larger flux.

\subsection{Background Diffuse Gamma Rays}

The Galactic diffuse gamma rays originate primarily from the interactions
of high energy charged particles contained in cosmic rays with the nuclei
in the interstellar medium and the associated radiation fields of the 
charged particles, via a few of the following mechanisms.

\begin{itemize}
\item[{\rm (i)}]  
Gamma-rays coming from the $\pi^0$ decay, which was originated from
the interactions of the cosmic rays with the nucleons in the interstellar
medium.  

\item[{\rm (ii)}] 
Inverse Compton scattering occurs when high energy $e^\pm$ collide
with the photons of the interstellar medium, such as
CMB, star-light, and far-infrared photons.

\item[{\rm (iii)}] 
Bremsstrahlung photons occur when high energy $e^\pm$ are deflected
by the Coulomb field of the interstellar medium.

\item[{\rm (iv)}]
Those point sources that have been identified by ${\it Fermi}$-LAT in the 
low-latitude region \cite{fermi-lat}.

\item[{\rm (v)}]
Synchrotron radiation occurs when high energy $e^\pm$ are deflected
by Galactic magnetic field.  
However, synchrotron radiation only gives a very weak flux 
in the photon energy range collected by $Fermi$-LAT. We would not include
synchrotron radiation in the background flux.

\item[{\rm (vi)}] 
An extragalactic background (EGB), which is expected to be isotropic
and receives contributions from many sources including 
unresolved point sources (PS), diffuse emission from 
large scale structure formation 
and from interactions between ultra-high energy cosmic ray background (CRB),
and relic photons, etc.
This background is the least determined and so a fairly large uncertainty
is associated with it. 
Following $Fermi$-LAT we use a parameterization for the photon flux
\begin{equation}
\label{fit}
  E^2 \frac{ d \Phi}{d E} = A \left( \frac{E}{0.281 \; {\rm GeV}} 
\right)^\delta \;,
\end{equation}
where $A$ and $\delta$ are fitted parameters (the power-law index is 
$\gamma=\delta - 2$) to fit the extra-galactic background (EGB) by
minimizing the $\chi^2_{\rm bkdg}$ from the data:
\begin{equation}
\chi^2_{\rm bkgd} = \sum_i \left( 
\frac{ \Phi_{\rm SAB} (E_i) + \Phi_{\rm EGB} (E_i) + \Phi_{\rm CRB}( E_i)
  +\Phi_{\rm PS} (E_i) - \Phi_{\rm data}(E_i) }{ \sigma_{\rm total}(E_i) } 
\right )^2 \;,
\end{equation}
where $\sigma^2_{\rm total} = \sigma_{\rm CRB}^2 + \sigma_{\rm PS}^2 +
\sigma_{\rm data}^2$. 
Hence, we used the best-fitted point for the standard astrophysical background
(SAB).
The normalization $A$ will be varied freely when combining with 
the contribution from DM annihilation.

\end{itemize}
All the above sources (i) to (iii) are referred as the standard astrophysical
background (SAB).  We include the SAB, point sources, and the EGB as the
background photon flux in our analysis.
The dominant uncertainty comes from the propagation parameters inside
GALPROP \cite{GALPROP}.
We use the best-fit model from GALPROP group \cite{bestfit},
in which they fitted to a number of isotopic ratios such as 
B/C, Be$^{10}$/Be$^{9}$, and so on. 
We employ the NFW profile with the caution that the halo uncertainty can give 
as much as a factor of $O(10)$ change to the photon flux. Since the 
annihilation cross section scales as either $1/\Lambda^4$ or 
$1/\Lambda^6$, this translates to the uncertainty
within about 50\% of the lower limit of $\Lambda$.

\begin{figure}[th!]
\centering
\includegraphics[width=3.2in]{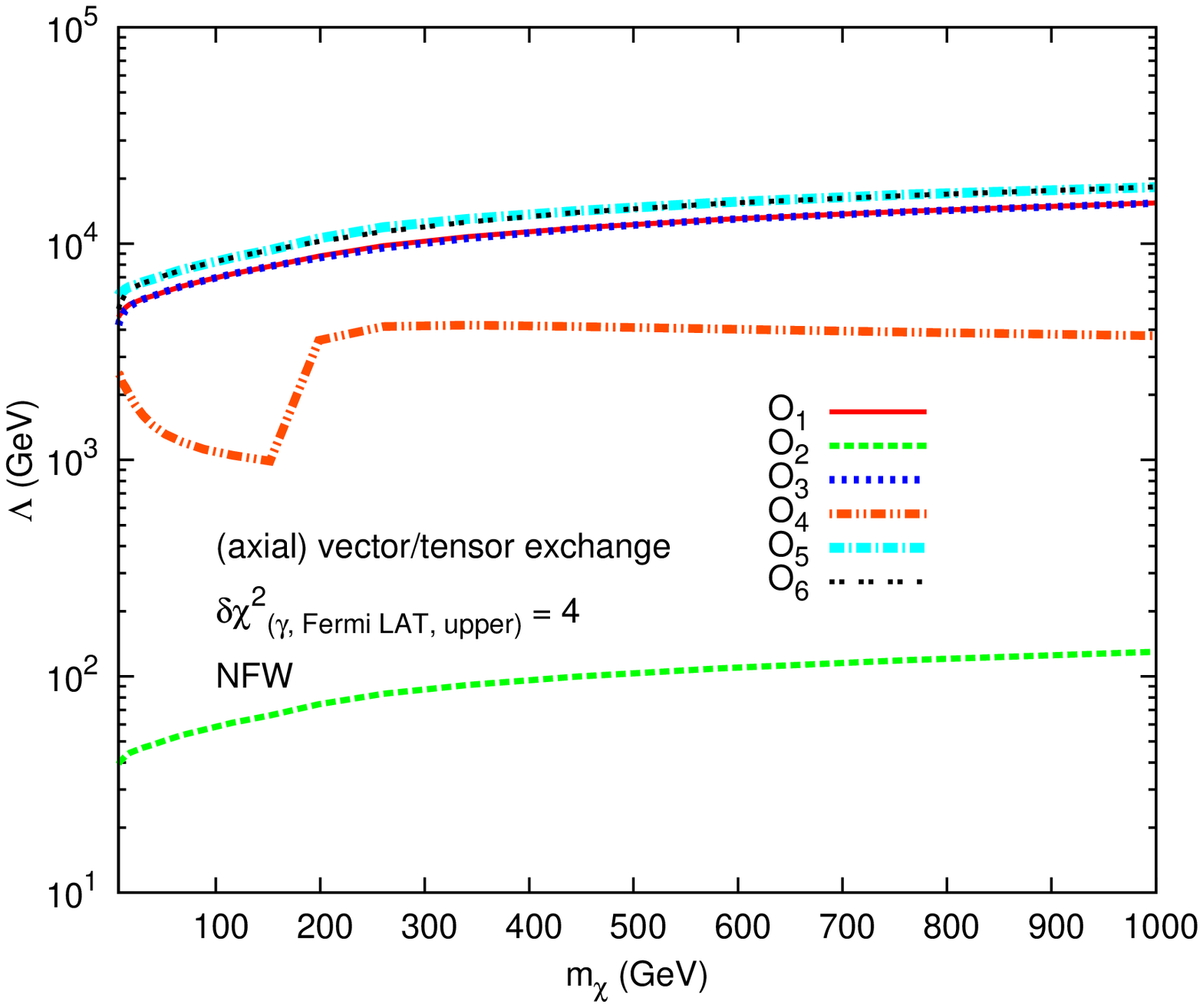}
\includegraphics[width=3.2in]{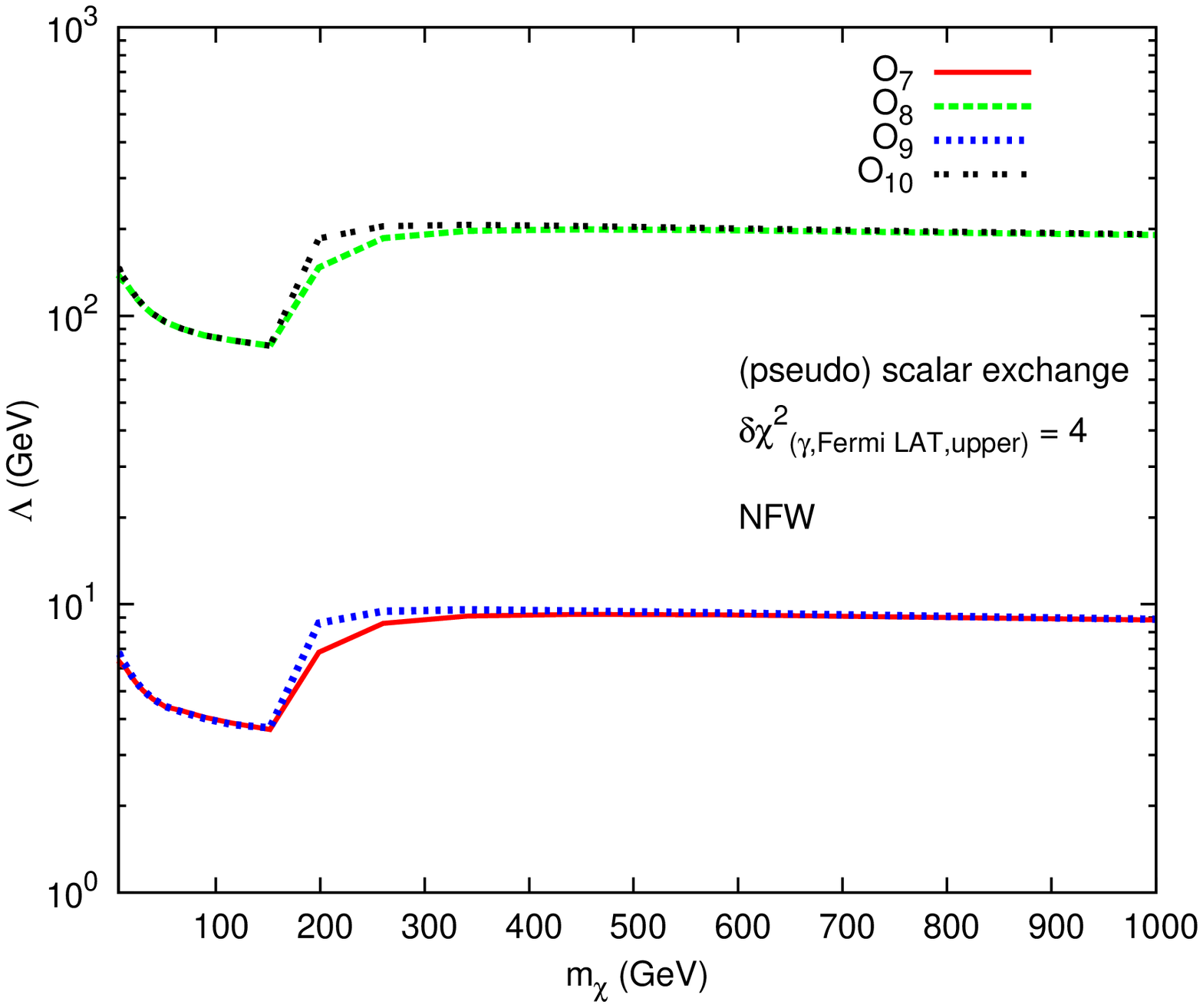}
\includegraphics[width=3.2in]{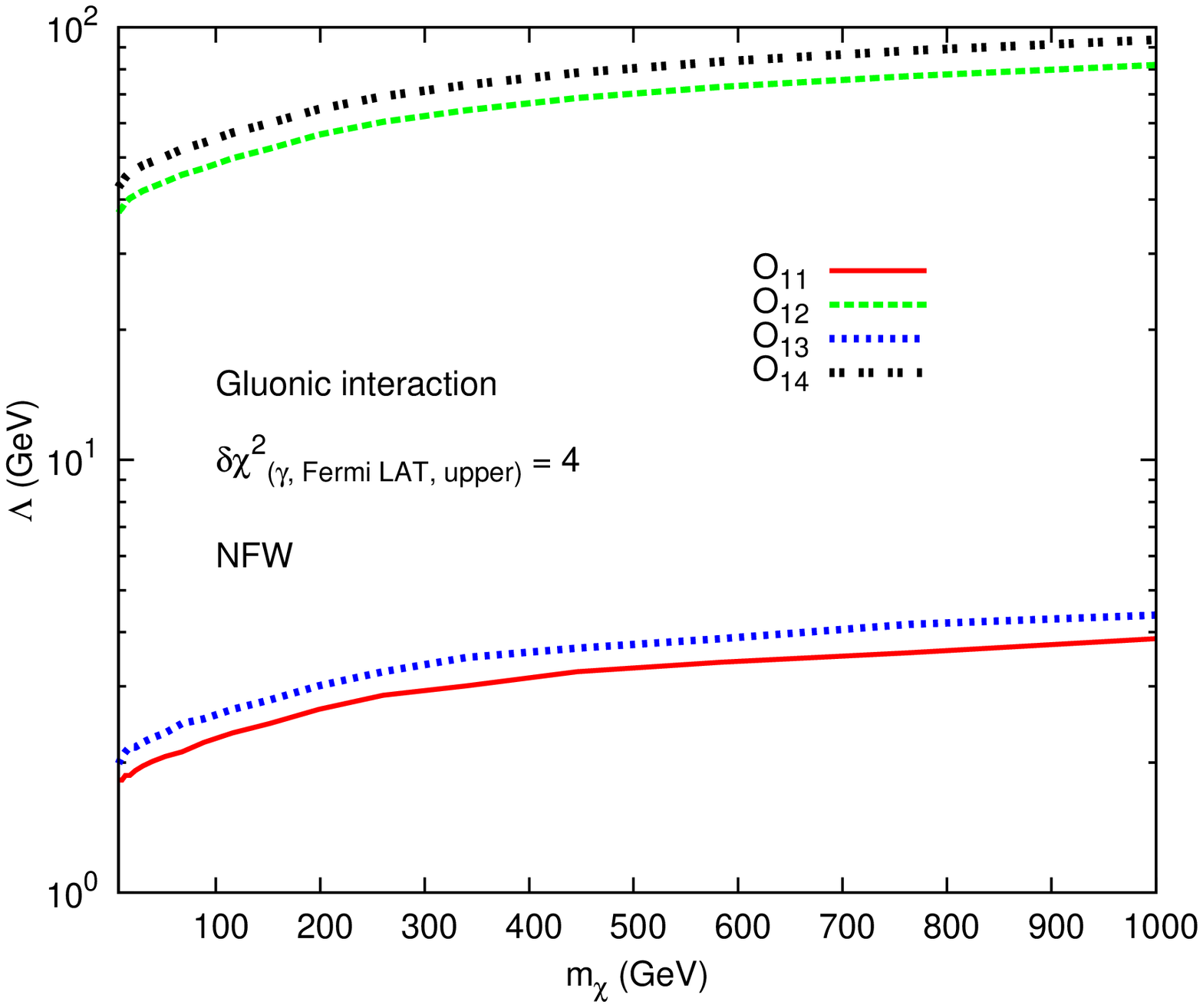}
\includegraphics[width=3.2in]{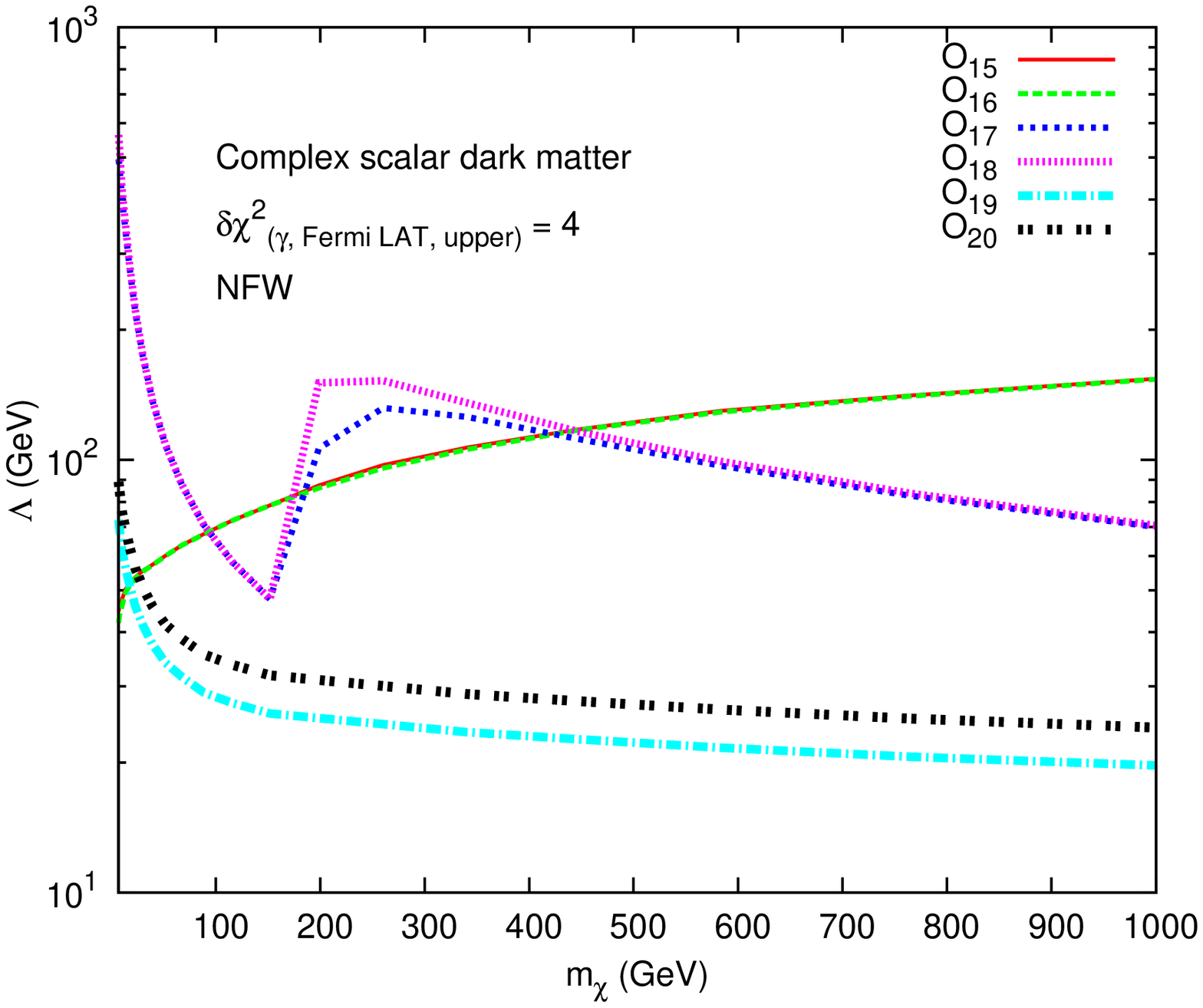}
\caption{\small \label{gam}
The $2\sigma$ lower limits on $\Lambda$ for each operator due to
the low-latitude gamma-ray-flux data of ${\it Fermi}$-LAT \cite{fermi-lat}. 
}
\end{figure}

\subsection{Dark Matter Annihilation}

In Ref.~\cite{gamma-4}, monochromatic photon-line flux was 
calculated via a loop with fermions running in it and photons being 
attached to the internal fermion line. Although the photon-line would be a 
smoking-gun signal to compare with the data, the rate is suppressed 
because of the
loop factor. On the other hand, photons may come from the decay of
neutral pions, which are originated from the fragmentation of the quarks
in the annihilation of the dark matter. The chance that an energetic 
quark fragments into neutral pions is high and the branching ratio of
a neutral pion into two photons is 98.823\%.
Therefore, the amount
of photons coming from the quark fragmentation is much larger than 
those coming off a loop process.  Nevertheless, the spectrum of such
photons is continuous and in general have no structure, 
except for a cutoff due to the mass of the dark matter.
In this work, we focus on the continuous gamma-ray flux spectrum
coming from the fragmentation of quarks into neutral pions, followed
by their decays into photons, in the annihilation of the dark matter.
In addition, the quarks can also fragment into charged
  pions, which subsequently decay into muons and eventually electrons. The
  dark matter can also directly annihilate into taus, muons, and
  electrons.  The taus and muons will eventually decay into electrons.
  All these electrons undergo the inverse Compton scattering and
  bremsstrahlung, which may give rise to photons too.  We include all these 
  effects in calculating the gamma-ray flux from DM annihilation.
Such annihilation of DM will give rise to an additional source
of diffuse gamma-rays other than the known backgrounds.
If the experimental measurement is consistent with the known gamma-ray
background estimation, then one could use the data to constrain the
amount of gamma-ray flux coming from the dark matter annihilation, thus
constraining the effective interactions between the dark matter and
the fermions.

We modified DarkSUSY \cite{darksusy} for the effective DM interactions
under consideration to generate the photon spectrum $dN_\gamma/d
E_\gamma$ and $e^\pm$ spectrum $d N_{e^\pm} / dE_{e^\pm}$
of a DM annihilation operator of Sec. II for a particular DM
mass and a selected $\Lambda$, say $\Lambda = 300$ GeV.  
\footnote{
  Inside DarkSUSY \cite{darksusy}
  there are some PYTHIA tables, which were generated
  using PYTHIA \cite{pythia} 
  to simulate quark or gluon fragmentation into pions and kaons with a
  central energy ($2m_\chi$) and then decay/annihilate to $\gamma,
  e^+, \bar{p}$, and so on.  After collecting a large number of events
  (say $\sim 10^7$), histograms of $dN/dE(m_\chi,\,E_p)$ vs $E_p$ are
  tabulated for recycle uses.  As described in the manual, the
  uncertainties should be less than a factor of 2.}
Note that the DM annihilation cross section scales as either $1/\Lambda^4$
or $1/\Lambda^6$ depending on operators.  The photon and $e^\pm$ spectra
are then fed into GALPROP with the same running parameter of 
the best-fit model \cite{bestfit}.
The output photon flux then includes $\pi^0$ decays,
bremsstrahlung, and inverse-Compton scattering. This DM flux is then
added to all the other astrophysical background fluxes and compared to
the data as
\begin{equation}
\chi^2_{\rm DM} = \sum_i \left( 
\frac{ \Phi_{\rm DM} + \Phi_{\rm SAB} (E_i) + \Phi_{\rm EGB} (E_i) + 
 \Phi_{\rm CRB}( E_i) +\Phi_{\rm PS} (E_i) - \Phi_{\rm data}(E_i) }
  { \sigma_{\rm total}(E_i) }  \right )^2 \;
\end{equation}
where $\sigma^2_{\rm total} = \sigma_{\rm CRB}^2 + \sigma_{\rm PS}^2 +
\sigma_{\rm data}^2$.
Both the normalization $A$ of the EGB component and the scale 
$\Lambda$ of the DM contribution are allowed to vary freely in the fit.

 Here we adopt a simple statistical measure to quantify the effect
of each DM operator.  We calculate the $2\sigma$ 
limit on each scale $\Lambda_i$ while allowing the normalization $A$ to
vary, until we obtain a chi-square difference of 
$\Delta \chi^2_{\rm DM} \equiv \chi^2_{\rm DM} - {\rm min}( \chi^2_{DM} ) = 4$  
($2\sigma$). For each operator we repeat the procedures for each DM mass.
We show the results in Fig.~\ref{gam}. 
For those unsuppressed operators the limit is of
order $O({\rm TeV})$. But for those operators suppressed by the 
velocity of the DM, 
light fermion masses or strong coupling constant, the limit is significantly 
weaker of order $0.01 - 0.1$ TeV. The effects due to the onset of the 
heavy top quark in the final state are discernible by the cusps seen at 
some of these curves in Fig.~\ref{gam}.

\begin{figure}[t!]
\centering
\includegraphics[width=2.5in]{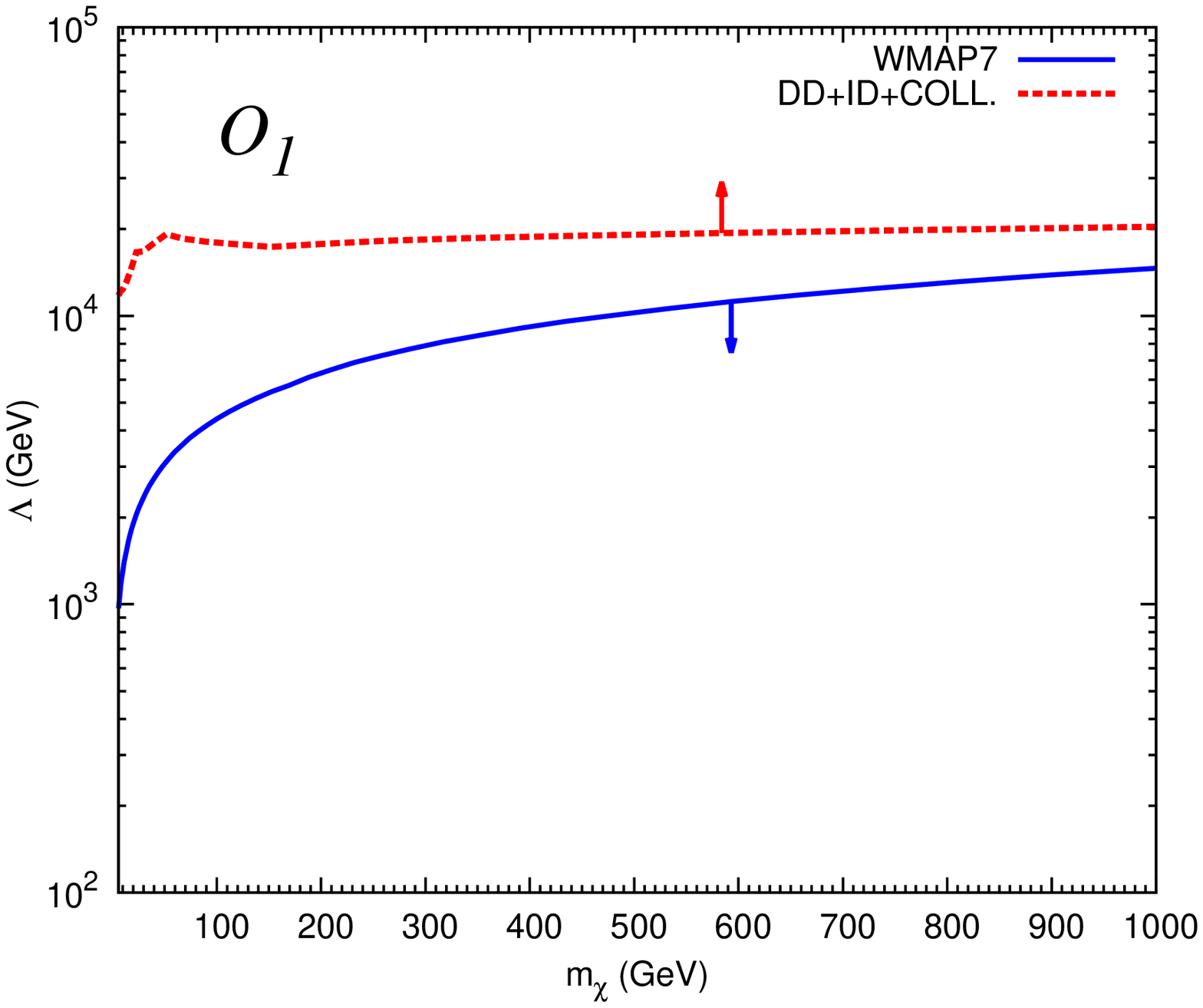}
\includegraphics[width=2.5in]{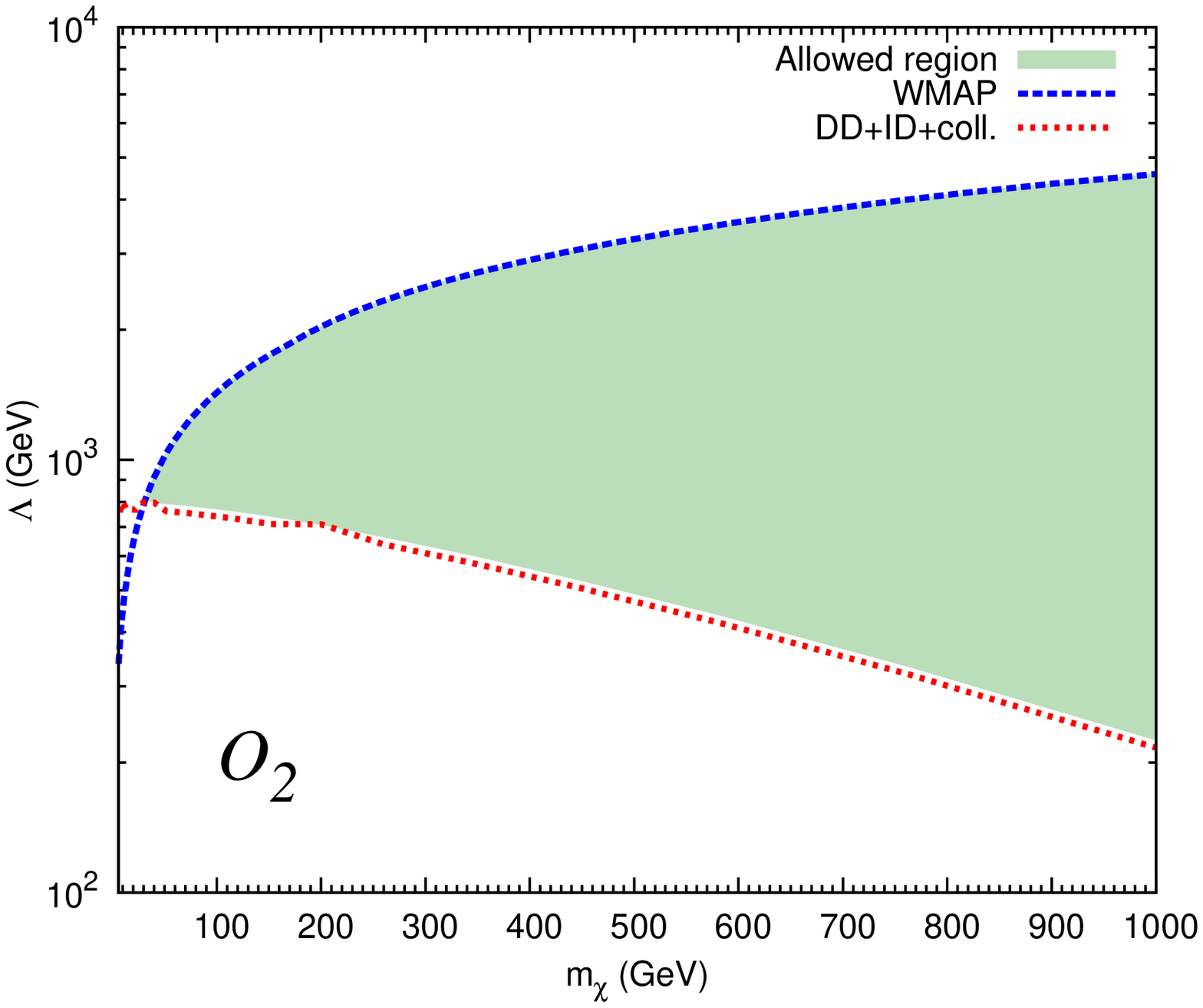}
\includegraphics[width=2.5in]{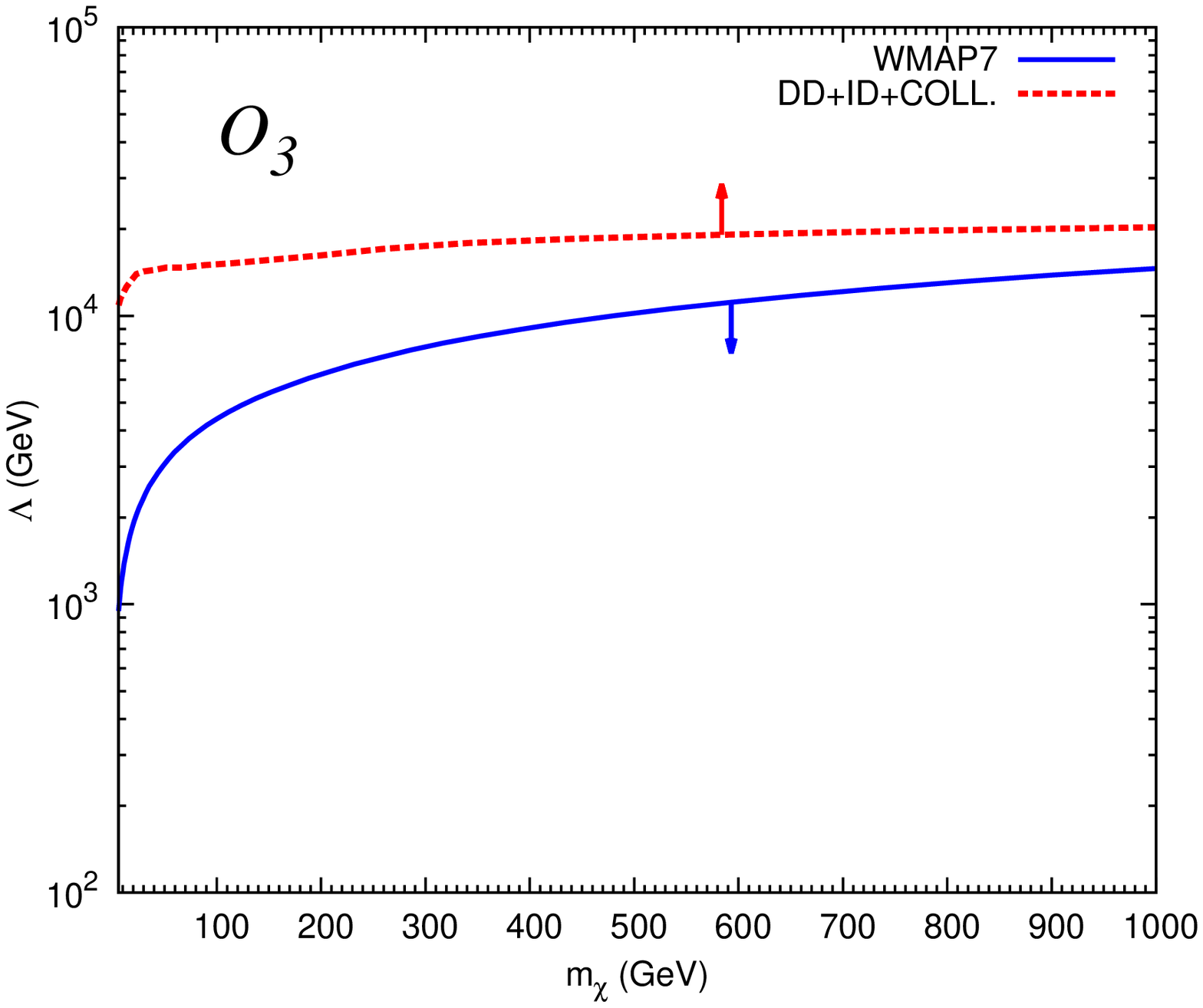}
\includegraphics[width=2.5in]{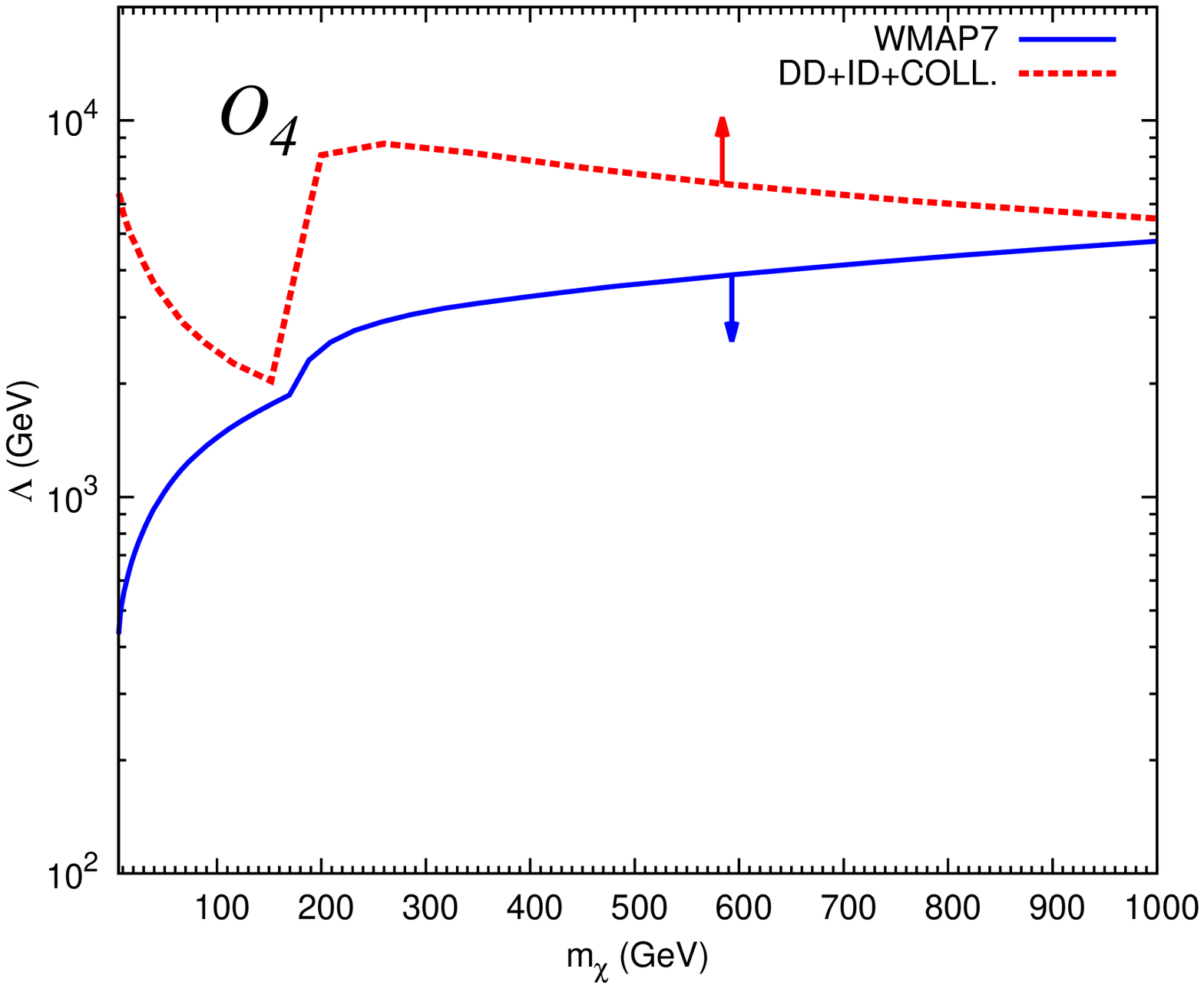}
\includegraphics[width=2.5in]{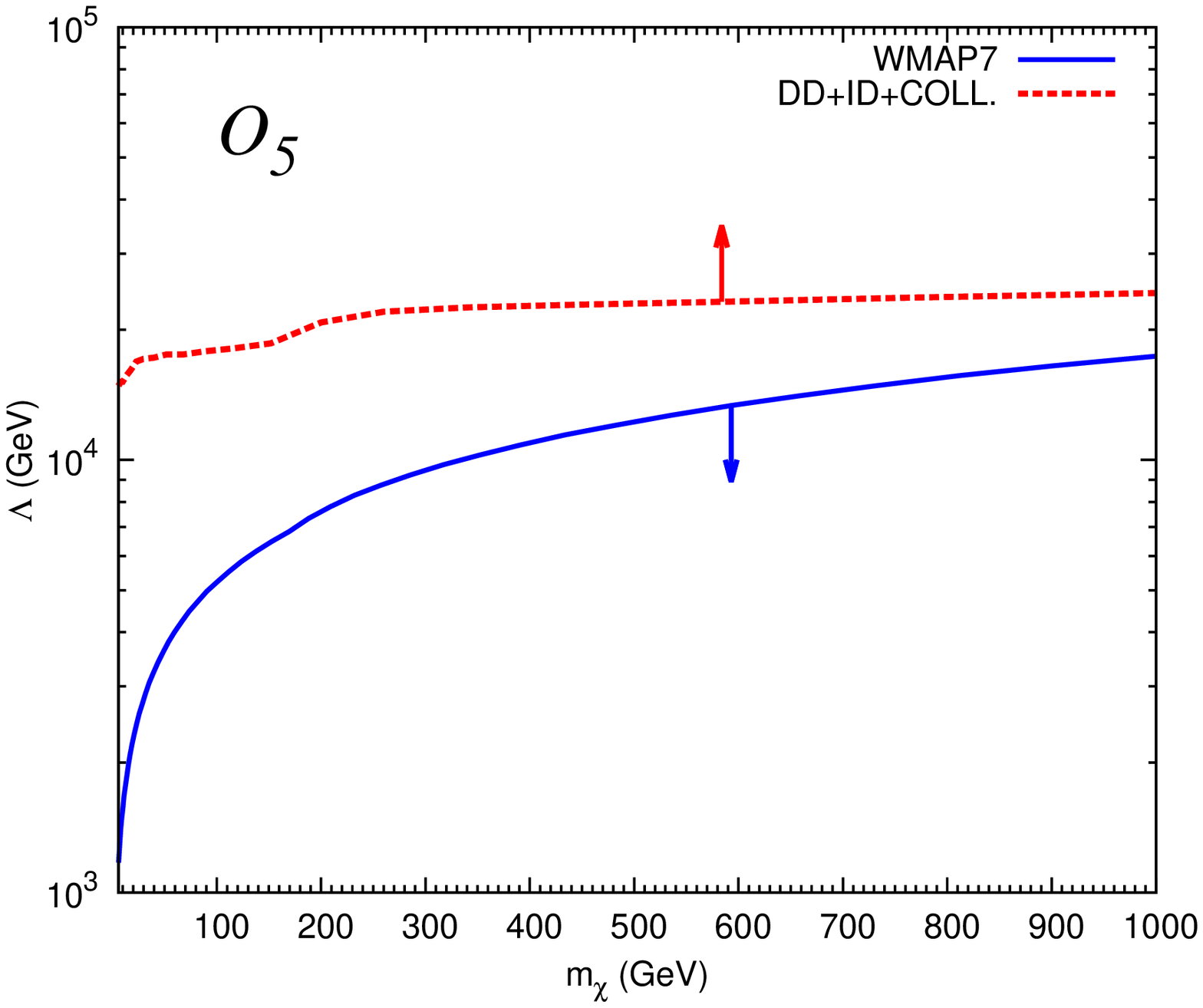}
\includegraphics[width=2.5in]{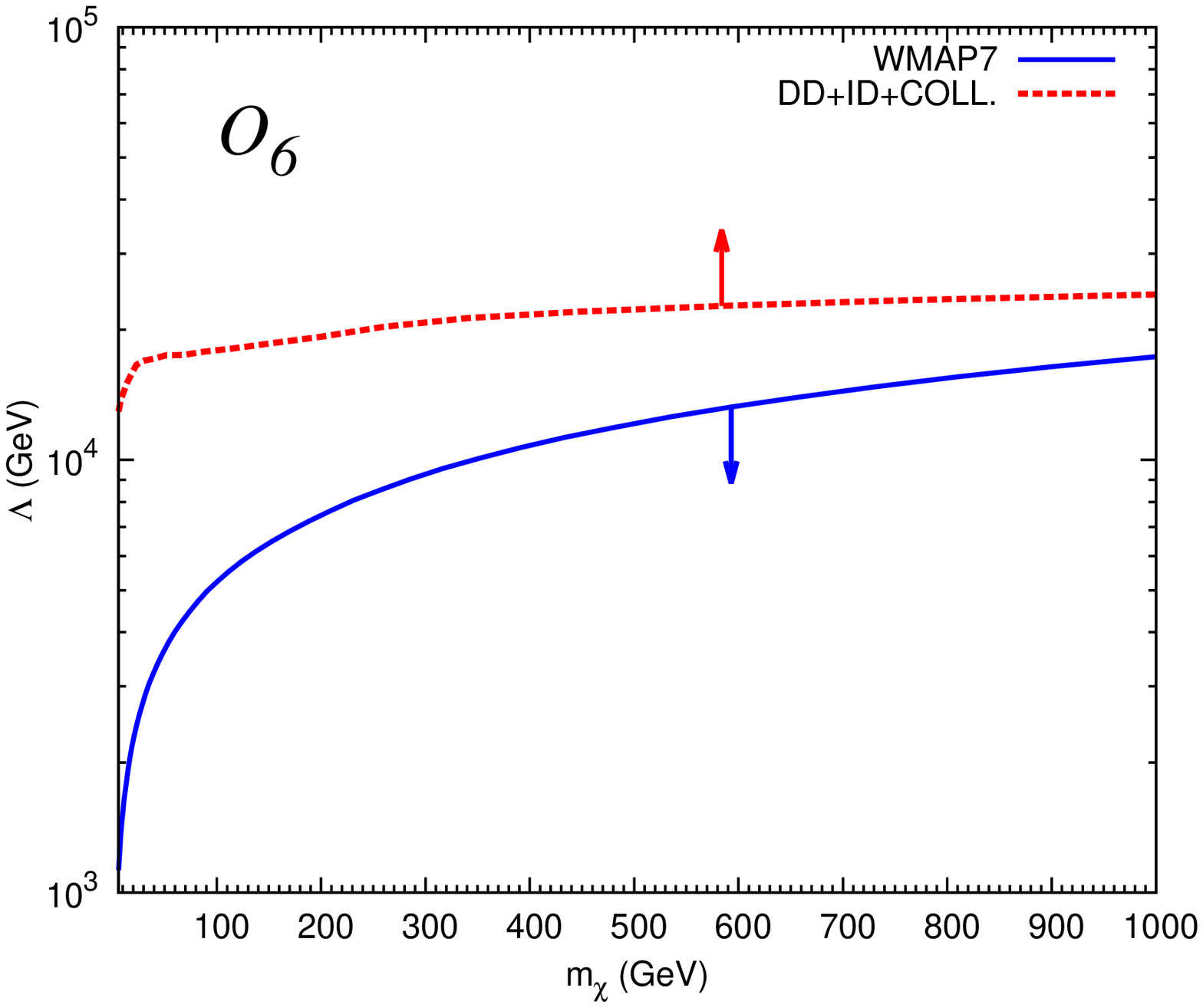}
\caption{\small \label{combined-O1-O6}
The combined analysis for $O_{1}$, $O_{2}$, $O_{3}$, $O_{4}$, $O_{5}$ and $O_6$.  
In each panel, the WMAP7 data requires the area below the blue 
curve (indicated by the blue arrow)
while all the other data requires the area above the red curve 
(indicated by the red arrow).  The allowed region is shaded for $O_2$.
}
\end{figure}

\begin{figure}[t!]
\centering
\includegraphics[width=2.5in]{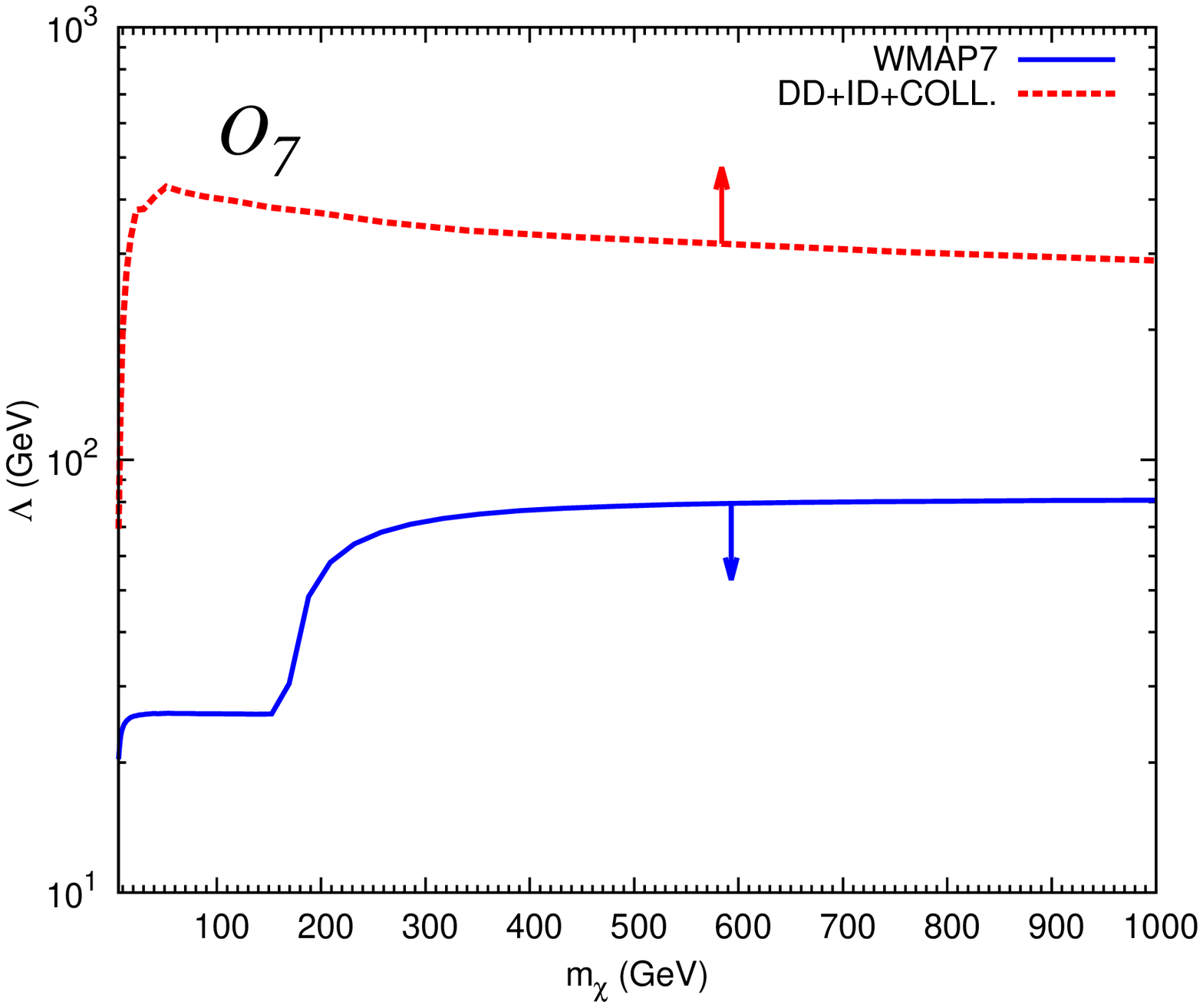}
\includegraphics[width=2.5in]{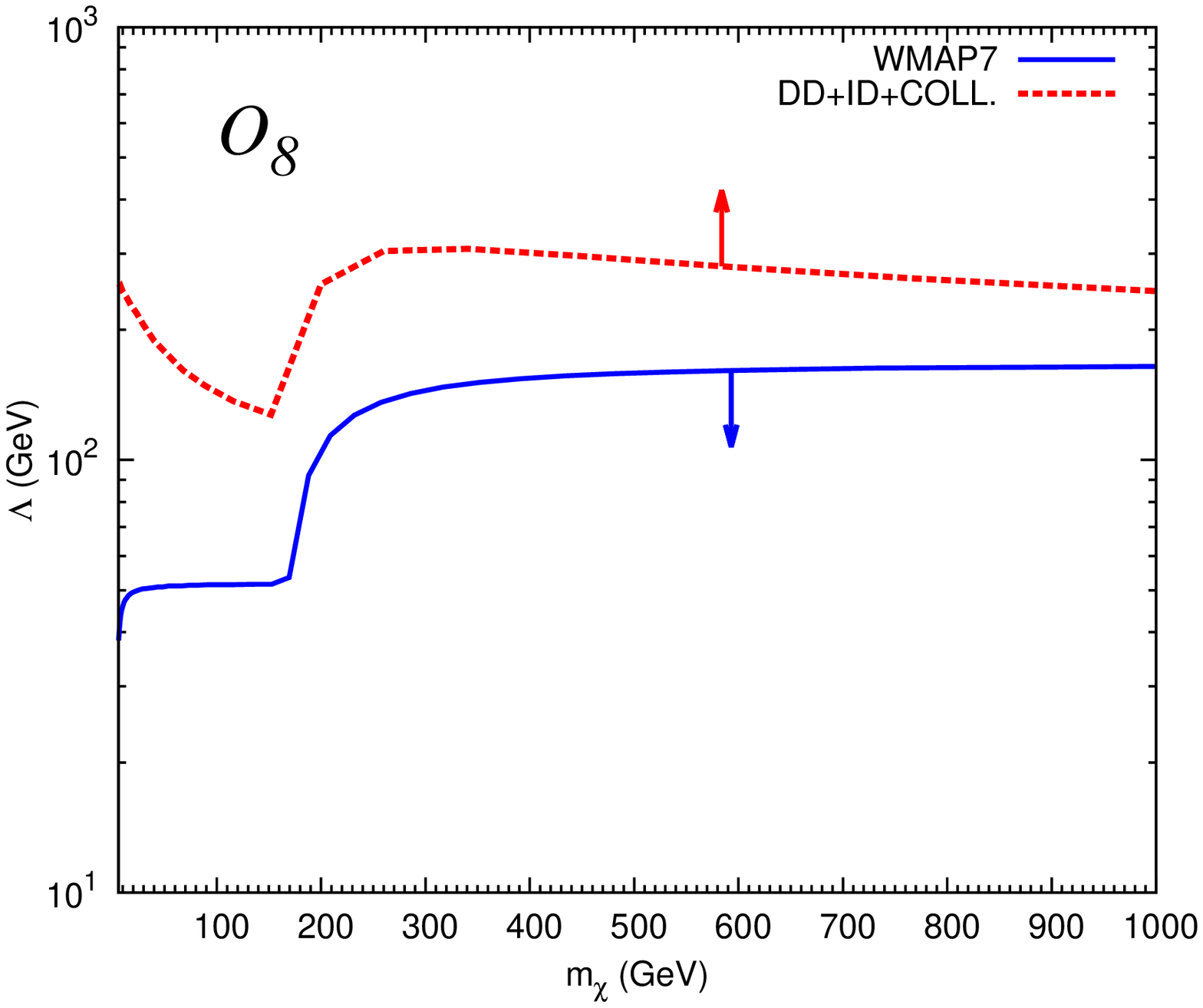}
\includegraphics[width=2.5in]{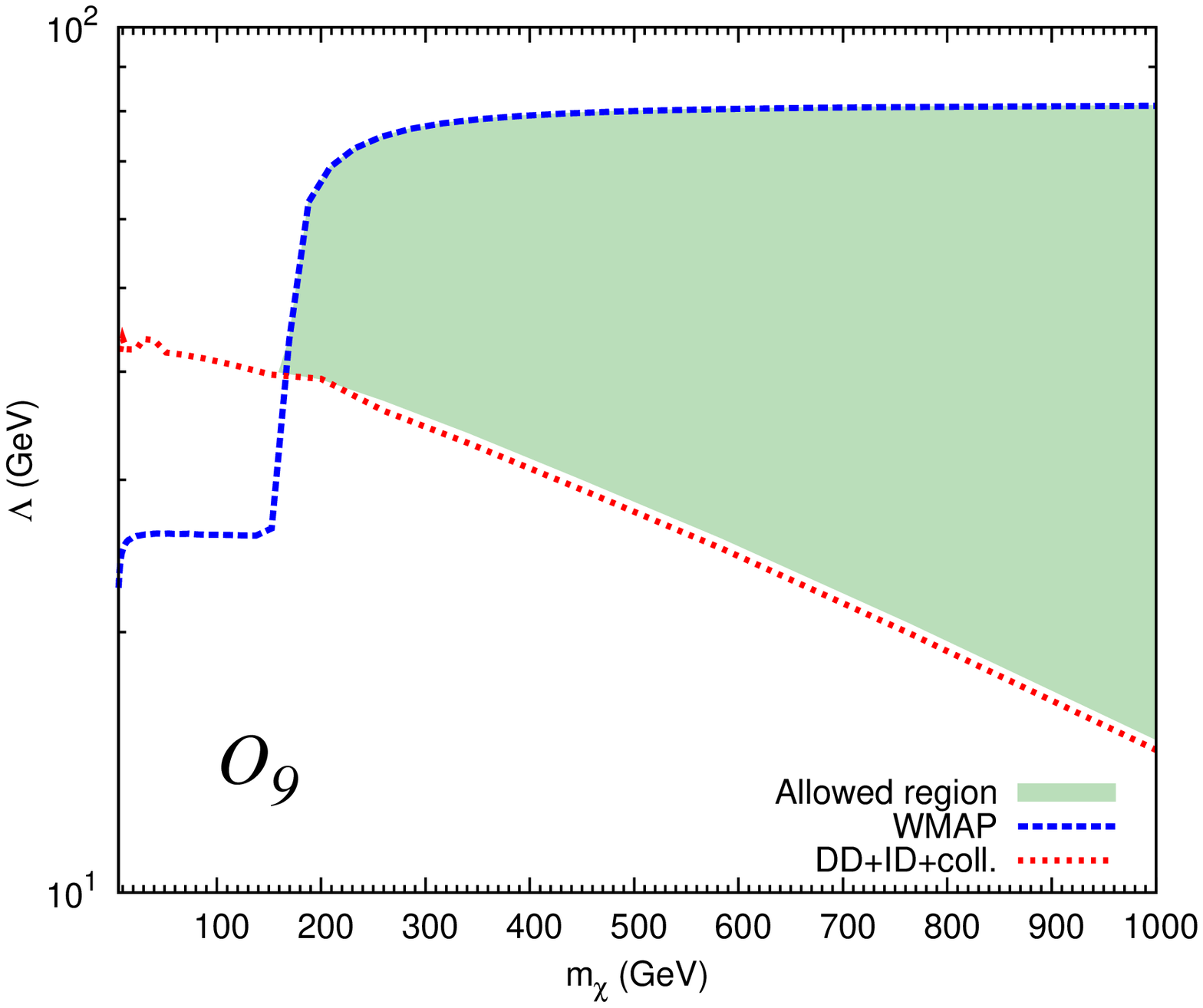}
\includegraphics[width=2.5in]{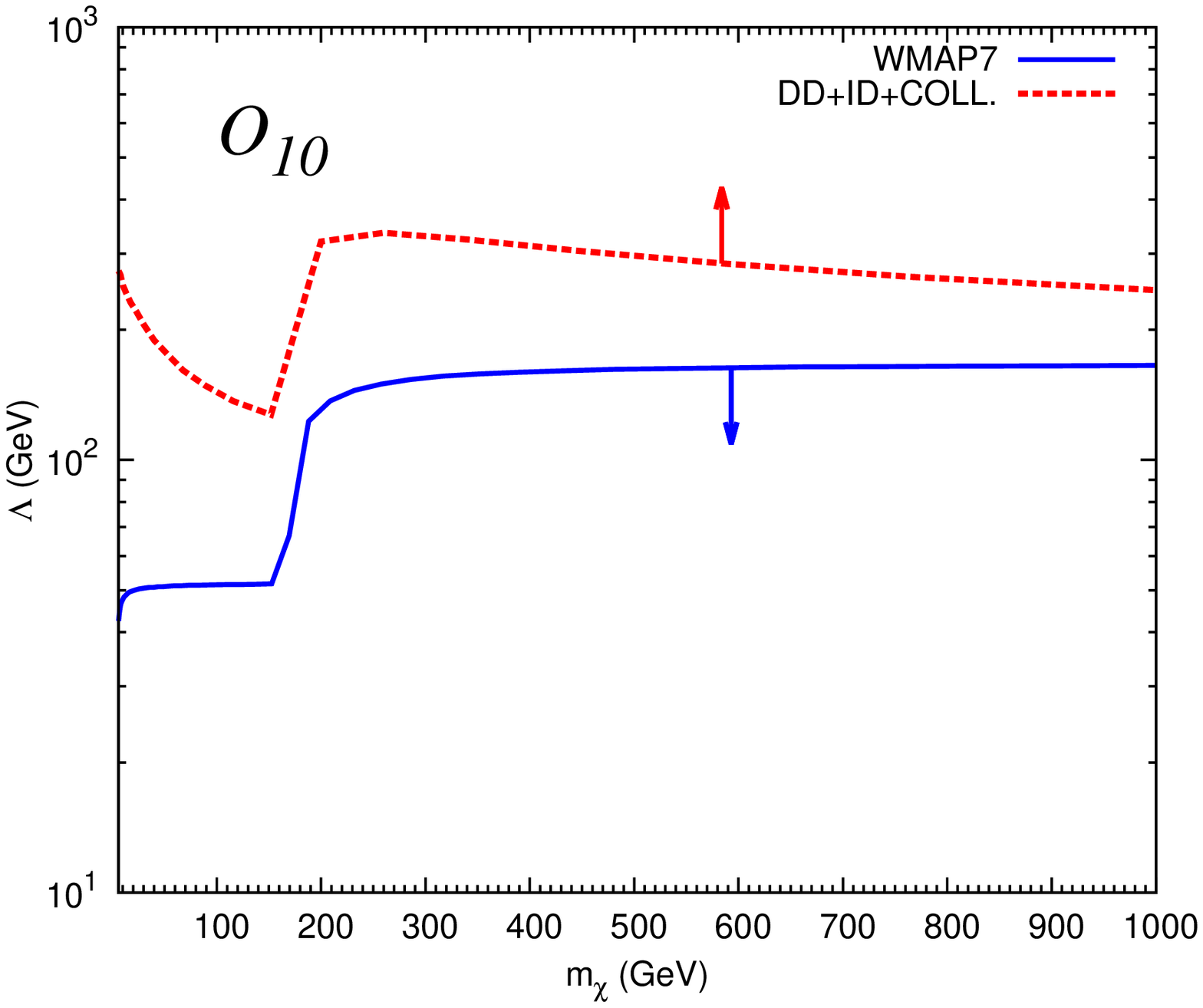}
\caption{\small \label{combined-O7-O10}
The combined analysis for $O_{7}$, $O_{8}$, $O_{9}$ and $O_{10}$.  
In each panel, the WMAP7 data
requires the area below the blue curve (indicated by the blue arrow)
while all the other data requires the area above the red curve 
(indicated by the red arrow).  The allowed region is shaded for $O_9$.
 }
\end{figure}

\begin{figure}[t!]
\centering
\includegraphics[width=2.5in]{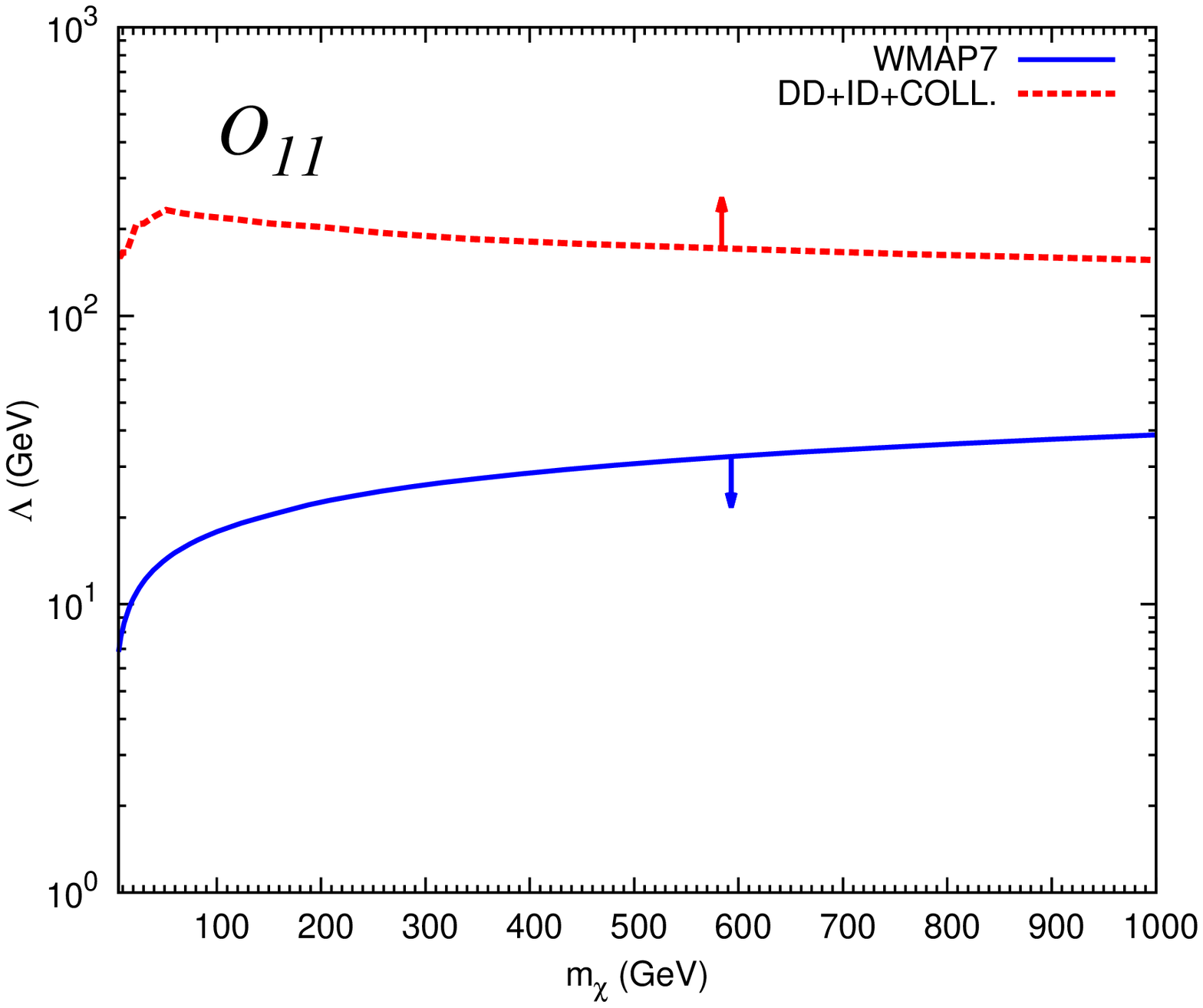}
\includegraphics[width=2.5in]{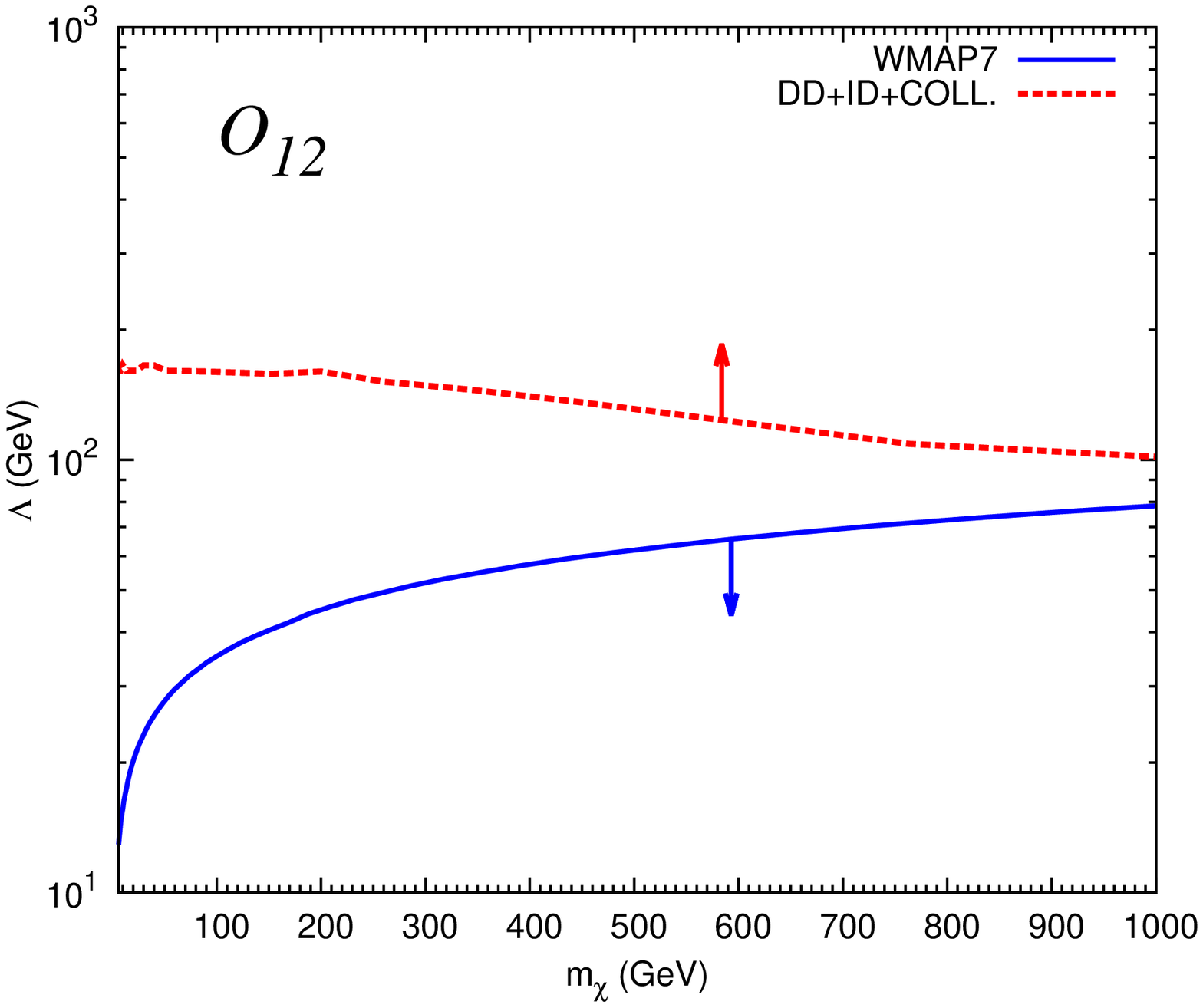}
\includegraphics[width=2.5in]{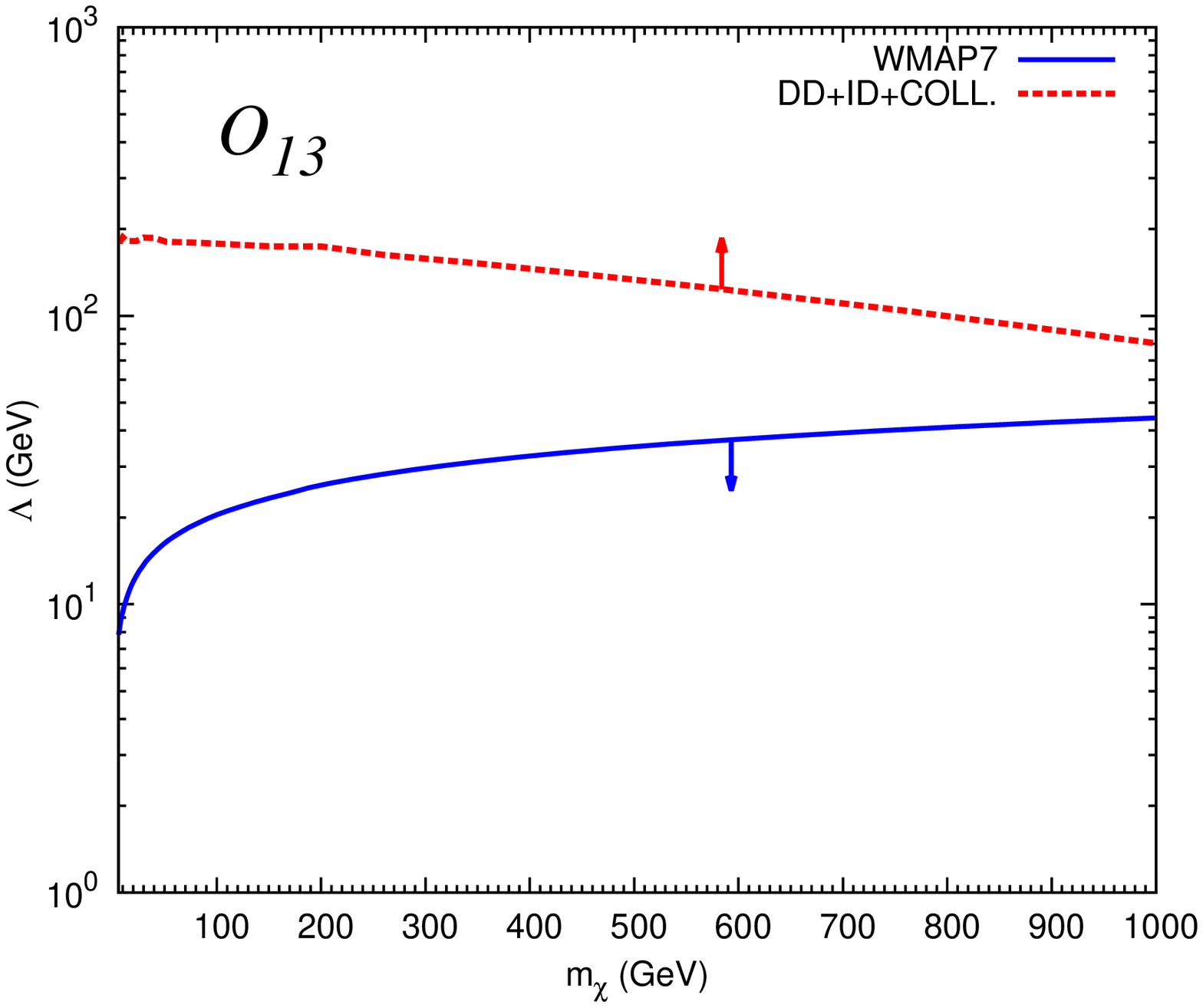}
\includegraphics[width=2.5in]{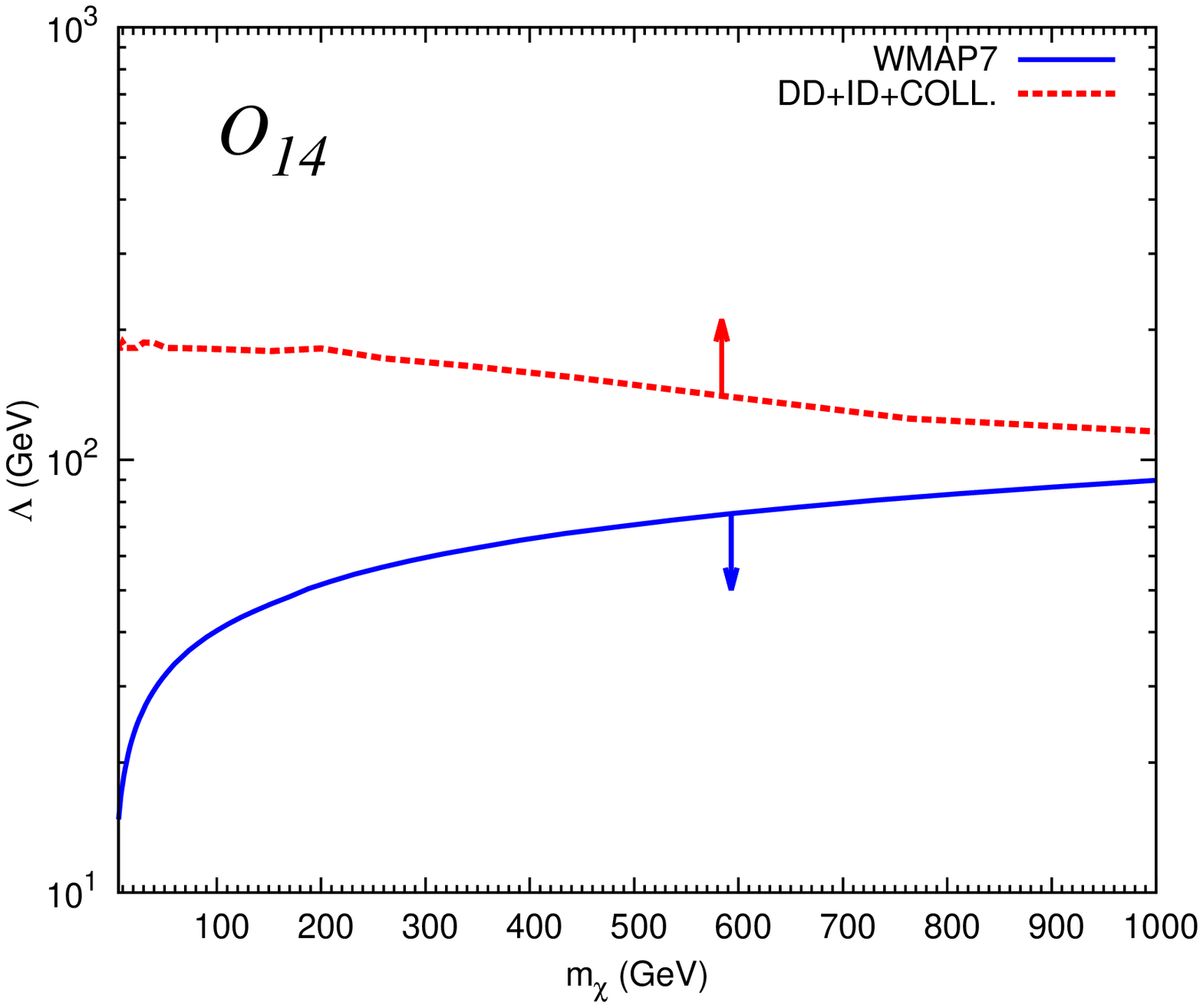}
\caption{\small \label{combined-O11-O14}
The combined analysis for operators $O_{11}$, $O_{12}$, $O_{13}$ and $O_{14}$.  
In each panel, the WMAP7 data requires the area below the blue curve 
(indicated by the blue arrow)
while all the other data requires the area above the red curve 
(indicated by the red arrow).  Allowed region for these operators do not exist.
 }
\end{figure}

\begin{figure}[t!]
\centering
\includegraphics[width=2.5in]{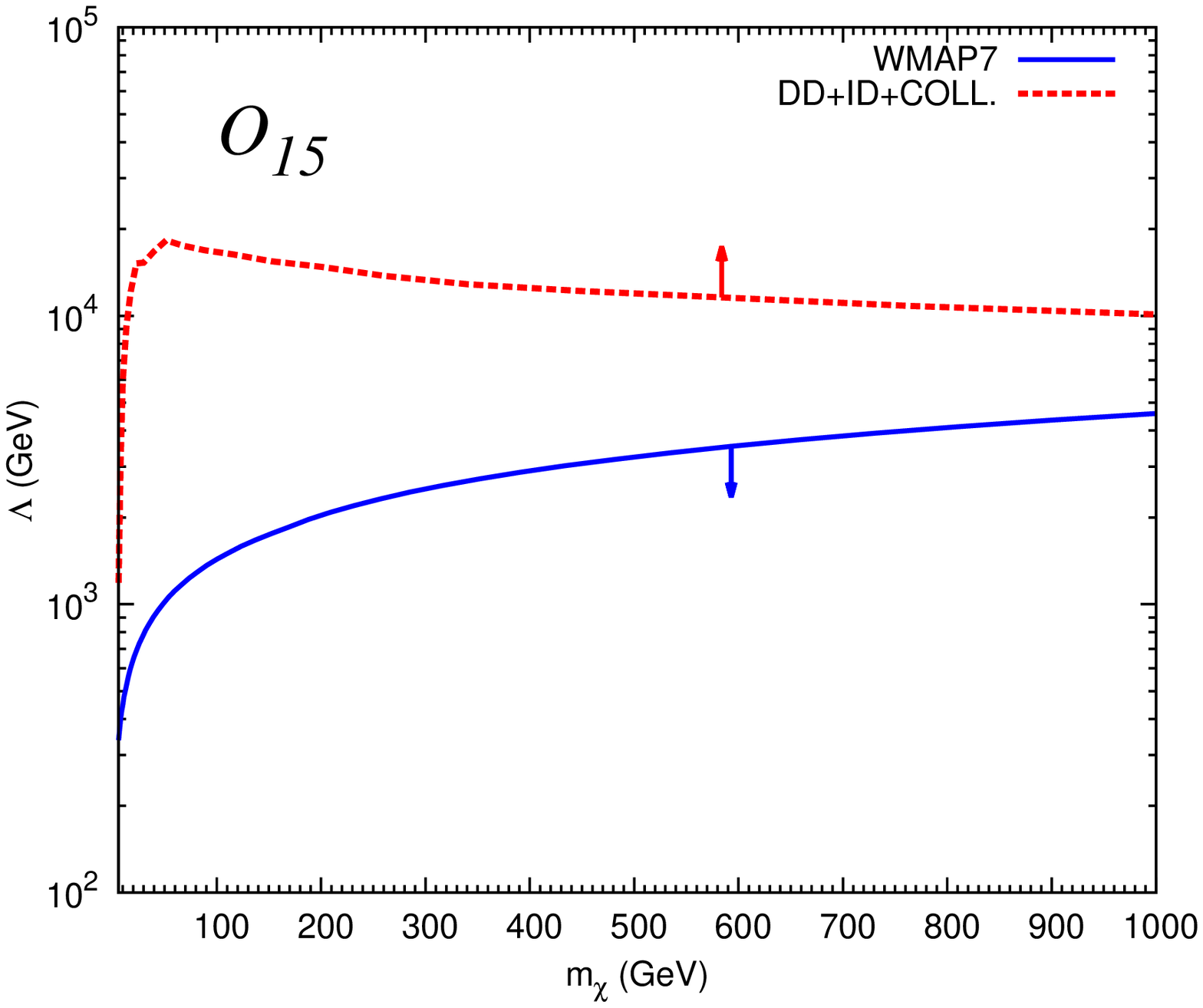}
\includegraphics[width=2.5in]{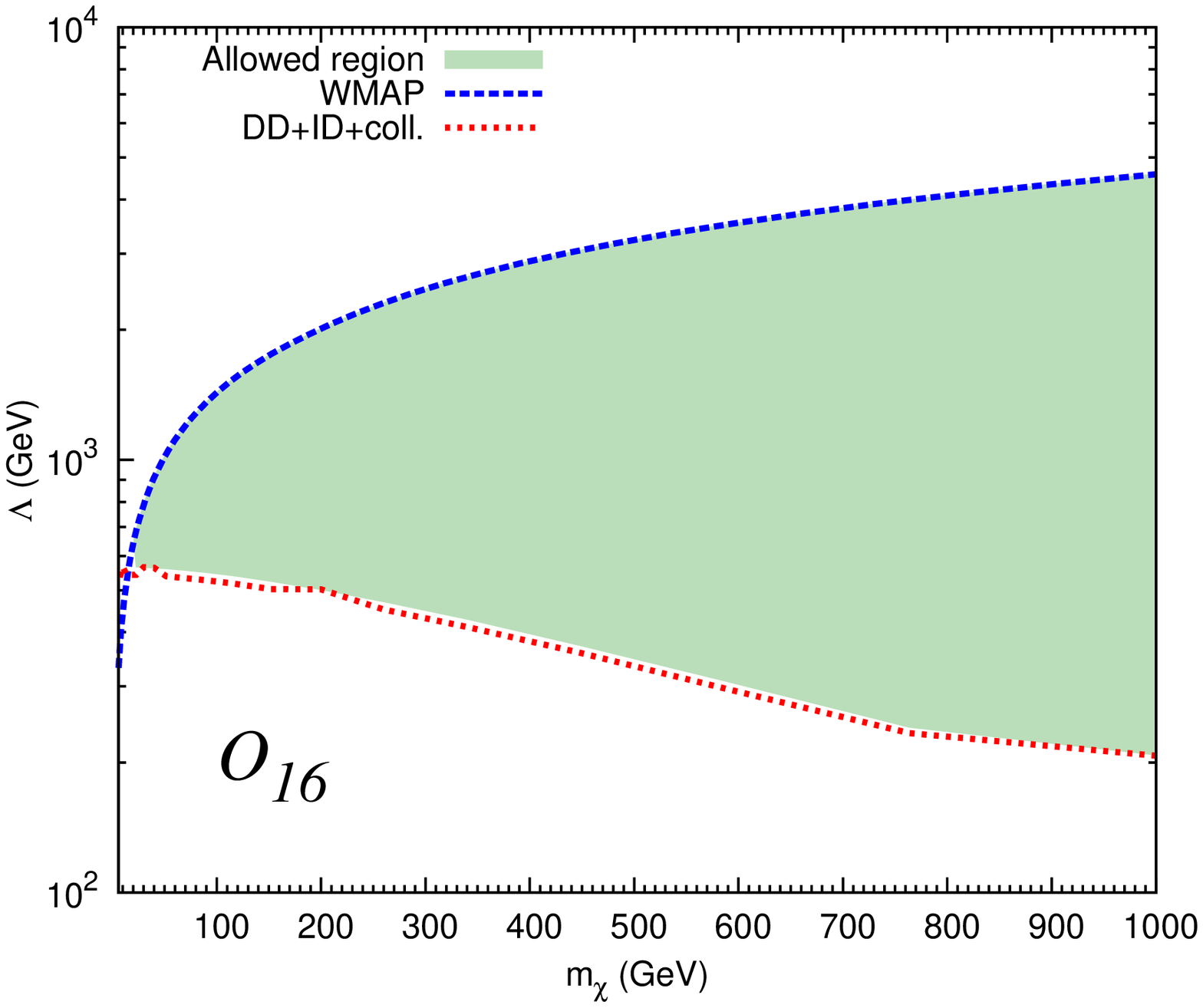}
\includegraphics[width=2.5in]{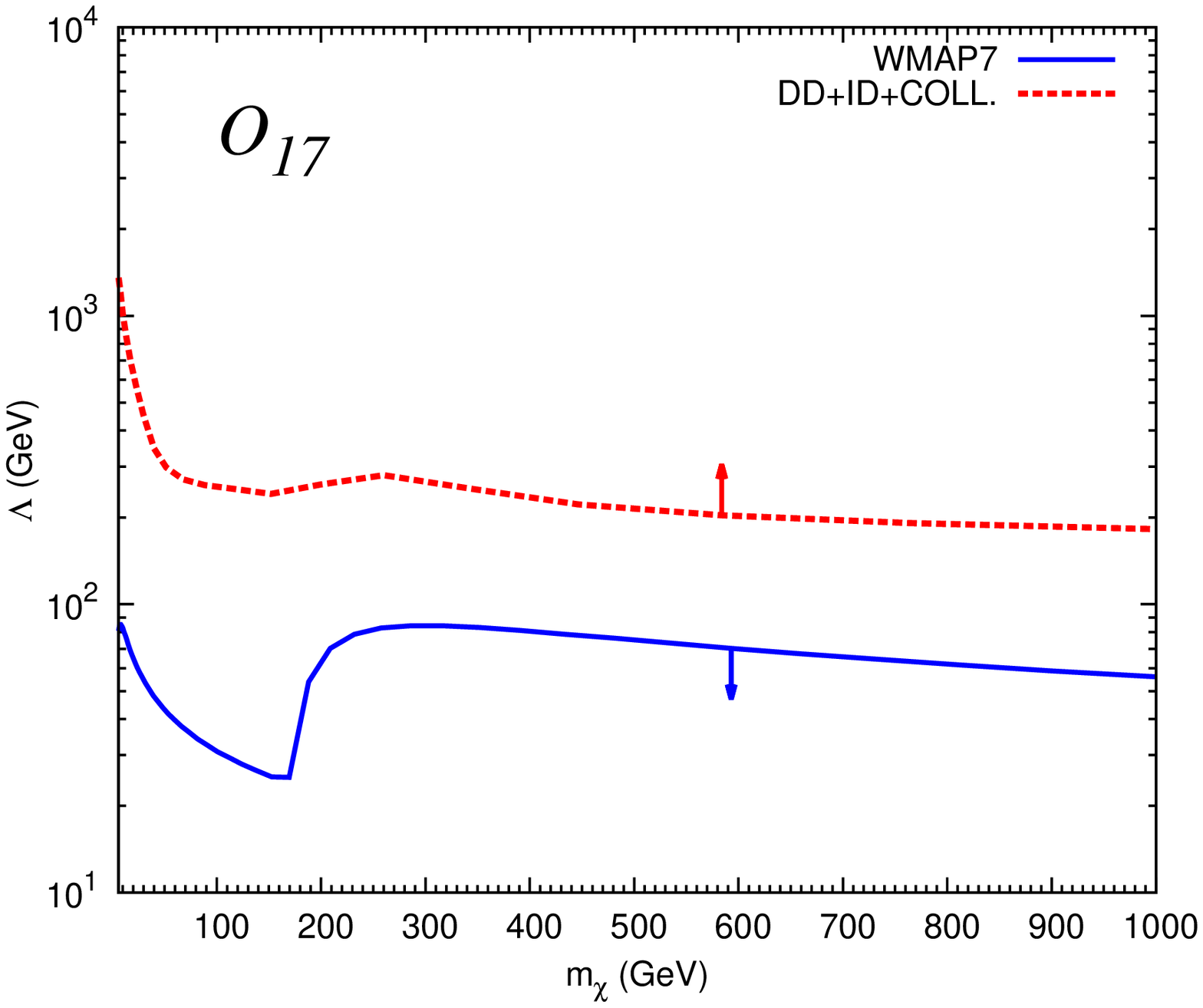}
\includegraphics[width=2.5in]{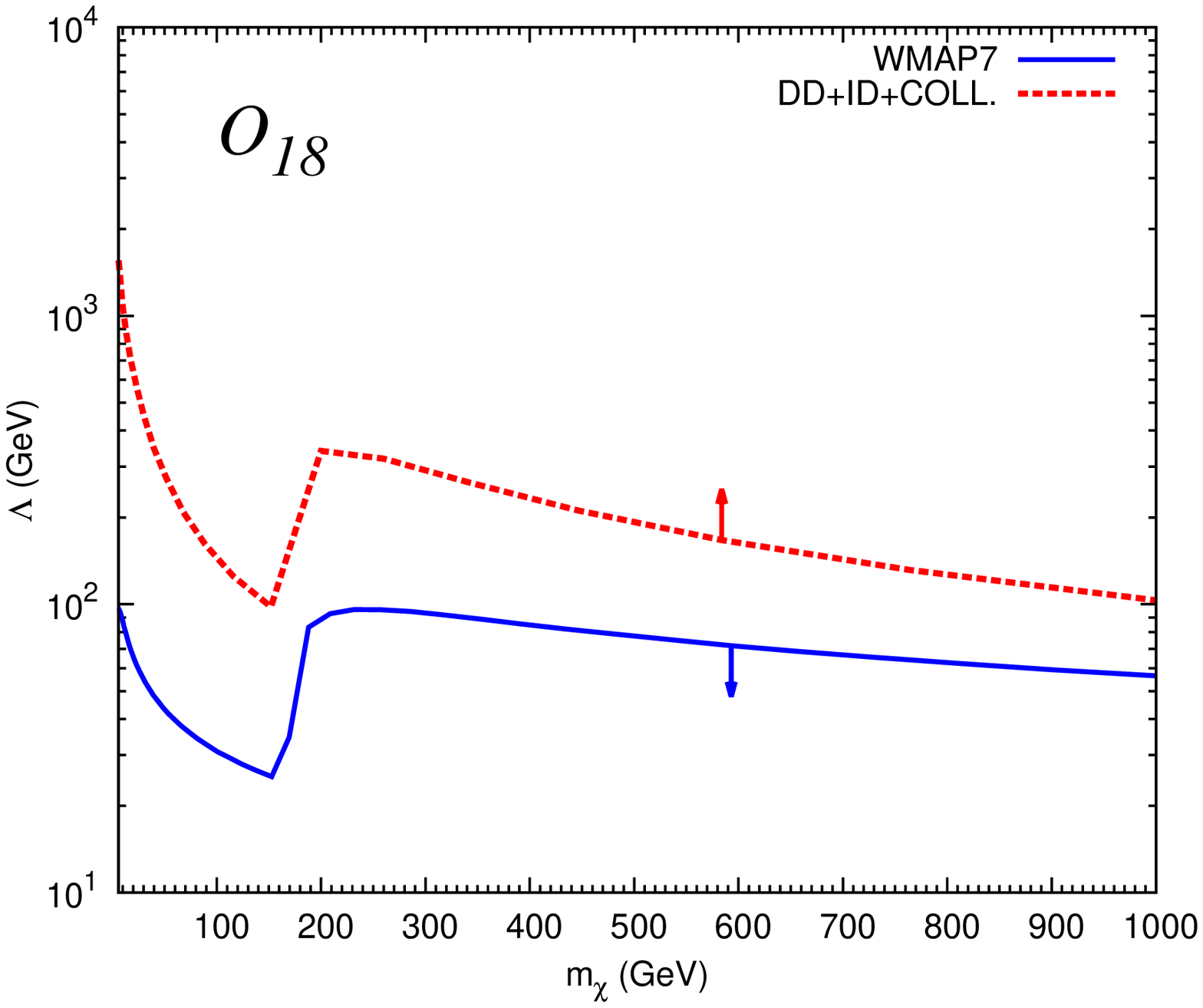}
\includegraphics[width=2.5in]{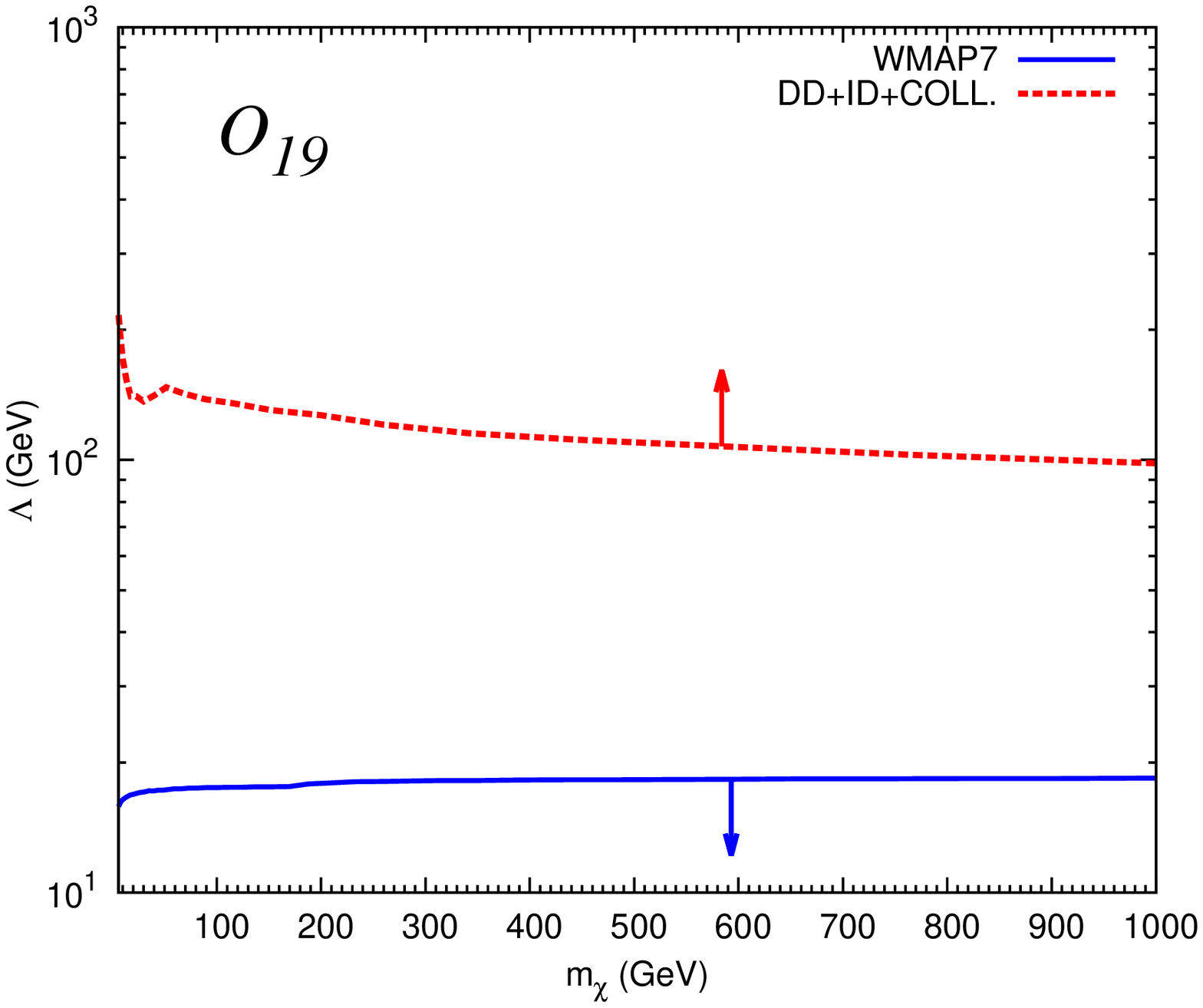}
\includegraphics[width=2.5in]{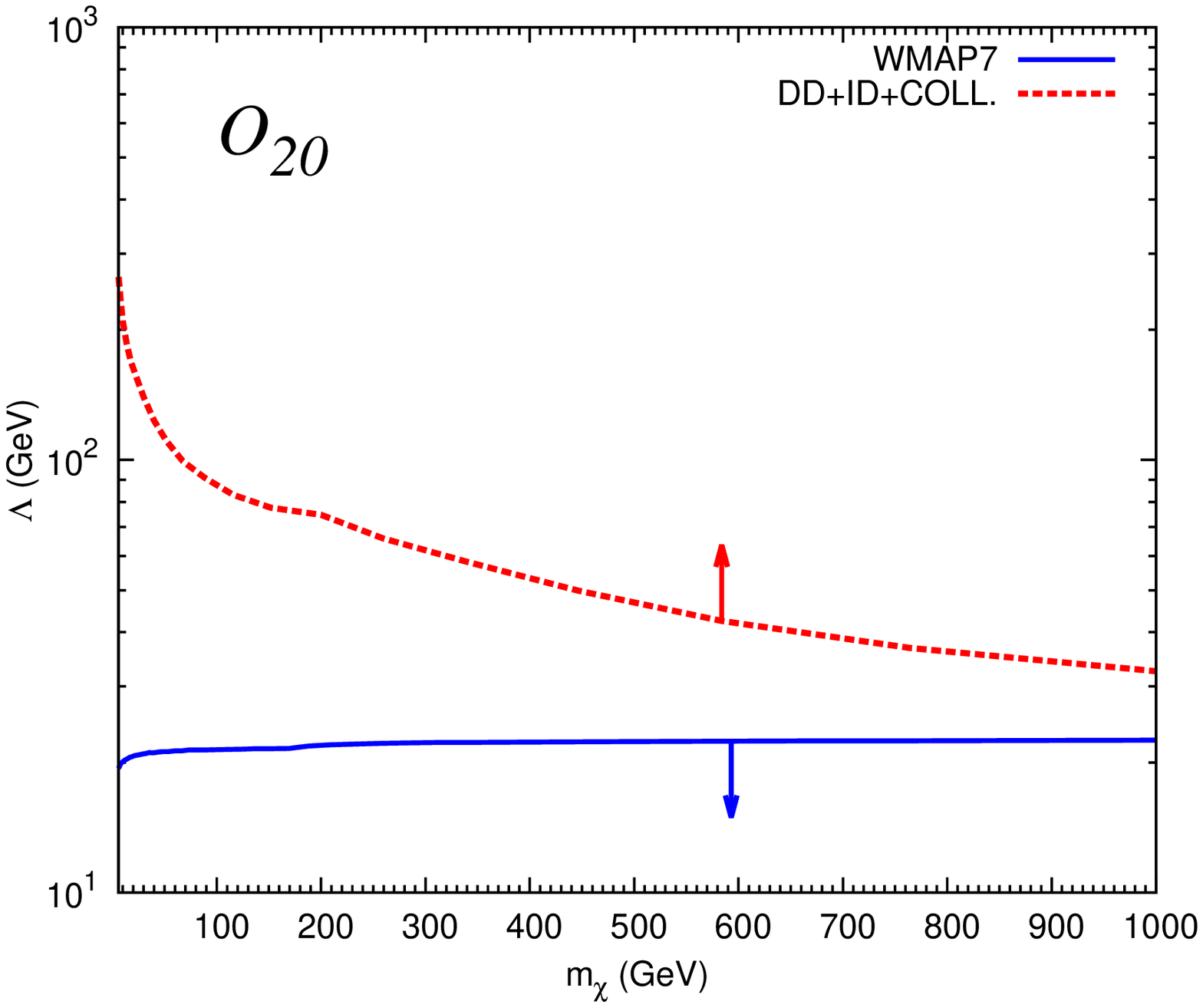}
\caption{\small \label{combined-O15-O20}
The combined analysis for $O_{15}$, $O_{16}$, $O_{17}$, $O_{18}$, $O_{19}$ and 
$O_{20}$.  
In each panel, the WMAP7 data
requires the area below the blue curve (indicated by the blue arrow)
while all the other data requires the area above the red curve 
(indicated by the red arrow).  The allowed region is shaded for $O_{16}$. 
}
\end{figure}

There is another data set on the gamma rays emitted from 
Dwarf spheroidal satellite galaxies (dSphs) of the Milky Way
collected by the $Fermi$-LAT Collaboration \cite{dsphs}.  They
derived 95\% C.L. limits on the WIMP annihilation cross sections
for a number of channels.  The upper limits on the 
annihilation cross sections for the most stringent channel $b\bar b$ 
are $(1.7 - 68) \times 10^{-26}\, {\rm cm}^3 \,{\rm s}^{-1}$ for 
DM mass from $10-1000$ GeV.  If we convert these limits to the
limits on $\Lambda$ for the operator $O_1$ with the $b\bar b$ final state
only, the limits are $0.5-2$ TeV for DM mass from $10-1000$ GeV.
Including the other channels, $\tau^+ \tau^-$, $\mu^+ \mu^-$, $W^+W^-$, 
would only lower the limits of $\Lambda$ mildly within a factor of 2.  
If we compare the dSphs limits to the limits in our summary figures 
\ref{combined-O1-O6} to \ref{combined-O15-O20}
for $O_1$, the dSphs limits are much less stringent than ours.
Therefore, including the dSphs data would not affect our current 
results significantly.

\section{Indirect Detection: Antiproton flux}

\begin{figure}[t!]
\centering
\includegraphics[height=2.5in,width=3.0in]{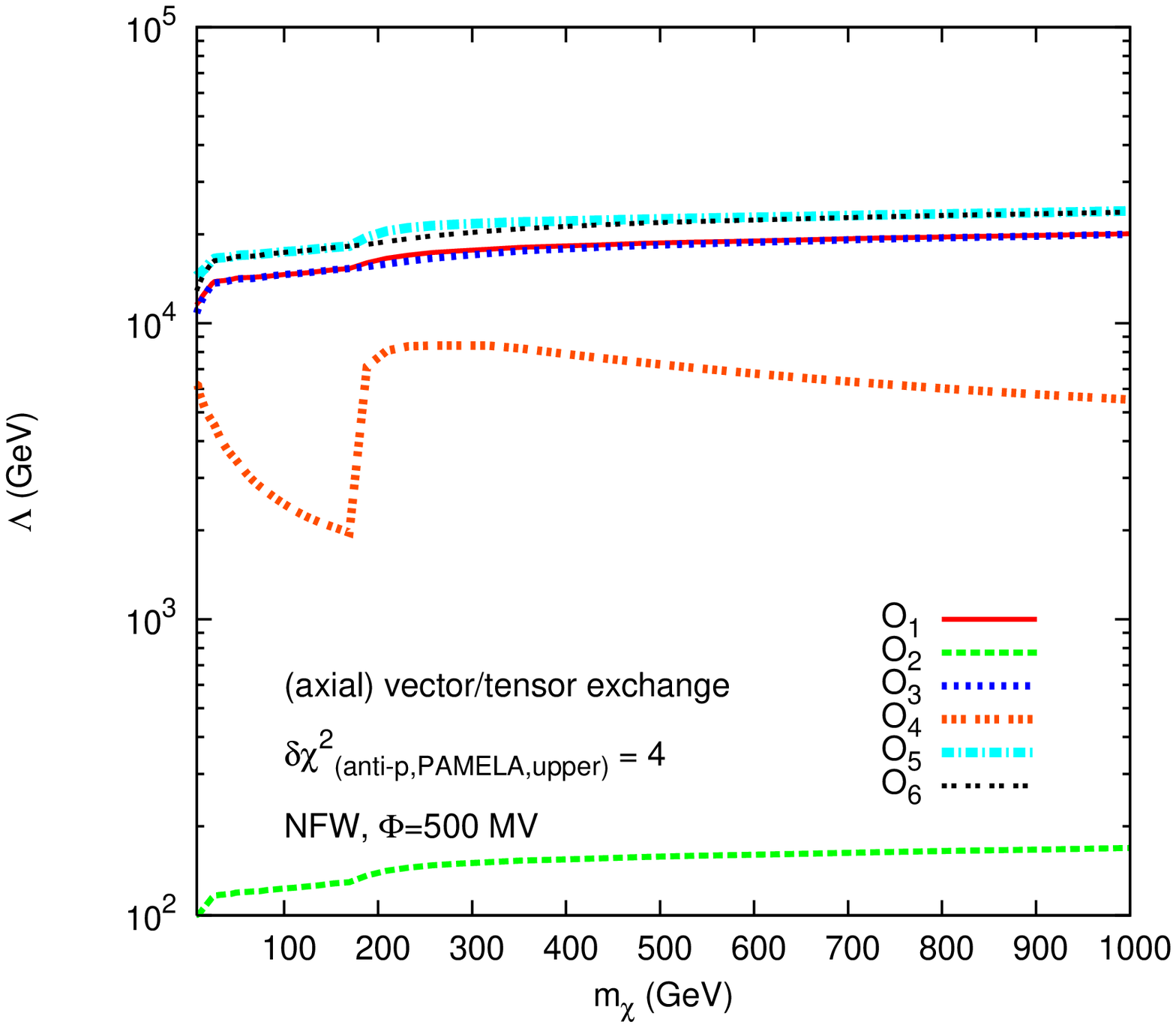}
\includegraphics[height=2.5in,width=3.0in]{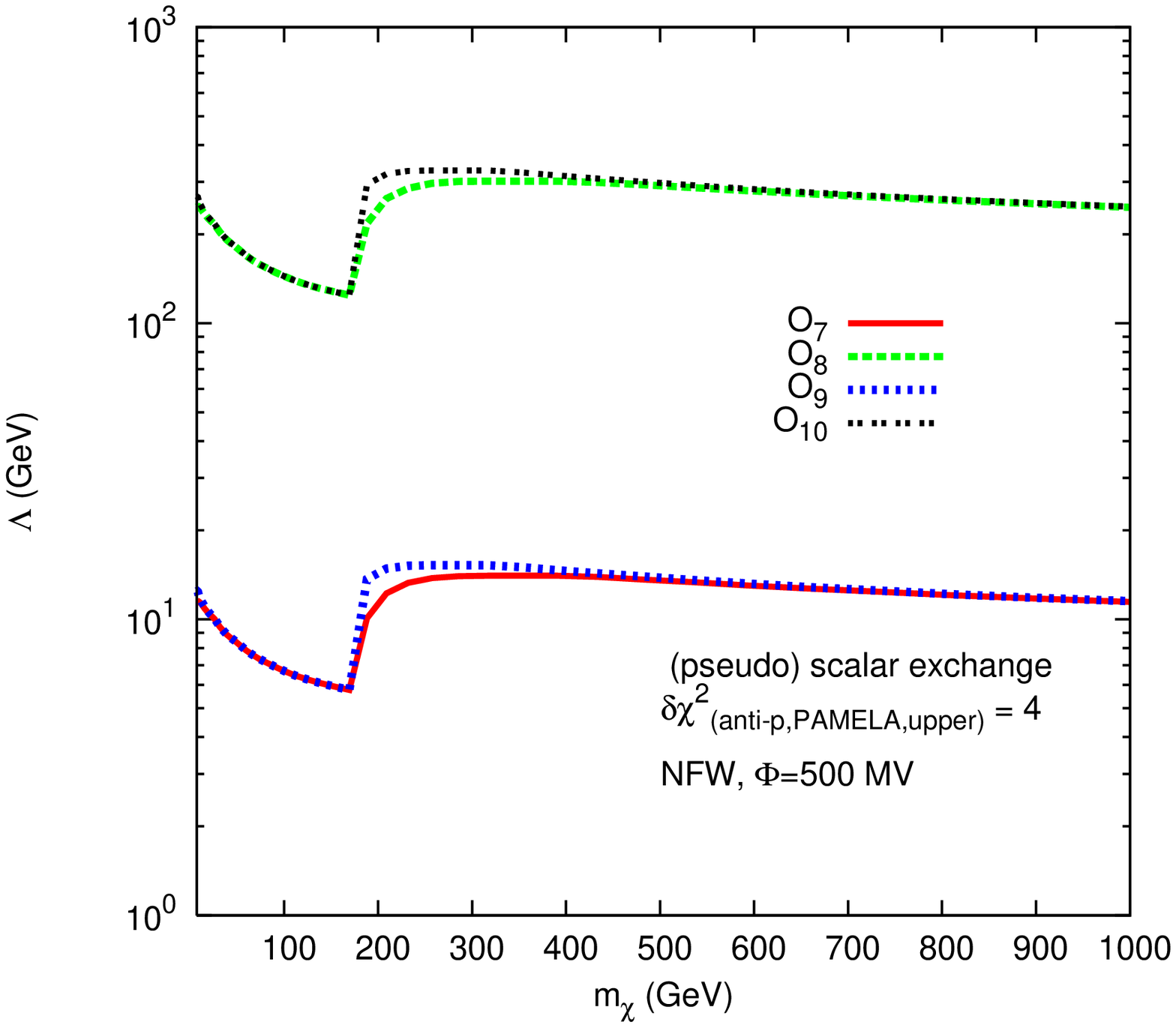}
\includegraphics[height=2.5in,width=3.0in]{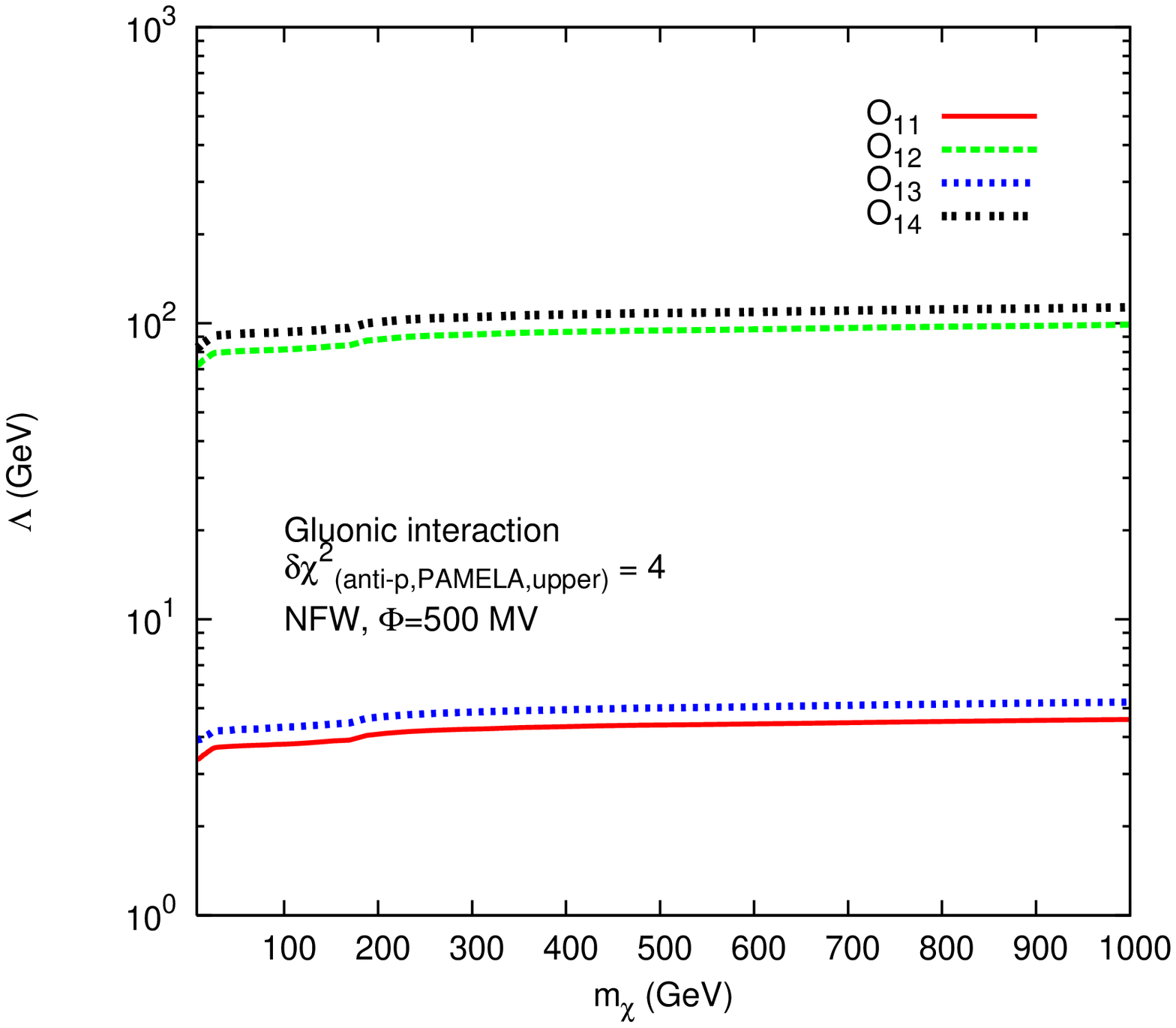}
\includegraphics[height=2.5in,width=3.0in]{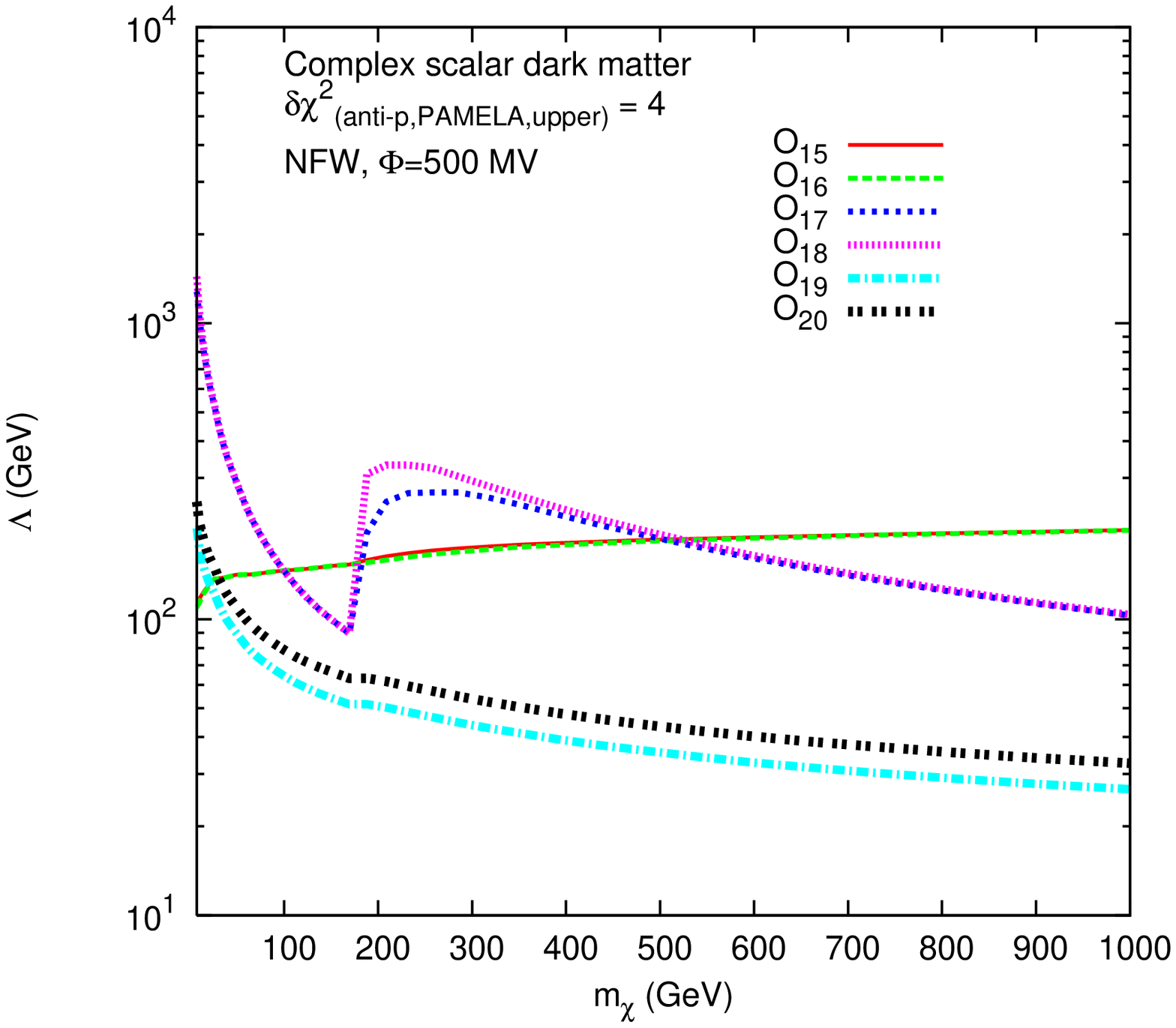}
\caption{\small \label{pb}
The $2\sigma$ lower limits on $\Lambda$ for each operator due to
the antiproton flux data of PAMELA \cite{pamela-p}.
}
\end{figure}

Annihilation of dark matter particles 
may give rise to large enough signals of antimatter,
such as positron and antiproton,
that can be identified by a number of antimatter search experiments. 
The most recent ones come from PAMELA \cite{pamela-e,pamela-p}, 
which showed a spectacular rise in the positron spectrum but an expected 
spectrum for antiproton 
compared with their backgrounds. 
The rise in the positron spectrum may be due to nearby pulsars 
or dark matter annihilation or decays.
If it is really due to dark matter annihilation, the dark matter
would have very strange properties, because it only gives positrons in 
the final products but not antiprotons. Here we adopt a conservative 
approach. We use the observed antiproton spectrum as a constraint
on the annihilation products in $\chi\overline{\chi}$ annihilation.

The analysis performed here is similar to that of gamma ray given in the
previous section. We modified DarkSUSY \cite{darksusy} 
for our effective DM interactions to 
generate the antiproton-flux spectrum $d N_{\bar p} / d T_{\bar p}$
(here conventionally the kinetic energy $T$ is used.)
The source term for solving the diffusion equation to obtain
the antiproton spectrum is given by
\begin{equation}
Q_{\rm ann} = \eta \left( \frac{\rho_{\rm CDM} }{M_{\rm CDM}} \right )^2 
\, \sum \langle \sigma v \rangle_{\bar p} \, \frac{d N_{\bar p}}{ d T_{\bar p} }
 \;,
\end{equation}
where $\eta =1/2\;(1/4) $ for (non-)identical initial state.
This source term is then fed into GALPROP with the running 
parameters of the best fit model \cite{bestfit}.  
The same NFW profile is employed as in the previous gamma-ray case.

The data set in this analysis comes from PAMELA in Ref.~\cite{pamela-p}.
We construct the chi-square using the 46 data points from PAMELA
antiproton flux and the ratio $\bar p/p$. The data point at the 
lowest energy is ignored because it does not have a central value.
We included the solar modulation effect because it is important for the 
data points of the low-energy region. We used a modulation of $500$ MV.

We found that the background estimation of $\bar p/p$ using the 
best-fit model parameters of the GALPROP (which only took into account 
the isotopic ratios B/C, Be$^{10}$/Be$^{9}$, Oxygen, and Carbon but not the 
$\bar p/p$ ratio) fits well to the data points.
We, therefore, calculate the $2\sigma$ limit on each scale $\Lambda_i$ 
based on the fact that the independent background estimation agrees
well with the data points, by a chi-square difference of 
$\Delta \chi^2 \equiv \chi^2 - {\rm min}(\chi^2_{\rm bkgd}) = 4$ 
($2\sigma$).  
We show the resulting limits for each operator in Fig.~\ref{pb}.
We note that the limits are both qualitatively and quantitatively similar to 
those obtained in the gamma-ray case.

In principle, one can use the PAMELA positron spectrum to constrain
the interactions. However, the uprising $e^+$ spectrum observed could
be an indication of DM annihilation if there are no other known
sources. In order to fit the $e^+$ spectrum the size of annihilation
cross section $\sigma \cdot v \sim 10^{-24} - 10^{-23} \, {\rm cm}^3
\; {\rm s}^{-1}$, the range of which depends on the dark matter mass. 
It corresponds to $\Lambda_1 = 0.84$ TeV for the operator $O_1$ when
$\sigma \cdot v = 5 \times 10^{-24}\,{\rm cm}^3 \, {\rm s}^{-1}$ and
$m_\chi = 200$ GeV. 
Given this is the fitted value, the limit obtained would be slightly worse 
than that.  It is obvious that this is negligible compared with the
limit from antiproton data (Fig.~\ref{pb}). Therefore, even including 
the $e^+$ data would not improve our results in a significant way.

\section{Combined Analysis}

In this section, we do a combined chi-square analysis from all the 
experimental data sets on each effective operator.  Note that the relic density
from WMAP7 constrains $\Lambda$ from above, while all the other 
experiments constrain $\Lambda$ from below.  Therefore, we combine
the chi-squares from (i) direct detection, (ii) collider, (iii) 
gamma-ray, and (iv) antiproton:
\begin{equation}
\chi^2({\rm total})  = \chi^2({\rm direct}) + \chi^2({\rm collider}) 
     + \chi^2 ({\rm gamma}) + \chi^2({\rm antiproton}) \;\; .
\end{equation}
We vary the input parameter $\Lambda$ until the increase in chi-square 
is 4 units from the minimum value, i.e.,
\begin{equation}
\Delta \chi^2 \equiv \chi^2({\rm total}) - \chi^2 ({\rm total})_{\rm min} = 4\;.
\end{equation}
The limit on $\Lambda$ thus obtained is a $2\sigma$ lower limit. 
Together with the upper limit due to the WMAP7 data, we show the results
for all the operators in 
Figs.~\ref{combined-O1-O6}, \ref{combined-O7-O10}, \ref{combined-O11-O14}, 
and \ref{combined-O15-O20}.  
For each operator there are
two curves: one from WMAP7 bounded from above and one from all other 
experimental data sets bounded from below. We indicate the allowed
region by an arrow for each curve.  
Except for operators $O_2$, $O_9$, and $O_{16}$,
the two arrows in each panel are pointing away from each other, and therefore
no region is allowed for all other operators. 
The working assumption here is that the effective interaction between
the DM and SM particles thermalized the DM particles in
equilibrium in the early Universe and later decoupled 
the DM particles
according to the standard Boltzmann equation,
and there are no other sources for the DM.  Under this assumption most of
the effective operators, except for $O_2$, $O_9$, and $O_{16}$, cannot 
give a smaller interaction constrained by direct detection, indirect 
detection and collider, while at the same time provide a larger 
interaction allowed by the
WMAP7 data.  This is the main result of this work. 

Very little parameter space is allowed for most of the operators
because we take the assumption
that {\it only} one operator exists for the early universe and for 
present day experiments.  If there are more than one operators exist 
at the same time, then the lower limit obtained by all the detection 
experiments (collider, indirect, and direct) will be stronger; 
on the other hand, the upper limit due to the relic density will be 
weaker.  Therefore, there would be more allowable regions.
Also note that if we further extends to larger masses for $m_\chi$ 
in almost all of the operators, there could be some allowed regions.

\section{Discussion and Conclusions}

Since we have performed the analysis for each experimental data set and the 
combined analysis, we can easily see which data set dominates for each
operator.  They are summarized as follows.
\begin{itemize}
\item
 dominated by direct detection: $O_7, O_{15}$
\item
 by collider: $O_2,O_9,O_{13},O_{14},O_{16}$
\item 
by indirect detection ($\bar p$ and $\gamma$-ray): 
 $O_1, O_3, O_4, O_5, O_6, O_8, O_{10}, O_{17},O_{18}, O_{20}$
\item
 by collider at low $m_\chi$ and direct detection at high $m_\chi$: $O_{11}$
\item
 by collider at low $m_\chi$ and indirect detection at high $m_\chi$: $O_{12}$
\item
 by indirect detection at low $m_\chi$ and direct detection at high 
$m_\chi$: $O_{19}$
\end{itemize}
The operators $O_2$, $O_9$, and $O_{16}$ that have allowed regions of 
parameter space are dominantly constrained by the collider data only.
This is because these operators in the nonrelativistic limit (e.g.
the present Universe) are highly suppressed and thus cannot contribute
at any significant level to direct and indirect detection.

In this work, we have exhausted all recent experimental data sets from
WMAP7, direct/indirect detection, gamma-ray flux, antiproton flux and
collider to obtain important constraints on the effective interactions
of the dark matter with the SM particles.  We found that almost all
effective operators, except for $O_2$, $O_9$, and $O_{16}$, cannot
give a smaller interaction that was constrained by the direct and
indirect detection as well as collider data, while at the same time
provide a large enough interaction required by the WMAP7 data for the
relic density to avoid the Universe over closed by the DM.  The result
has interesting implications to model buildings, especially those with
a heavy mediator between the dark sector and the SM sector. A lot of
possibilities shown in this work will not work if we allow at the same
time the new physics to give a thermal relic density and to be
consistent with the existing data from direct and indirect detections
as well as collider data.

\section*{Acknowledgments}
We thank Eiko Yu and Joel Heinrich for a discussion on treatment of data.
This work was supported in parts by the National Science Council of
Taiwan under Grant Nos. 99-2112-M-007-005-MY3 and
98-2112-M-001-014-MY3 as well as the
WCU program through the KOSEF funded by the MEST (R31-2008-000-10057-0). 
Y.S.T. is funded by the Welcome Programme of the Foundation for Polish Science.

\appendix

\section{Reductions to effective operators}
\subsection{Darkon Model}

The darkon model \cite{darkon} consists of a real-scalar boson $D$,
which is hidden from the SM interactions, except for a Higgs-portal
type interaction with the SM Higgs boson
\[
  {\cal L}_D = - \frac{\lambda_D}{4} D^4 - \frac{m_D^2}{2} D^2 - 
 \lambda D^2 H^\dagger H \;.
\]
After the Higgs field develops a VEV $v$, the interactions between the
physical Higgs boson and the darkon are given by
\[
{\cal D}_D = - \frac{\lambda_D}{4} D^4 - \frac{m_D^2 + \lambda v^2}{2} D^2
 - \frac{\lambda}{2} D^2 h^2 - \lambda v D^2 h \;,
\]
such that the interactions of the darkon $D$ proceed via the Higgs
boson.  We can write down the amplitude for $D D \to f \bar f$ as
\[
 {\cal L} = \frac{g m_f}{2 m_W} \bar f f \; \frac{1}{(2 m_D)^2 - m_h^2 } \;
 \lambda v D^2  \;.
\]
In the limit $m_h \gg m_D$, the amplitude becomes
\[
 {\cal L} = C \frac{m_f}{\Lambda^2}\;  ( \bar f f ) D^2  \;,
\]
where $C/\Lambda^2 = g \lambda v / (2 m_W m_h^2 )$.  This is very similar
to the operator $O_{17}$ in the case of a real scalar DM, with the 
explicit $m_f$ dependence.

\subsection{Higgs Portal model for fermionic DM} 
The hidden sector consists of a fermion $\chi$ as the DM and a scalar boson
$\phi$, which can mix with the SM Higgs field $H$:
\[
{\cal L} = \lambda_1 \phi \bar \chi \chi + \lambda_2 ( \phi^\dagger \phi ) 
(H^\dagger H) \;.
\]
The amplitude for $\bar \chi \chi \to f \bar f$, after boson mixing,
 is given by
\[
  {\cal L} \sim (\lambda_1 \bar \chi \chi ) \;\frac{1}{ (2 m_\chi)^2 - m_h^2 }
 \; \frac{ g m_f}{2 m_W} ( \bar f f ) \;,
\]
which becomes, in the limit $m_\chi \ll m_h$, 
\[
{\cal L} \sim C \frac{m_f }{\Lambda^3}\, (\bar \chi \chi) \, ( \bar f f) \;,
\]
which is exactly the same as $O_7$ with the explicit dependence on $m_f$.

\subsection{$Z$-$Z'$ portal model}
The hidden sector consists of a fermionic DM $\chi$ and a gauge boson
$Z'$, which then mixes with the SM $Z$ boson via the kinetic mixing or
Stueckelberg-type mixing \cite{ZZ}. The hidden sector does not have
any SM interactions originally, but via the mixing with the SM $Z$
boson, some level of interactions with the SM particles is
possible. Suppose the interactions of the DM $\chi$ and the SM
fermions are given by, after mixing and integrating out the heavy $Z'$
boson,
\[
{\cal L} = \bar \chi \gamma^\mu ( g_v^\chi - g_a^\chi \gamma^5 ) \chi \, Z_\mu
 + \bar f \gamma^\mu (  g_v^f - g_a^f \gamma^5 ) f \, Z_\mu \;,
\]
then the amplitude for the process $\bar \chi \chi \to f \bar f$ 
can be written as
\[
{\cal L} = \bar \chi \gamma^\mu ( g_v^\chi - g_a^\chi \gamma^5 ) \chi \;
 \frac{1}{ (2m_\chi)^2 - m_Z^2 } \; 
 \bar f  \gamma_\mu (  g_v^f - g_a^f \gamma^5 ) f \;.
\]
In the limit of $m_\chi \ll m_Z$, the amplitude becomes
\begin{eqnarray}
 {\cal L} &\sim&   \frac{g_v^\chi g_v^f}{m_Z^2} \;
   (\bar \chi \gamma^\mu \chi )\, (\bar f \gamma_\mu f ) 
  + \frac{g_v^\chi g_a^f}{m_Z^2} \;
   (\bar \chi \gamma^\mu \chi )\, (\bar f \gamma_\mu \gamma^5 f )  \nonumber \\
 && + \frac{g_a^\chi g_v^f}{m_Z^2} \;
   (\bar \chi \gamma^\mu \gamma^5 \chi )\, (\bar f \gamma_\mu f ) 
   + \frac{g_a^\chi g_a^f}{m_Z^2} \;
   (\bar \chi \gamma^\mu \gamma^5 \chi )\, (\bar f \gamma_\mu \gamma^5 f ) 
 \nonumber
\end{eqnarray}
The couplings of the $Z$ boson to the SM fermions are all of the same
order and the coupling to $\chi$ is unknown, we can, to a crude
approximation, take the overall couplings to be similar. Thus, we
arrive at the operators $O_1$ to $O_4$.  Note that the coefficients
for different SM fermions may as well be different, though they are
highly model dependent, they should not differ from one another too
much.  Again, we take the crude approximation that they are similar
and thus reduce a large number of parameters.

A more thorough discussion on deriving various effective operators
from various particle exchanges can be found in Ref.~\cite{chacko}.

\section{Annihilation cross section formulas}

Here we list all the differential cross section formulas $d\sigma_i / dz$
for the dark matter annihilation of the operators $O_i \; (i$ =1 to 20).  
\begin{eqnarray}
\frac{d\sigma_1}{dz} &=& \frac{1}{\Lambda_1^4} \frac{N_C}{16 \pi s}
\frac{\beta_f}{\beta_\chi} \left[ u_m^2 + t_m^2 + 2s (m_\chi^2 +m_f^2)\right ]
                     \;, \\
\frac{d\sigma_2}{dz} &=& \frac{1}{\Lambda_2^4} \frac{N_C}{16 \pi s}
   \frac{\beta_f}{\beta_\chi} \left[ u_m^2 + t_m^2 + 2s (m_f^2 - m_\chi^2)
          - 8 m_f^2 m_\chi^2 \right ] \;,  \\
\frac{d\sigma_3}{dz} &=& \frac{1}{\Lambda_2^4} \frac{N_C}{16 \pi s}
   \frac{\beta_f}{\beta_\chi} \left[ u_m^2 + t_m^2 + 2s (m_\chi^2 - m_f^2)
          - 8 m_f^2 m_\chi^2 \right ] \;,  \\ 
\frac{d\sigma_4}{dz} &=& \frac{1}{\Lambda_4^4} \frac{N_C}{16 \pi s}
   \frac{\beta_f}{\beta_\chi} \left[ u_m^2 + t_m^2 - 2s (m_\chi^2 + m_f^2)
          +16 m_f^2 m_\chi^2 \right ] \;,  \\  
\frac{d\sigma_5}{dz} &=&  \frac{1}{\Lambda_5^4} \frac{N_C}{4 \pi s}
  \frac{\beta_f}{\beta_\chi} \left[2(u_m^2 + t_m^2) + 2s (m_\chi^2 +m_f^2) 
   + 8 m_f^2 m_\chi^2 - s^2\right ] \;, \\
\frac{d\sigma_6}{dz} &=&  \frac{1}{\Lambda_6^4} \frac{N_C}{4 \pi s}
  \frac{\beta_f}{\beta_\chi} \left[2(u_m^2 + t_m^2) + 2s (m_\chi^2 +m_f^2) 
   -16 m_f^2 m_\chi^2 - s^2\right ] \;, 
\end{eqnarray}
\begin{eqnarray}
\frac{d\sigma_7}{dz} &=&  \frac{m_f^2}{\Lambda_7^6} \frac{N_C}{32 \pi} 
  s \beta_\chi \beta_f^3 \;, \\
\frac{d\sigma_8}{dz} &=&   \frac{m_f^2}{\Lambda_8^6} \frac{N_C}{32 \pi} 
   \frac{s \beta_f^3 }{\beta_\chi} \;, \\
\frac{d\sigma_9}{dz} &=&  \frac{m_f^2}{\Lambda_9^6} \frac{N_C}{32 \pi} 
  s \beta_\chi \beta_f \;, \\
\frac{d\sigma_{10}}{dz} &=& \frac{m_f^2}{\Lambda_{10}^6} \frac{N_C}{32 \pi} 
   \frac{s \beta_f }{\beta_\chi} \;, 
\end{eqnarray}   
\begin{eqnarray}
\frac{d\sigma_{11}}{dz} &=&   \frac{\alpha_s^2}{\Lambda_{11}^6} \frac{1}{4608 \pi^3} 
  s^2 \beta_\chi \;, \\
\frac{d\sigma_{12}}{dz} &=&   \frac{\alpha_s^2}{\Lambda_{12}^6} \frac{1}{4608 \pi^3} 
  \frac{s^2}{\beta_\chi} \;, \\
\frac{d\sigma_{13}}{dz} &=&  \frac{\alpha_s^2}{\Lambda_{13}^6} \frac{1}{2048 \pi^3} 
  s^2 \beta_\chi \;, \\
\frac{d\sigma_{14}}{dz} &=&  \frac{\alpha_s^2}{\Lambda_{14}^6} \frac{1}{2048 \pi^3} 
  \frac{s^2}{\beta_\chi} \;, \\
  \end{eqnarray}
\begin{eqnarray}
\frac{d\sigma_{15}}{dz} &=&  \frac{1}{\Lambda_{15}^4}\frac{N_C}{4\pi s}
   \frac{\beta_f}{\beta_\chi} ( u t - m_f^2(u+t) - m_\chi^4 +m_f^4 ) \;, \\
\frac{d\sigma_{16}}{dz} &=&  \frac{1}{\Lambda_{16}^4}\frac{N_C}{4\pi s}
   \frac{\beta_f}{\beta_\chi} \left( u t - (m_\chi^2 - m_f^2 )^2 \right) \;, \\
\frac{d\sigma_{17}}{dz} &=&  \frac{m_f^2}{\Lambda_{17}^4}\frac{N_C}{16\pi}
   \frac{\beta_f^3}{\beta_\chi} \;, \\
\frac{d\sigma_{18}}{dz} &=&  \frac{m_f^2}{\Lambda_{18}^4}\frac{N_C}{16\pi}
   \frac{\beta_f}{\beta_\chi} \;, \\ 
\frac{d\sigma_{19}}{dz} &=&   \frac{\alpha_s^2}{\Lambda_{19}^4} \frac{1}{2304 \pi^3} 
  \frac{s}{\beta_\chi} \;, \\
\frac{d\sigma_{20}}{dz} &=&  \frac{\alpha_s^2}{\Lambda_{20}^4} \frac{1}{1024 \pi^3} 
  \frac{s}{\beta_\chi} \;,
\end{eqnarray}
where $s$, $t$ and $u$ are the usual Mandelstam variables,
$z$ is the cosine of scattering angle, $u_m = u - m_\chi^2 -m_f^2$, $t_m = t - m_\chi^2 - m_f^2$, 
$\beta_\chi = (1 - 4 m_\chi^2/s)^{1/2}$, $\beta_f = (1-4 m_f^2/s)^{1/2}$,
and $N_C$ is the color factor (3 for quarks and 1 for leptons).  
We have absorbed the coefficients $C_i$ into $\Lambda_i$ in these formulas. 
The nonrelativistic limits of $\sigma_i v = \sigma_i \cdot (2 \beta_\chi)$ are listed at the last column of Table \ref{prop}.

\end{document}